# Generalized selfish bin packing

György Dósa*     Leah Epstein†

**Abstract**

Standard bin packing is the problem of partitioning a set of items with positive sizes no larger than 1 into a minimum number of subsets (called bins) each having a total size of at most 1. In bin packing games, an item has a positive weight, and given a valid packing or partition of the items, each item has a cost or a payoff associated with it. We study a class of bin packing games where the payoff of an item is the ratio between its weight and the total weight of items packed with it, that is, the cost sharing is based linearly on the weights of items. We study several types of pure Nash equilibria: standard Nash equilibria, strong equilibria, strictly Pareto optimal equilibria, and weakly Pareto optimal equilibria. We show that any game of this class admits all these types of equilibria. We study the (asymptotic) prices of anarchy and stability (PoA and PoS) of the problem with respect to these four types of equilibria, for the two cases of general weights and of unit weights. We show that while the case of general weights is strongly related to the well-known FIRST FIT algorithm, and all the four PoA values are equal to 1.7, this is not true for unit weights. In particular, we show that all of them are strictly below 1.7, the strong PoA is equal to approximately 1.691 (another well-known number in bin packing) while the strictly Pareto optimal PoA is much lower. We show that all the PoS values are equal to 1, except for those of strong equilibria, which is equal to 1.7 for general weights, and to approximately 1.611824 for unit weights. This last value is not known to be the (asymptotic) approximation ratio of any well-known algorithm for bin packing. Finally, we study convergence to equilibria.

## 1 Introduction

We study a class of games which originate in bin packing [28, 9, 11, 10], a well-known and basic combinatorial optimization problem. In this problem, a set of items, each of size in $(0, 1]$, is given. The goal is to partition (or pack) the items into a minimum number of subsets, called bins. Each bin has unit capacity, and the load of a bin is defined to be the total size of items packed into it. Thus, the goal is to find a packing of the items into a minimum number of bins, such that the load of each bin is at most 1. The problem is NP-hard, and research has concentrated on the study and development of approximation algorithms which design nearly optimal solutions. A bin packing algorithm is called *online* if it receives the items one by one, and must assign each item to a bin immediately and irrevocably without any information on subsequent items. If the input is given as a set then the problem is called *offline*, in which case an algorithm is expected to use polynomial time.

Consider a minimization problem and a set $X$ of instances for it. Assume that the cost associated with every feasible solution for every $I \in X$ is a positive integer. Let $A$ be a set of solutions which contains a solution for each $I \in X$, and let $cost(A(I))$ (we simply use $cost(A)$ if $I$ is clear from the context) denote the cost of such a solution (if there are multiple solutions for $I$ then one of them is chosen). The set $A$ is a class of solutions with specific properties which could possibly be a set of outputs of a given algorithm. Let OPT denote an optimal algorithm and let $\text{OPT}(I)$ (or OPT) denote the minimum cost for the input $I$, that is, the cost of an optimal algorithm on this input. We define the (asymptotic) *approximation ratio* of $A$ to be:

$$\mathcal{R}(A) = \limsup_{N \to \infty} \max_I \{cost(A(I))/\text{OPT}(I) \,|\, \text{OPT}(I) = N\}.$$

---

*Department of Mathematics, University of Pannonia, Veszprém, Hungary, dosagy@almos.vein.hu.
†Department of Mathematics, University of Haifa, 31905 Haifa, Israel. lea@math.haifa.ac.il.

This term is used for any type of algorithm or set $A$ (of solutions). In the literature, this term is often used for polynomial time approximation algorithms, and the term (asymptotic) competitive ratio is usually used for online algorithms (but has the same meaning). In what follows we only deal with asymptotic approximation ratios and omit the word *asymptotic*.

Following a recent trend we study bin packing from the point of view of algorithmic game theory. We define the game theoretical concepts required for the definition of the bin packing game. A *strategic game* consists of a finite set of players, and a finite, non-empty set of strategies (or actions) that the set of players can perform. Each player has to choose a strategy (possibly independently from other players). Each player has a payoff associated with each one of the possible situations or outcomes (sets of strategies of all players, containing one strategy for each player). Each outcome also has a social cost associated with it.

A *Nash equilibrium* (NE) [36] is a kind of solution concept of a game with at least two players, where no player can gain anything (i.e., decrease its payoff) by changing only its own strategy unilaterally. That is, if each player has chosen a strategy and no player can benefit by changing its strategy while the other players keep their unchanged, then the current set of strategy choices and the corresponding payoffs result in a Nash equilibrium. We focus on pure Nash equilibria, where the actions of each players is chosen in a deterministic way. We are interested in the approximation ratio of specific sets of equilibria. A solution which is a Nash equilibrium is not necessarily socially optimal.

A much stronger concept of a stable solution is a strong equilibrium (SNE) [3, 39, 27, 1, 17, 18], which is a NE where not only single players cannot benefit from changing their strategy but no non-empty subset of players can form a coalition, where a coalition means that a subset of players change their strategies simultaneously, all gaining from the change. The grand coalition is defined to be a coalition composed of the entire set of players. A solution is called *weakly Pareto optimal* if there is not alternative solution to which the grand coalition can deviate simultaneously and every player benefits from it. A solution is called *strictly Pareto optimal* if there is not alternative solution to which the grand coalition can deviate simultaneously, at least one player benefits from it, and no player has a larger cost as a result. The last two concepts are borrowed from welfare economics. The two requirements, that a solution is both (strictly or weakly) Pareto optimal and a NE results in two additional kinds of NE, Strictly Pareto optimal NE(SPO-NE) and Weakly Pareto optimal NE(WPO-NE) [14, 8, 16, 4]. By these definitions, every WPO-NE is a NE, every SPO-NE is a WPO-NE, and every SNE is a WPO-NE. Strictly Pareto optimal points are of particular interest in economics, as stated in a textbook: "The concept of Pareto optimality originated in the economics equilibrium and welfare theories at the beginning of the past century. The main idea of this concept is that society is enjoying a maximum ophelimity when no one can be made better off without making someone else worse off" [32]. Even though these concepts are stronger than NE, still for many problems a solution which is a SNE, a SPO-NE, or a WPO-NE is not necessarily socially optimal.

The price of anarchy (PoA) (see [30, 38, 37, 12, 33]) of a game $G$ is the ratio between the maximum social cost of any NE, and the minimum social cost of any situation. Similarly, we define the strong price of anarchy (SPoA) [1, 22, 15] (as the ratio between the maximum social cost of any SNE and the minimum social cost of any solution), the strictly Pareto optimal PoA(SPO-PoA) and the weakly Pareto optimal PoA(WPO-PoA). The price of stability (PoS) (see [2]), the strong price of stability (SPoS), the strictly Pareto optimal PoS (SPO-PoS), and the weakly Pareto optimal PoS (WPO-PoS) are defined analogously, taking into account the equilibria of each kind with the minimum social cost. Note that in order to use these definitions, it is required to show that a game $G$ admits the relevant kind of equilibrium, that is, to consider the PoA and the PoS, one has to prove a NE exists, to use the definitions of SPoA and SPoS, one has to prove first that $G$ admits a SNE, and similarly for the SPO-PoA, SPO-PoS, a SPO-NE must exist, and for WPO-PoAand WPO-PoS, a WPO-NE must exist.

The bin packing problem, where for each input and packing every item has a payoff associated with it, can be seen as a class of games in which the social cost for an input $I$ is simply OPT($I$). If indeed every such game admits a SNE and a SPO-NE (and thus also a NE), then each one of the measures is the approximation ratio of a class of solutions: the PoA is the approximation ratio of the set of worst NE (where every game is represented by one worst NE), the PoS is the approximation ratio of the set of best equilibria, the SPoA is the approximation ratio of the set of worst strong equilibria, and the SPoS is the approximation ratio of the set of best strong equilibria, the WPO-PoA, SPO-PoA, WPO-PoS, and SPO-PoS, are the approximation ratios of the sets of set of worst WPO-NE, worst SPO-NE, best WPO-NE and best SPO-NE, respectively.

We now define a game based on bin packing problems. Let $I = \{1, 2, \ldots, n\}$ be a set of items. For an item $i \in I$, we let $s_i \in (0, 1]$ denote its size and $w_i > 0$ denotes its weight. Thus, for a set of items $B \subseteq I$ packed into a bin, we say that $B$ is a valid bin if $\sum_{i \in B} s_i \leq 1$. For $C \subseteq I$, we define the size of $C$ by $s(C) = \sum_{i \in C} s_i$ and the weight of $C$ by $w(C) = \sum_{i \in C} w_i$ (where $w(C) > 0$ if $C \neq \emptyset$). The strategy of a player is the bin in which it is packed. Changing the strategy means that it moves to be packed in a different (existing or new) bin. Such a deviation is possible only if the empty space in the bin (into which it moves) is sufficient, that is, that the deviation results in it being packed into a valid bin. We always let the cost (or payoff) of an item which is not packed into a valid bin be infinite. For an item packed into a valid bin, we define its cost to be its proportional share of the bin according to the weights. That is, if an item of weight $w$ is packed into a bin where the total weight of packed item is $W$, then its cost is $w/W$[1]. Note that the cost of an item cannot be smaller than its weight. Similarly (and more generally) for deviations of coalitions, a deviation is possible if in the resulting solution all bins are valid.

Bin packing in general, and more specifically the bin packing game have a number of applications [6, 17, 18]. The bin packing game with proportional weights ($w_i = s_i$) was introduced by Bilò [6], who was the first to study the bin packing problem from a game theoretic perspective. He proved that every game in this class admits a NE and provided bounds on the PoA, a lower bound of $\frac{8}{5}$ and an upper bound of $\frac{5}{3}$. He also proved that any bin packing game converges to a NE after a finite (but possibly exponentially long) sequence of selfish improving steps, starting from any initial configuration of the items. This last result implies that the PoS is equal to 1. The time of convergence was also studied in [34, 35]. The quality of NE solutions was further investigated in [17], where nearly tight bounds for the PoA were given; an upper bound of 1.6428 and a lower bound of 1.6416. Interestingly, the PoA is not equal to the approximation ratio of any natural algorithm for bin packing. Yu and Zhang [40] later designed a polynomial time algorithm to compute a packing that is a NE. The SPoA and SPoS were also analyzed in [17], and it was shown that these two measures are equal. Moreover, it was shown that all SNE are outputs of an exponential time algorithm which at each time picks a subset of items of largest total size (of at most 1) to be packed into a bin. This algorithm was previously studied by Graham [25] and by Caprara and Pferschy [7]. In the paper [18], the exact SPoA was determined, and it was shown that its value is approximately 1.6067. In the same article, the parametric problem where the size of every item is upper bounded by a parameter is studied. A two dimensional class of bin packing games was considered in [21].

The case of unit weights ($w_i = 1$, which is equivalent to the case of equal weights) was studied by Han et al. [26]. They show that for this case the process of convergence into a NE is much faster, and requires $O(n^2)$ steps (and thus a NE always exists). As a result, since an item never migrates to an empty bin, any polynomial time algorithm can be adapted to output a NE solution without increasing the approximation ratio, and therefore there exists a polynomial time approximation scheme (using [13]) and even a fully polynomial approximation scheme (using [29]) which give NE solutions as outputs. Additionally, the greedy algorithm NEXT FIT INCREASING (NFI), which sorts the items by non-decreasing size and applies NEXT FIT (NF), that is, uses one active bin at a time, and replaces it with a new active bin if an item does not fit is studied in [26]. The authors show that this algorithm creates a NE whose approximation ratio is known to be approximately 1.69103 [23]. In fact, this is the approximation ratio of a number of algorithms, such as NFD (which sorts the items by non-increasing size and applies NF), and the limit of the sequence of approximation ratios of a class of online algorithms, called the HARMONIC ALGORITHMS [31], which partition items into classes according to size and pack each class independently. Finally, it is shown that any NE packing is the output of a run of FIRST FIT (FF) which always packs an item into the lowest index bin where it can fit [28], and thus the PoA is at most 1.7. An example is provided where an optimal solution uses 10 bins, while a NE solution uses 17 bins.

---

[1] We do not allow zero weights as this leads to degenerate instances where an item is oblivious to its packing.

**Our results.** We show that all kinds of Nash equilibria defined above (NE, SNE, SPO-NE, and WPO-NE) exist for bin packing games in the general setting. We further show that given a packing, a process in which at each time one item can perform a move to another bin (where its cost is reduced) always converges for the case of general weights. Since it is never beneficial for an item to move into an empty bin, the number of bins in the resulting output cannot increase. The number of steps may be exponential for general weights, but for the case of unit weights we find an improved tight bound as a function of $n$. This last bound is $\Theta(n^{\frac{3}{2}})$ (see Theorem 7.1 for the exact function of $n$). Our results for the PoA, SPoA, SPO-PoA, WPO-PoA, PoS, SPoS, SPO-PoS, and WPO-PoS are given in Table 1. For arbitrary weights all prices of anarchy are 1.7, which equals to the approximation ratio of FF. All prices of stability are 1 except for the SPoS which is 1.7 as well. This resolves the general case, while the results for unit weights reveal some interesting properties. We encounter some bounds which are the approximation ratios of well-known algorithms, as well as some new bounds. All the prices of anarchy are strictly below 1.7, and the SPoS is equal to approximately 1.69103 (the approximation ratio of NFI). The SPoS in this case is equal, however, to approximately 1.611824, a new number in bin packing. We prove that the PoA and WPO-PoA are equal, and moreover, that they are just slightly lower than 1.7 (recall that for proportional weights, the PoA is below 1.65 [17]).

|         | Unit weights lower bound | Unit weights upper bound | Arbitrary weights lower and upper bound |
|---------|--------------------------|--------------------------|------------------------------------------|
| PoA     | 1.696646                 | 1.6993996                | 1.7                                      |
| PoS     | 1                        | 1                        | 1                                        |
| SPoA    | 1.69103                  | 1.69103                  | 1.7                                      |
| SPoS    | 1.611824                 | 1.611824                 | 1.7                                      |
| WPO-PoA | 1.696646                 | 1.6993996                | 1.7                                      |
| WPO-PoS | 1                        | 1                        | 1                                        |
| SPO-PoA | 1.61678                  | 1.628113                 | 1.7                                      |
| SPO-PoS | 1                        | 1                        | 1                                        |

Table 1: Results for prices of anarchy and stability.

## 2 Preliminaries

In this section we show the existence of all types of NE, discuss properties of NE packings, and state some relations between the different types of equilibria and relations between equilibria and optimal solutions.

It was shown by [26] that for unit weights, every packing $A$ there exists a packing $A'$ with at most the same number of bins such that $A'$ is a NE. In particular, this implies that for unit weights a NE always exists, there always exists an optimal solution which is a NE, and the PoS is 1. We can show that the PoS is 1 for every weight function, by generalizing the last result. This will also show that a NE always exists for arbitrary weights.

**Proposition 2.1** *Given an input $I$ with general weights, for every packing $A$ there exists a packing $A'$ such that $A'$ is a NE and $cost(A') \leq cost(A)$.*

**Proof.** For an input set $I$, recall that $n = |I|$. Let $\mathcal{B} = \{I' \subseteq I | s(I') \leq 1\}$, and let $\mathcal{W} = \{s(I') | I' \in \mathcal{B}\}$. That is, $\mathcal{W}$ is a set of non-negative numbers which represents all possible total weights of subsets of items of total size of at most 1, i.e., $\mathcal{W}$ contains every possible total weight of items that can be packed into a bin. Note that $\mathcal{W}$ is a set and not a multiset, that is, if multiple subsets have the same weight, then this weight only appears once in $\mathcal{W}$. Since $0 \in \mathcal{W}$ and we always have $w_1 \in \mathcal{W}$ (the first item has an arbitrary positive weight), we have $|\mathcal{W}| \geq 2$. Let $\Omega = \max_{x \in \mathcal{W}} x$ and $\omega = \min_{x_1, x_2 \in \mathcal{W}} |x_1 - x_2|$. Let $w_{\min} > 0$ denote the minimum weight of any item (then since $0 \in \mathcal{W}$, we have $w_{\min} \geq \omega$).

Consider a dynamic where as long as the packing is not a NE, one item is chosen to move from its bin to another bin where its cost becomes strictly smaller. We claim that after a sequence of at most $\frac{w(I)^2}{2\omega^2}$ such moves, the resulting packing is a NE. Since no item benefits from moving to an empty bin (no matter if it were packed in a dedicated bin or with at least one additional item), the number of bins does not increase. We define a potential function $\Phi = \sum_{i \geq 1} (w(B_i))^2$, where $B_1, B_2, \ldots$ are the packed bins. Let $\Phi_0$ be the value of $\Phi$ for the original packing, and $\Phi_j$ denote its value after $j$ steps of deviation of an item. Assume that in step $j$ an item moves from a bin of total weight $W$ to a bin of total weight $\bar{W}$ (before the deviation). We have $\bar{W} + w_j > W$, and since $\bar{W} + w_j \in \mathcal{W}$, $\bar{W} + w_j \geq W + \omega$. Since there is no change for any other bin, we get $\Phi_j - \Phi_{j-1} = (\bar{W} + w_j)^2 + (W - w_j)^2 - \bar{W}^2 - W^2 = 2\bar{W}w_j - 2Ww_j + 2w_j^2$. Using $\bar{W} \geq W + \omega - w_j$, we find $2\bar{W}w_j - 2Ww_j + 2w_j^2 \geq 2(W + \omega - w_j)w_j - 2Ww_j + 2w_j^2 = 2\omega w_j \geq 2\omega \cdot w_{\min}$. The claim follows since the value of $\Phi$ never exceeds $w(I)^2$, and $\Phi_0 > 0$.

Note that another upper bound on the number of steps is $n^n$, since $n$ items require at most $n$ bins, this is an upper bound on the number of different assignments of items to bins, and since the potential increases for each step, one assignment cannot be reached twice in the process. ∎

**Corollary 2.2** *A NE always exists for any set of weights. In particular, there always exists a solution of optimal social cost which is a NE. Thus, the PoS is equal to 1 for general weights.*

Next, we discuss Pareto optimal solutions. We can show that every optimal solution is strictly Pareto optimal (and therefore also weakly Pareto optimal). Thus, an optimal solution which is a NE is strictly (and weakly) Pareto optimal.

Recall that the cost of an item $i$ in a given packing $A$ is defined by $\frac{w_i}{W}$, where $W$ is the total weight of the bin of $i$ in $A$. Let this cost for a packing $A$ be denoted by $c_i^A$. Thus $cost(A) = \sum_{i \in I} c_i^A$.

**Proposition 2.3** *Every optimal packing is strictly Pareto optimal. Thus, there always exists an optimal packing which is a SPO-NE and a WPO-NE. Therefore the WPO-PoS and the SPO-PoS are also equal to 1 for general weights and for unit weights.*

**Proof.** Assume that a packing $A$ is optimal and assume by contradiction that $A$ is not strictly Pareto optimal. Let $A'$ be an alternative packing which results from a deviation of all items. We have $cost(A') = \sum_{i \in I} c_i^{A'}$. However, for every item $i$, $c_i^{A'} \leq c_i^A$, and there exists an item $\iota^*$ for which $c_{\iota^*}^{A'} < c_{\iota^*}^A$ holds. Thus $cost(A') < cost(A)$ contradicting the optimality of $A$. By Proposition 2.1, there exists an optimal packing which is a NE, and we showed that this packing is strictly Pareto optimal. Thus both the SPO-PoS and the WPO-PoS are equal to 1. ∎

Another interesting property is the relation between WPO-PoA and PoA.

**Proposition 2.4** *The WPO-PoA is equal to the PoA for any set of weights.*

**Proof.** Clearly, the WPO-PoA cannot exceed the PoA. To prove the other direction, we show that if there exists an instance $I$ and a NE packing $A_1$, then there exists an instance $I'$ and a NE packing $A'_1$ for it, which is weakly Pareto optimal, and so that $\text{OPT}(I) \leq \text{OPT}(I') \leq \text{OPT}(I) + 1$, and $cost(A'_1(I')) = cost(A_1(I)) + 1$.

Given $I$, augment this input with one item of size 1 and weight 1. Let $A'_1$ for $I'$ be identical to the packing $A_1$ for $I$, with the additional item (of $I' \setminus I$) packed into a dedicated bin. No item can migrate to the new bin, and the new item cannot migrate from it, so $A'_1$ remains a NE. We show that it is also a WPO-NE. The new item must be packed into a dedicated bin in every packing, and so its cost is 1 in every packing. Thus, there is no alternative packing where all items strictly reduce their costs, and thus $A'_1$ is weakly Pareto optimal. ∎

Next, we discuss strong equilibria. We start with giving a generic algorithm which finds a SNE. This algorithm is a special case of the well-known greedy algorithm for (weighted or unweighted) set cover. The unweighted version of the algorithm is used for the case of unit weights.

**Greedy Set Cover (GSC)** Given a set of items $I$, repeatedly find a maximum weight subset of items that can be packed into a bin, pack it, and remove it from $I$.

The proof that every output of GSC is a SNE and every SNE can be found as an output of GSC is similar to the proof for proportional weights [17]. Since the algorithm needs to consider subsets of items, it may require exponential running time.

**Proposition 2.5** *For every set of weights, every output of the algorithm is a SNE (thus a SNE always exists). Every SNE can be found by an execution of the algorithm (with some tie-breaking policy).*

**Proof.** We start with proving the first claim. Assume by contradiction that an output $P$ of the algorithm is a packing which does not satisfy the conditions of a SNE. Let $B_1, B_2, \ldots$ denote the bins created by $P$ (in this order). Consider a coalition $X$ which can benefit from a deviation. Let $i$ be a minimum index of a bin of the original packing which contains an item of $X$ and let $j$ be such an item. Note that all items packed into $B_1, \ldots, B_{i-1}$ remain in their bins. We first show that no item of $X$ moves into one of these bins. Assume by contradiction that an item $j'$ moves from a bin of index at least $i$ into bin $B_{i'}$ for $i' < i$. Then at the time that the bin $B_{i'}$ is created by the algorithm, the set $B_{i'} \cup \{j'\}$, which can be packed into a bin, is available for packing, and its weight exceeds the weight of $B_{i'}$, contradicting the definition of the algorithm.

Similarly, since no items of $X$ are packed into bins $B_1, \ldots, B_{i-1}$, all of them were available at the time of the creation of $B_i$. We consider the bin into which $j$ is packed in the packing resulting from the deviation. No matter if a new bin is created, or $j$ moves to an existing bin of index larger than $i$ (where some items may have possibly left this bin), let $X'$ denote the set of items which are packed into the resulting bin of $j$. Since in both cases $X'$ is a subset of the items previously packed into bins $B_i, B_{i+1} \ldots$, then by the definition of the algorithm $w(X') \leq W(B_i)$, contradicting the fact that $j$ benefits from joining the coalition.

Next, consider a SNE $A$. Sort the bins of $A$ by non-increasing total weight: $\mathcal{B}_1, \mathcal{B}_2, \ldots, \mathcal{B}_\ell$. We prove by induction that after $k < \ell$ bins were created, $\mathcal{B}_{k+1}$ is a valid choice for the algorithm. As a consequence, after $\ell$ bins are created, the algorithm packs all items and outputs the packing $A$. For the base case, we show that $w(\mathcal{B}_1)$ is the maximum total weight of items that can fit into a bin. Assume by contradiction that a subset $Y$ such that $w(Y) > w(\mathcal{B}_1)$ can fit into a bin, then $Y$ is a coalition for $A$ that can benefit from a deviation where all items of $Y$ move into a new bin together, which is a contradiction. Consider the case that $k > 0$ bins were created. Similarly to the base case, assume by contradiction that there exists a set $Z \subseteq \cup_{i=k+1}^{\ell} \mathcal{B}_i$, where $w(Z) > w(\mathcal{B}_{k+1})$. All the items of $Z$ are packed into bins with at most a total weight of $w(\mathcal{B}_{k+1})$ in $A$, so $Z$ is a is a coalition for $A$ that can benefit from a deviation where all items of $Z$ move into a new bin together, which is a contradiction. ∎

# 3 General weights

In this section we show that the value of the PoA, SPoA, SPoS, WPO-PoA, and SPO-PoA for general weights is exactly 1.7. We have shown that the PoS, WPO-PoS, and SPO-PoS are equal to 1 and therefore this section resolves the case of general weights.

To show an upper bound, we generalize the results of [26, 17, 7] (which were given for the cases of unit weights and for proportional weights) and prove that every NE can be obtained by an execution FF, which implies that the PoA is no larger than the approximation ratio of FF [28]. To show a lower bound, we use a set of items which is similar to those of the lower bound example for FF [28], and define an appropriate set of weights.

**Theorem 3.1** *The* PoA, SPoS, SPoA, WPO-PoA, *and* SPO-PoA *are equal to 1.7.*

**Proof.** For the upper bound it is sufficient to consider the PoA. Let $I$ be an input and consider a fixed NE packing. Let $B_1, B_2, \ldots, B_k$ denote the packed bins and assume that they are sorted so that $w(B_1) \geq w(B_2) \geq \ldots \geq w(B_k)$. We define the following sorted input for FF. The input consists of all the items of $I$ where the items of $B_1$ appear first, then the items of $B_2$ and so on (that is, the items of $B_j$ appear just before the items of $B_{j+1}$ for $1 \leq j \leq k-1$). We claim that an application of FF on this input results exactly in the bins $B_1, B_2, \ldots, B_k$. We prove this by induction. Assume that $j$ bins have been created with exactly the contents of $B_1, \ldots, B_j$, for some $0 \leq j \leq k-1$. The next items are of $B_{j+1}$, and none of them can fit into any of the bins $B_1, \ldots, B_j$, since otherwise for such an item $i \in B_{j+1}$ that can be packed into an earlier bin $1 \leq \ell \leq j$, we get $w(B_\ell) + w_i > w(B_{j+1})$ (because $w(B_\ell) \geq w(B_{j+1})$ and $w_i > 0$), which would imply that this packing is not a NE. Since the items of $B_{j+1}$ can fit into one bin, FF packs them into one bin. As the (asymptotic) approximation ratio of FF is 1.7 [28], the upper bound on the PoA follows.

Next, we adapt the lower bound example of FF [28] by slightly modifying the item sizes and by defining weights for the items. Let $\ell$ be a positive integer, let $0 < \varepsilon < \frac{1}{120}$ be a small value and let $\delta < \frac{\varepsilon}{3^{\ell+4}}$. The instance consists of $30\ell$ items. We describe the items and a packing $A$ where the items are packed into $17\ell$ bins: $B_1, \ldots, B_{17\ell}$. The weights of all items will be negative powers of 3.

The first $10\ell$ items are denoted by $a_{i,p}$ for $i = 1, \ldots, 10$ and $p = 1, \ldots, \ell$, and their sizes are defined as follows. $a_{i,p}$ has size $\frac{1}{6} + \frac{\varepsilon}{3^p} - \delta$ for $1 \leq i \leq 3$, $\frac{1}{6} + \frac{\varepsilon}{3^p} - 2\delta$ for $4 \leq i \leq 5$, $\frac{1}{6} - \frac{\varepsilon}{3^{p+1}} - \delta$ for $6 \leq i \leq 7$, and $\frac{1}{6} - \frac{\varepsilon}{3^{p+1}} - 2\delta$ for $8 \leq i \leq 10$. The weight of $a_{i,p}$, where $i = 1, 2, 3, 6, 7$ is $\frac{3}{9^p}$, and the weight of $a_{i,p}$, where $i = 4, 5, 8, 9, 10$ is $\frac{1}{9^p}$. Note that all item sizes are in $(\frac{1}{7}, \frac{1}{5})$ since the smallest item has size $\frac{1}{6} - \frac{\varepsilon}{9} - 2\delta > \frac{1}{7}$ as $\frac{\varepsilon}{9} + 2\delta < \varepsilon < \frac{1}{120}$, and the largest item has size $\frac{1}{6} + \frac{\varepsilon}{3} - \delta < \frac{1}{5}$.

Every ten items are packed into two bins as follows. The items $a_{1,p}, a_{2,p}, a_{3,p}, a_{6,p}, a_{7,p}$ are packed into one bin and $a_{4,p}, a_{5,p}, a_{8,p}, a_{9,p}, a_{10,p}$ are packed into another bin. We call these bins $B_{2p-1}$ and $B_{2p}$. The total size of items in $B_{2p-1}$ is $\frac{5}{6} + \frac{7\varepsilon}{3^{p+1}} - 5\delta$ and the total size of items in $B_{2p}$ is $\frac{5}{6} + \frac{3\varepsilon}{3^{p+1}} - 10\delta$. The total weight in $B_{2p-1}$ is $\frac{15}{9^p}$ and the total weight in $B_{2p}$ is $\frac{5}{9^p}$. Note that the least loaded bin out of the first $2\ell$ bins has load of $\frac{5}{6} + \frac{\varepsilon}{3^\ell} - 10\delta > \frac{5}{6} + \frac{\varepsilon}{3^{\ell+1}}$, since $10\delta < \frac{10\varepsilon}{3^{\ell+4}} < \frac{2\varepsilon}{3^{\ell+1}}$.

The next $10\ell$ items are denoted by $b_{i,p}$ for $i = 1, \ldots, 10$ and $p = 1, \ldots, \ell$. Their sizes are defined as follows. The size of $b_{i,p}$ for $1 \leq i \leq 5$ is $\frac{1}{3} + \frac{\varepsilon}{3^{p-1}} - i\delta$ and for $6 \leq i \leq 10$ it is $\frac{1}{3} - \frac{\varepsilon}{3^p} - (i-5)\delta$. The weight of $b_{i,p}$ is $\frac{1}{3^{2\ell+5(p-1)+i}}$ for $1 \leq i \leq 5$, and $\frac{1}{3^{2\ell+5(p-2)+i}}$ for $6 \leq i \leq 10$. Note that the largest weight of any item in this set is $\frac{1}{3^{2\ell+1}}$, while the smallest weight of any item of the previous set is $\frac{1}{3^{2\ell}}$. The minimum weight of any item in the current set is $\frac{1}{3^{7\ell}}$.

$5\ell$ bins are created from these items, bins $B_{2\ell} + 5(p-1) + j$, where $1 \leq j \leq 5$ contains items $b_{j,p}$ and $b_{j+5,p}$. The total size of items in each such bin is $\frac{2}{3} + \frac{2\varepsilon}{3^p} - 2j\delta$ and the total weight is $\frac{2}{3^{2\ell+5(p-1)+j}}$. The least loaded bin has a load of $\frac{2}{3} + \frac{2\varepsilon}{3^\ell} - 10\delta > \frac{2}{3} + \frac{2\varepsilon}{3^\ell} - \frac{10\varepsilon}{3^{\ell+4}} > \frac{2}{3} + \frac{\varepsilon}{3^\ell}$.

The last $10\ell$ items are denoted by $c_i$ for $i = 1, 2, \ldots, 10\ell$, each of these has size of $\frac{1}{2} + \delta$ and weight $\frac{1}{3^{7\ell+i}}$. These items are packed into the dedicated bins $B_{7\ell+j}$ for $1 \leq j \leq 10\ell$.

Using the result of [28], these items can be packed into $10\ell + O(1)$ bins. For completeness, and since the item sizes are slightly adapted, we give this packing below.

For $i = 1, \ldots, 5$ and $p = 1, \ldots, \ell$ there is a bin containing $\{a_{i,p}, b_{5+i,p}, c_{5(p-1)+i}\}$. For $i = 1, \ldots, 5$ and $p = 3, \ldots, \ell$ there is a bin containing $\{a_{5+i,p-2}, b_{i,p}, c_{5(p+\ell-3)+i}\}$. This gives a total of $10\ell - 10$ bins. The remaining items are packed into 12 additional bins: five bins containing $\{c_{10(\ell-1)+i}, b_{i,1}\}$ for $i = 1, \ldots, 5$, five bins containing $\{c_{10(\ell-1)+5+i}, b_{i,2}\}$ for $i = 1, \ldots, 5$, and two bins with five items each, a bin with $\{a_{6,\ell}, a_{7,\ell}, a_{8,\ell}, a_{9,\ell}, a_{10,\ell}\}$, and a bin with $\{a_{6,\ell-1}, a_{7,\ell-1}, a_{8,\ell-1}, a_{9,\ell-1}, a_{10,\ell-1}\}$.

We will show that $A$ is a the unique SNE and that it is strictly Pareto optimal (this shows that not only the SPoS is equal to 1.7 but so is the SPoS). The first property implies that it is a NE, and from both properties imply that $A$ is a WPO-NE (it is also possible to deduce the claim regarding the WPO-PoA from the claim regarding the PoA, using Proposition 2.4). We let $A_k$ denote the set of items packed into the first $k$ bins, and $F_k$ the complement set of items. The next claim is based on the property that FF can pack exactly the bins of $A$ if the items are given in the order of the bins.

**Claim 3.2** *Consider an item $X$ which is packed into the k-th bin. For each earlier bin, there is no sufficient space for $X$.*

**Proof.** If $X = c_i$, then since all bins are packed with load above $\frac{1}{2}$, the claim holds.

Assume $X = b_{i,p}$. The first $2\ell$ bins have total size larger than $\frac{5}{6}$, while the minimum size of this item can be $\frac{1}{3} - \frac{\varepsilon}{3^p} - 5\delta > \frac{1}{4}$ since $\delta < \varepsilon < \frac{1}{120}$. Moreover, the bins $B_{2\ell+1}, \ldots, B_{2\ell+5p}$ cannot accommodate such items since $(\frac{2}{3} + \frac{2\varepsilon}{3^p} - 10\delta) + (\frac{1}{3} - \frac{\varepsilon}{3^p} - 5\delta) > 1$, since $15\delta < \frac{15\varepsilon}{3^{\ell+4}} < \frac{\varepsilon}{3^{\ell+1}} < \frac{\varepsilon}{3^p}$.

Finally, assume $X = a_{i,p}$. The size of $X$ is at least $\frac{1}{6} - \frac{\varepsilon}{3^{p+1}} - 2\delta$. A bin of the set $\{B_1, \ldots, B_{2p}\}$ which is not the bin of $X$ cannot accommodate $X$, since the least full such bin contains a total size of items of at least $\frac{5}{6} + \frac{\varepsilon}{3^p} - 10\delta$, and since $(\frac{1}{6} - \frac{\varepsilon}{3^{p+1}} - \delta) + (\frac{5}{6} + \frac{\varepsilon}{3^p} - 10\delta) = 1 + \frac{2\varepsilon}{3^{p+1}} - 9\delta > 1$, since $9\delta < \frac{\varepsilon}{3^{\ell+2}} < \frac{2\varepsilon}{3^{p+1}}$ holds using $p \leq \ell$. ∎

The following claim will assist us in proving that $A$ satisfies the required properties.

**Claim 3.3** *Let $0 \leq k \leq 17\ell - 1$. Consider a packing $A'$ in which the bins of the items of $A_k$ are identical to the set of first $k$ bins of $A$. Assume that at least one item $Z$ packed in bin $B_{k+1}$ for $A$ has at most the same cost in $A'$ as it has in $A$. Then the $(k+1)$-th bin of $A$ is a bin of $A'$. Moreover, the unique possible bin of maximum weight which can be packed using items of $F_k$ is $B_{k+1}$.*

**Proof.** If $k \geq 7\ell$, then $A_k$ contains all items of size at most $\frac{1}{2}$, and $F_k$ contains only items of size above $\frac{1}{2}$, and $Z$ is such an item (of maximum weight), so $Z$ is packed in a dedicated bin in $A$. It cannot be combined in a bin with another item of $F_k$, and by Claim 3.2 it cannot be packed into a bin containing items of $A_k$, so the bin $B_{k+1}$ must exist in $A'$ as well. Since $Z$ is the unique item of maximum weight in $F_k$, the second claim follows.

If $2\ell \leq k \leq 7\ell - 1$, then $Z = b_{i,p}$. Let $\omega$ be its weight. Then the item packed with $Z$ in $A$ (in bin $B_{k+1}$) has the same weight, and any further item has weight at most $\frac{\omega}{3}$. By claim 3.2 applied for $k + 1$, if the two items of weight $\omega$ are packed together in $A'$, then no additional item of $F_{k+1}$ (which is equal to $F_k$ excluding the two items of weight $\omega$) can be packed into the bin. Therefore, the bin of $Z$ in $A'$ is identical to $B_{k+1}$. Otherwise, since all items of $F_k$ have size above $\frac{1}{4}$, $Z$ can be combined in a bin with at most two items. The weight of this bin will be at most $\omega + 2\frac{\omega}{3} < 2\omega$, so $Z$ has a cost above $\frac{1}{2}$ in this bin (but a cost of exactly $\frac{1}{2}$ in $B_{k+1}$), which is a contradiction. Any subset of items which can be packed into a bin, and which does not contain any item of $B_{k+1}$, has a total weight of at most $3 \cdot \omega$ (while $w(B_{k+1}) = 2\omega$), and the second claim follows as well.

Finally, if $0 \leq k \leq 2\ell - 1$, then $Z = a_{i,p}$. Let $\omega$ be its weight. In the set $F_k$ there are four additional items of weight $\omega$, and all other items have weights of at most $\frac{\omega}{3}$. Once again, if $Z$ is packed with all four items which are packed with it in $A$, by Claim 3.2 (applied to $k + 1$), no other item of $F_k$ can fit into the bin the bin is identical to $B_{k+1}$. Otherwise, since all item sizes are above $\frac{1}{7}$, the bin can contain at most six items, one of which is $Z$, and at most three are of weight $\omega$. Thus the total weight of the bin is no larger than $4\omega + 2\frac{\omega}{3} < 5\omega$, so $Z$ has a higher cost than $\frac{1}{5}$, which is its cost in the bin $B_{k+1}$, which is a contradiction. Any subset of items which can be packed into a bin, and which does not contain any item of $B_{k+1}$ has a total weight of at most $5\omega/3$, and so the second claim follows as well. ∎

To prove that $A$ is strictly Pareto optimal, assume by contradiction that there is an alternative packing $\bar{A}$ where no item has larger cost and at least one item has a lower cost. Let $i$ be the first bin of $A$ that does not exist in $\bar{A}$ (if $i$ does not exist then $A = \bar{A}$). Let $Z$ be an item of bin $i$ of $A$. By Claim 3.3, for any $Z \in B_i$, the bin of $Z$ in $\bar{A}$ is identical to $B_i$, which is a contradiction.

We prove that $A$ is the unique strong equilibrium. After $k \geq 0$ bins were created, by Claim 3.3, GSC has a single choice, which is exactly $B_{k+1}$. Thus, the unique output of GSC is the packing $A$. ∎

We generally assume that two items of the same size can have different weights. However, in the lower bound proof this option is not used and therefore the theorem holds even if $w_i$ is a function of $s_i$.

## 4 The PoA and WPO-PoA for unit weights

Recall that by Proposition 2.4, the WPO-PoA is equal to the PoA, and therefore we consider the PoA in this section. We define a $k$-bin to be a bin of $A$ which has exactly $k$ items. We define a $k^+$-bin to be a bin of $A$ which has at least $k$ items. For unit weights, the deviation of an item $j$ packed in a $k_1$-bin $B_1$ (where $j$ is included in the number of items of $B_1$) to a $k_2$-bin $B_2$ (where $j$ is not included in the number of items of $B_2$) is possible and beneficial if $s(B_2) + s_j \leq 1$ and $k_2 \geq k_1$. In what follows we use the following sequence which is common in bin packing. We start with a lower bound on the PoA, and show an upper bound afterwards.

### 4.1 Lower bound

Let $t_1 = 2$, and for $i > 1$, $t_i = t_{i-1} \cdot (t_{i-1} - 1) + 1$. Thus $t_i - 1$ is divisible by $t_{i-1} - 1$. By induction it is divisible by $t_j - 1$ for all $1 \leq j \leq i - 1$. It is not difficult to show by induction that $t_i \geq 2$ and the sequence is strictly increasing. Moreover, $t_2 = 3$, and for $i \geq 3$, we have $t_i \geq 7$ and $t_i \geq t_{i-1} + 4$. Thus for $i \geq 3$, $t_i - 2 > t_{i-1} + 1$. The sequence satisfies $\sum_{i=1}^{\infty} \frac{1}{t_i} = 1$, $\sum_{i=1}^{r} \frac{1}{t_i} = 1 - \frac{1}{t_{r+1}-1}$, and $\sum_{i=1}^{\infty} \frac{1}{t_i - 1} \approx 1.69103$ [31, 24]. We let $T_\infty = \sum_{i=1}^{\infty} \frac{1}{t_i - 1}$.

**Theorem 4.1** *The* PoA *is at least* $1.696646$.

**Proof.** Let $r \geq 3$ be an integer. Let $0 < \varepsilon < \frac{1}{2}$ be a small number. We define $\delta_i = (\frac{\varepsilon}{(t_{r+1}+1)^4})^{r-i+1}$ for $1 \leq i \leq r$ (so $(t_{r+1}+1)^4 \delta_i = \varepsilon \delta_{i+1}$, $\delta_{i+1} > \delta_i$, and $\delta_i \leq \frac{\varepsilon}{(t_r+1)^4} \leq \frac{\varepsilon}{(t_i+1)^4}$).

We have $r$ classes of items, defined as follows. The first class contains a simple type of items $a_1 = \frac{1}{t_1} + \delta_1 = \frac{1}{2} + \delta_1$. For $2 \leq i \leq r$, in class $i$ there are five types of items:

$$a_i^1 = \frac{1}{t_i} + \delta_i, \quad a_i^2 = \frac{1}{t_i} - (t_i - 1)\delta_i + \delta_{i-1}, \quad a_i^3 = \frac{1}{t_i} + (t_i - 1)^2 \delta_i - (t_{i-1} - 1)^2 \delta_{i-1},$$

$$b_i^1 = \frac{1}{t_{i+1} - 1} - \delta_i - (t_{i-1} - 1)^2 \delta_{i-1}, \quad b_i^2 = \frac{1}{t_{i+1} - 1} + (t_i - 1)\delta_i - ((t_{i-1} - 1)^2 + 1)\delta_{i-1}.$$

We show that these sizes are positive and do not exceed 1. To show that the sizes are positive we show the following stronger properties: for $2 \leq i \leq r$, $a_i^1 > \frac{1}{t_i}$, $a_i^2 > \frac{1}{t_{i+1}}$, $a_i^3 > \frac{1}{t_i}$, $b_i^1 > \frac{1}{t_{i+1}}$, and $b_i^2 > \frac{1}{t_{i+1}-1}$ (and thus $a_i^j > \frac{1}{t_{i+1}}$ for $j = 1, 2, 3$ and $b_i^j > \frac{1}{t_{i+1}}$ for $j = 1, 2$).

The property holds for $a_i^1$ since $\delta_i > 0$. We have $a_i^2 > \frac{1}{t_i} - (t_i - 1)\delta_i$, where $\delta_i \leq \frac{\varepsilon}{(t_i+1)^3} < \frac{1}{(t_i-1)t_i(t_i+1)}$, so $a_i^2 > \frac{1}{t_i} - \frac{t_i-1}{(t_i-1)t_i(t_i+1)} = \frac{1}{t_i+1}$. We have

$$a_i^3 = \frac{1}{t_i} + (t_i - 1)^2 \delta_i - (t_{i-1} - 1)^2 \delta_{i-1} > \frac{1}{t_i} + (t_i - 1)^2 (\delta_i - \delta_{i-1}) > \frac{1}{t_i},$$

using $\delta_i > \delta_{i-1}$ and $1 < t_{i-1} < t_i$. We have $b_i^1 = \frac{1}{t_{i+1}-1} - \delta_i - (t_{i-1}-1)^2 \delta_{i-1} > \frac{1}{t_{i+1}-1} - t_{i-1}^2 \delta_i$ (using $\delta_i > \delta_{i-1}$ and $t_{i-1} > 1$), and using $\delta_i \leq \frac{1}{t_{i-1}^2(t_{i+1}-1)t_{i+1}}$ (since $t_{r+1} \geq t_{i+1} > t_{i-1}$) we have $b_i^1 > \frac{1}{t_{i+1}}$. Finally, $\delta_i > (t_{i-1}+1)^2 \delta_{i-1} > ((t_{i-1}-1)^2+1)\delta_{i-1}$, so $b_i^2 = \frac{1}{t_{i+1}-1} + (t_i-1)\delta_i - ((t_{i-1}-1)^2+1)\delta_{i-1} > \frac{1}{t_{i+1}-1}$ using $t_i \geq 2$.

Next, we show that the sizes are no larger than 1. $a_1 < \frac{1}{2} + \varepsilon < 1$ and similarly $a_i^j < \frac{1}{2} + \varepsilon < 1$ for $j = 1, 2$. We have $a_i^3 < \frac{1}{2} + t_i^2 \delta_i < \frac{1}{2} + \frac{1}{t_{i+1}} < 1$, $b_i^1 < \frac{1}{t_{i+1}-1} < 1$, and $b_i^2 < \frac{1}{t_{i+1}-1} + (t_i-1)\delta_i < \frac{1}{2} + \varepsilon \leq 1$.

Let $M$ be a large integer. For $2 \leq j \leq r$, let $\pi_j = t_j(t_j-1)(t_j-2) + 2 = t_j^3 - 3t_j^2 + 2t_j + 2$. Since for $j \geq 2$, $t_j \geq 3$ holds we find $\pi_j \geq 2(t_j+1)$ and $\pi_j \geq t_j^2 - t_j + 2$. For $2 \leq i \leq r+1$, we define

$$\Delta_i = M \cdot (\prod_{j=1}^{i-1} t_j)(\prod_{j=i}^{r} \pi_j)(t_r - 1) \ ,$$

(so $\Delta_{r+1} = M \cdot (\prod_{j=1}^{r} t_j)(t_r - 1)$) and let $n_i = \frac{\Delta_i}{\pi_i}$ and $\lambda_i = \frac{\Delta_i}{\Delta_2}$. We have $\lambda_i = \prod_{j=2}^{i-1} \frac{t_j}{\pi_j}$ and $\frac{\Delta_{i+1}}{\Delta_i} = \frac{\lambda_{i+1}}{\lambda_i} = \frac{t_i}{\pi_i}$. Let $\mu_i = \frac{\lambda_i}{\pi_i} = \frac{\lambda_{i+1}}{t_i}$. For $2 \leq i \leq r$, $n_i$ is an integer divisible by $t_i - 1$. The number of items of size $a_1$ is $\Delta_2$. For $2 \leq r$, the number of items of size $a_i^1$ and $b_i^1$ is $n_i(\pi_i - t_i^2 + t_i)$. The number of items of size $a_i^2$ and $b_i^2$ is $n_i(t_i^2 - 2t_i)$. The number of items of size $a_i^3$ is $n_i t_i$.

Consider the following packing. Every item of size $a_1$ is packed into a dedicated bin. For $2 \leq i \leq r$, there are three types of bins for the items of class $i$. Every bin of the first type contains $t_i - 1$ items of size $a_i^1$. Every bin of the second type contains one item of size $a_i^3$ and $t_i - 2$ items of size $a_i^2$. Every bin of the third type contains $t_i$ items of size $b_i^2$ and $(t_i-1)^2 - 2$ items of size $b_i^1$, for a total of $t_i^2 - t_i - 1$ items. Using $i \geq 2$ we have $(t_i-1)^2 - 2 \geq 2$ and $t_i - 1 < t_i^2 - t_i - 1 < t_{i+1} - 1$. There are $\frac{n_i(\pi_i - t_i^2 + t_i)}{t_i - 1}$ bins of the first type, $n_i t_i$ bins of the second type, and $n_i(t_i - 2)$ bins of the third type. First, we show that all items are packed. This is clear for the items of size $a_i^1$, $a_i^2$, $b_i^2$, and $a_i^3$. For the items of size $b_i^1$ this holds since $\frac{\pi_i - t_i^2 + t_i}{t_i^2 - 2t_i - 1} = t_i - 2$.

Next, we show that this packing is a NE, and that the total size of items in each bin does not exceed 1. Consider a bin which contains an item of size $a_1$. This type of bin contains a single item so no item from another type of bin can benefit from moving to it, and another item of the same size cannot fit into this bin. Consider the first type of bin for some $i \geq 2$. This bin contains $t_i - 1$ items, so only items of bins with at most $t_i - 1$ items could benefit from moving into it. These are items packed into bins of classes $1, 2, \ldots, i-1$ and the first two types of bins of class $i$. The total size of items in this bin is $1 - \frac{1}{t_i} + (t_i - 1)\delta_i$. Every item of classes $1, 2, \ldots, i-1$ has size above $\frac{1}{t_i}$, and every item of class $i$, packed into the first two types of bins, has size of at least $a_i^2 = \frac{1}{t_i} - (t_i - 1)\delta_i + \delta_{i-1}$, so none of this items can migrate due to lack of space. The total size of items in the second type of bins of class $i$ is $(t_i-2)a_2^i + a_3^i = \frac{t_i - 2}{t_i} - (t_i-2)(t_i-1)\delta_i + (t_i-2)\delta_{i-1} + \frac{1}{t_i} + (t_i-1)^2 \delta_i - (t_{i-1}-1)^2 \delta_{i-1} = \frac{t_i-1}{t_i} + (t_i-1)\delta_i + (t_i-2-(t_{i-1}-1)^2)\delta_{i-1} = \frac{t_i-1}{t_i} + (t_i-1)\delta_i + ((t_{i-1}(t_{i-1}-1)+1) - 2 - (t_{i-1}-1)^2)\delta_{i-1} = \frac{t_i-1}{t_i} + (t_i-1)\delta_i + (t_{i-1}-2)\delta_{i-1}$. Since $i \geq 2$, $t_{i-1} \geq t_1 = 2$, the total size is at least as large as in the previous case. Since $\delta_i < \frac{1}{(t_r+1)^3} \leq \frac{1}{2t_i(t_i-1)}$, and $\delta_{i-1} < \frac{1}{(t_r+1)^3} \leq \frac{1}{2t_i(t_{i-1}-2)}$, the total size in both cases does not exceed 1.

It is left to consider the third type of bins. The total size of items in those bins is $t_i b_i^2 + ((t_i-1)^2 - 2)b_i^1 = t_i(\frac{1}{t_{i+1}-1} + (t_i-1)\delta_i - ((t_{i-1}-1)^2+1)\delta_{i-1}) + ((t_i-1)^2 - 2)(\frac{1}{t_{i+1}-1} - \delta_i - (t_{i-1}-1)^2 \delta_{i-1}) = \frac{t_i + (t_i-1)^2 - 2}{t_{i+1}-1} + (t_i(t_i-1) - (t_i-1)^2 + 2)\delta_i - (t_i((t_{i-1}-1)^2 + 1) + ((t_i-1)^2+1)(t_{i-1}-1)^2)\delta_{i-1} \geq \frac{t_i + (t_i-1)^2 - 2}{t_{i+1}-1} + (t_i+1)\delta_i - 2t_i^3 \delta_{i-1}$, since $t_i \geq (t_{i-1}-1)^2 + 1$. Using $\delta_i > (t_r+1)^4 \delta_{i-1} > 2t_i^3 \delta_{i-1}$, the total size is above $\frac{t_i(t_i-1)-1}{t_{i+1}-1} + t_i \delta_i = \frac{t_{i+1}-2}{t_{i+1}-1} + t_i \delta_i$. The total size does not exceed $\frac{t_{i+1}-2}{t_{i+1}-1} + (t_i+1)\delta_i$, and since $\delta_i < \frac{1}{(t_i+1)(t_{i+1}-1)}$, the total size does not exceed 1. As before, items of classes $i+1, \ldots, r$ would not benefit from moving to this bin, while items of classes $1, \ldots, i-1$, and the items of sizes $a_i^j$ have size of at least $\frac{1}{t_i+1} > \frac{1}{t_{i+1}-1}$. We have seen $b_i^2 > \frac{1}{t_{i+1}-1}$. Thus, it is left to show $b_i^1 + \frac{t_{i+1}-2}{t_{i+1}-1} + t_i \delta_i \geq 1$. We have $b_i^1 + \frac{t_{i+1}-2}{t_{i+1}-1} + t_i \delta_i = \frac{1}{t_{i+1}-1} - \delta_i - (t_{i-1}-1)^2 \delta_{i-1} + \frac{t_{i+1}-2}{t_{i+1}-1} + t_i \delta_i = 1 + (t_i - 1)\delta_i - (t_{i-1}-1)^2 \delta_{i-1} \geq 1 + \delta_i - t_{i-1}^2 \delta_{i-1}$. Using $\delta_{i-1} < \frac{\delta_i}{(t_r+1)^2}$ the claim is proved. The total number of bins in this packing is $\Delta_2 + \sum_{i=2}^r \left( \frac{\Delta_i}{\pi_i} (\frac{\pi_i - t_i^2 + t_i}{t_i - 1} + t_i + (t_i - 2)) \right)$.

Next, we describe the optimal solution for this input, which consists of $\Delta_2$ bins. Every bin contains exactly one item of size $a_1$, and we describe the additional items packed into each bin. For every $2 \leq i \leq r - 1$ there are two types of bins which contain exactly one item of each class $2, \ldots, i - 1$ and two items of class $i$, and in addition there are three types of bins containing one item of each class. Specifically, for every $2 \leq i \leq r$, there are $n_i(\pi_i - t_i^2 + t_i)$ bins which contain one item of each size $a_j^3$ for $2 \leq j \leq i-1$, one item of size $a_i^1$, and one item of size $b_i^1$, and there are $n_i(t_i^2 - 2t_i)$ bins which contain one item of each size $a_j^3$ for $2 \leq j \leq i-1$, one item of size $a_i^2$, and one item of size $b_i^2$. There are also $n_r t_r$ bins which contain one item of each size $a_j^3$ for $2 \leq j \leq r$.

We show that the total size of item in each bin is at most 1 and that all items are packed. The last type of bins has a total size of items of $a_1 + \sum_{i=2}^r a_i^3 = \frac{1}{t_1} + \delta_1 + \sum_{i=2}^r (\frac{1}{t_i} + (t_i-1)^2 \delta_i - (t_{i-1}-1)^2 \delta_{i-1}) = \sum_{i=1}^r \frac{1}{t_i} + (t_r - 1)^2 \delta_r = 1 - \frac{1}{t_{r+1}-1} + (t_r-1)^2 \delta_r$. Using $\delta_r \leq \frac{1}{(t_{r+1}+1)^4} < \frac{1}{(t_{r+1}-1)(t_r-1)^2}$, we find that the total is below 1.

For the first type of bin of class $i$, we have a total of $a_1 + \sum_{j=2}^{i-1} a_j^3 + a_j^1 + b_j^1 = \frac{1}{t_1} + \delta_1 + \sum_{j=2}^{i-1}(\frac{1}{t_j} + (t_j-1)^2 \delta_j - (t_{j-1}-1)^2 \delta_{j-1}) + (\frac{1}{t_i} + \delta_i) + (\frac{1}{t_{i+1}-1} - \delta_i - (t_{i-1}-1)^2 \delta_{i-1}) = \sum_{j=1}^i \frac{1}{t_j} + \frac{1}{t_{i+1}-1} = 1$, and the second type of bin of class $i$ contains a total of $a_1 + \sum_{j=2}^{i-1} a_j^3 + a_j^2 + b_j^2 = \frac{1}{t_1} + \delta_1 + \sum_{j=2}^{i-1}(\frac{1}{t_j} + (t_j-1)^2 \delta_j - (t_{j-1}-1)^2 \delta_{j-1}) + (\frac{1}{t_i} - (t_i-1)\delta_i + \delta_{i-1}) + (\frac{1}{t_{i+1}-1} + (t_i-1)\delta_i - ((t_{i-1}-1)^2 + 1)\delta_{i-1}) = \sum_{j=1}^i \frac{1}{t_j} + \frac{1}{t_{i+1}-1} = 1$.

The total number of bins is $\sum_{i=2}^r (n_i(\pi_i - t_i^2 + t_i) + n_i(t_i^2 - 2t_i)) + n_r t_r = \sum_{i=2}^r (\Delta_i \frac{\pi_i - t_i}{\pi_i}) + n_r t_r = \sum_{i=2}^r \Delta_i - \sum_{i=2}^r \frac{t_i}{\pi_i} \Delta_i + n_r t_r = \sum_{i=2}^r \Delta_i - \sum_{i=2}^r \Delta_{i+1} + n_r t_r = \Delta_2 - \Delta_{r+1} + n_r t_r = \Delta_2 - t_r \frac{\Delta_r}{\pi_r} + \frac{\Delta_r}{\pi_r} t_r = \Delta_2$, and thus all items of size $a_1$ are packed.

It can be seen that the number of bins which contains an item of one of the sizes $a_i^1, a_i^2, b_i^1, b_i^2$ is exactly as the number of such items, since there is only type of bin that contains such an item. We calculate the number of bins containing an item of size $a_i^3$, for $2 \leq i \leq r - 1$. Such an item is packed in every bin of every class $j > i$. The number of such bins is $\sum_{j=i+1}^r (n_j(\pi_j - t_j^2 + t_j) + n_j(t_j^2 - 2t_j)) + n_r t_r = \sum_{j=i+1}^r (\frac{\Delta_j}{\pi_j}(\pi_j - t_j)) + n_r t_r = \sum_{j=i+1}^r \Delta_j - \sum_{j=i+1}^r \frac{\Delta_j t_j}{\pi_j} + n_r t_r = \sum_{j=i+1}^r \Delta_j - \sum_{j=i+1}^r \Delta_{j+1} + n_r t_r = \Delta_{i+1} - \Delta_{r+1} + n_r t_r = \Delta_{i+1} = \frac{\Delta_i t_i}{\pi_i} = n_i t_i$. Note that the number of bins containing an item of size $a_r^3$ is the number of bins of the very last type, which is exactly the number of such items.

We are left with the calculation of the ratio between the numbers of bins:

$$\frac{\Delta_2 + \sum_{i=2}^r \left( \frac{\Delta_i}{\pi_i}(\frac{\pi_i - t_i^2 + t_i}{t_i - 1} + 2t_i - 2) \right)}{\Delta_2} = 1 + \sum_{i=2}^r \frac{\lambda_i}{\pi_i} \left( \frac{\pi_i}{t_i - 1} + t_i - 2 \right) = 1 + \sum_{i=2}^r \left( \frac{\lambda_i}{t_i - 1} + \mu_i(t_i - 2) \right).$$

Letting $r = 3$ the value of this expression is 1.6963443, for $r = 4$ the value of this expression is approximately 1.696646 (for $r = 5$ the value of this expression only increases by less than 0.000000004). ∎

## 4.2 Upper bound

Let $\varepsilon \leq \frac{1}{330}$ and let $\delta = \frac{\varepsilon}{5(1+\varepsilon)}$. We prove an upper bound of $1.7 - \delta$ on the PoA for the case of identical weights. Assume by contradiction that $1.7 - \delta$ is not an upper bound on the PoA. Thus, for any $C > 0$, there exists a game $G'$ and an NE $A'$ such that $cost(A') \geq (1.7 - \delta) \text{OPT}(G') + C$.

Consider a game with $C = 14$. We classify the items into huge, big, medium, and small items. The corresponding size intervals are $(\frac{1}{2}, 1]$, $(\frac{1}{3}, \frac{1}{2}]$, $(\frac{1}{6}, \frac{1}{3}]$, and $(0, \frac{1}{6}]$, respectively. The big items are split further into large and semi-big items, and the corresponding intervals are $(\frac{5}{12}, \frac{1}{2}]$ and $(\frac{1}{3}, \frac{5}{12}]$, respectively. The small items are split further into semi-small and tiny, and the corresponding intervals are $(\frac{1}{12}, \frac{1}{6}]$ and $(0, \frac{1}{12}]$, respectively.

If $A'$ contains a 1-bin with an item which is not huge (since $A'$ is a NE, there can be at most one such bin), we modify its size to be 1. Next, we remove from the instance every item packed in $A'$ in a bin together with some huge item. The resulting assignment is still a NE since the bins which now have additional empty space contain a single huge item; an item assigned to a $2^+$-bin has no incentive to move, while a huge item cannot join another huge item. We denote the resulting game $G$ and the resulting assignment $A$. Since the size of one item may have been increased, this item may require a new bin in an optimal solution, and we have $\text{OPT}(G) \leq \text{OPT}(G') + 1$ and $cost(A) = cost(A')$. In what follows we analyze $A$ and $G$, and we have $cost(A) \geq (1.7-\delta)\text{OPT}(G') + C \geq (1.7-\delta)(\text{OPT}(G)-1) + C > (1.7-\delta)\text{OPT}(G) + C - 2$. In the analysis of $A$ we assume that every 1-bin contains a huge item, and no $2^+$-bin contains a huge item.

For $k > 1$, we let $\rho_k \leq \frac{1}{k}$ denote the minimum size of any item packed in a $k$-bin in $A$ (if at least one such bin exists). The property $\rho_k \leq \frac{1}{k}$ must hold since otherwise a bin cannot contain $k$ items of size at least $\rho_k$. We let $\beta_k$ denote a $k$-bin which contains an item of size $\rho_k$, and call $\beta_k$ a *special* bin. The $k$-bins which are not $\beta_k$ are called *regular* $k$-bins.

We use the next lemmas regarding $A$.

**Lemma 4.2** *In the assignment $A$, for $k > 1$, every regular $k$-bin contains a total size of items which exceeds $\max\{1 - \rho_k, \frac{k}{k+1}\}$, every regular $k$-bin has at least one item of size strictly above $\frac{1}{k+1}$, and moreover, for a regular $k$-bin $\beta'$ (if it exists), $s(\beta_k) + s(\beta') > \frac{2k}{k+1}$.*

The lemma is proved for $k > 1$, however, note that every 1-bin has a huge item, and the total size of its items exceeds $\frac{1}{2}$, so these properties hold for $k = 1$ as well.

**Proof.** If there is at most one $k$-bin, then we are done since no regular $k$-bins exist. Otherwise, we start with proving the first claim. Since the item of size $\rho_k$ in the bin $\beta_k$ has no incentive to move to another $k$-bin, all these bins are occupied with a total size which exceeds $1 - \rho_k$. If $\rho_k \leq \frac{1}{k+1}$ then $1 - \rho_k \geq \frac{k}{k+1}$. Otherwise, every $k$-bin has $k$ items, each of size at least $\rho_k > \frac{1}{k+1}$, and thus the total exceed $\frac{k}{k+1}$. By averaging, a bin which is occupied by $k$ items of total size above $\frac{k}{k+1}$ has at least one item of size above $\frac{1}{k+1}$. Consider the regular $k$-bin $\beta'$. Since the item of size $\rho_k$ cannot migrate to the bin $\beta'$ we have $s(\beta') + \rho_k > 1$. An item of minimum size in the bin $\beta_k$ has size at most $s(\beta_k)/k$ so $\rho_k \leq s(\beta_k)/k$ and we have $s(\beta') + s(\beta_k)/k > 1$. An item of minimum size in $\beta'$ has size of at most $s(\beta')/k$. Since this item cannot migrate to the bin $\beta_k$, $s(\beta_k) + s(\beta')/k > 1$. Adding these two inequalities gives $(1 + \frac{1}{k})(s(\beta') + s(\beta)) > 2$. ∎

**Lemma 4.3** *In the assignment $A$, consider a value $k$ such that there is at least one $k$-bin. Then every $(k+1)^+$-bin has a total size of items which exceeds $1 - \rho_k$ (including special bins).*

**Proof.** As every item has an incentive to move to a bin with a larger or equal number of items, and since $A$ is a NE, such a move is not possible with respect to the total size of items, and thus every such bin has a total size of items exceeding $1 - \rho_k$. ∎

**Lemma 4.4** *In the assignment $A$, a regular 2-bin must contain at least one big item. The size of the other item is above $\frac{1}{4}$.*

**Proof.** If a 2-bin does not have a big item, since it has no huge item, its two items have sizes of at most $\frac{1}{3}$ which contradicts Lemma 4.2. If the smaller item has size of at most $\frac{1}{4}$, then $\rho_2 \leq \frac{1}{4}$, and the total size of the two items must exceed $\frac{3}{4}$. However, the size of the larger item is at most $\frac{1}{2}$, so the total size is at most $\frac{3}{4}$, which is a contradiction. ∎

**Lemma 4.5** *In the assignment $A$, if a regular 3-bin contains a small item, then the other two items are not small, and at least one of them is big.*

**Proof.** From the first condition and Lemma 4.2, $\rho_3 \leq \frac{1}{6}$ and all regular 3-bins are occupied with a total size above $\frac{5}{6}$. Since the bin contains a small item, then the total size of the two other items exceeds $\frac{2}{3}$, so at least one item is big. The third item must be either medium or big (since there are no huge items in $2^+$-bins), as the total size of two small items and one big item is at most $\frac{5}{6}$. ∎

**Lemma 4.6** *In the assignment A, if a regular 4-bin contains a small item, then the bin also contains either a big item or (at least) two medium items.*

**Proof.** If the bin contains a small item, then $\rho_4 \leq \frac{1}{6}$, and all regular 4-bins are occupied with a total size above $\frac{5}{6}$. The total size of the other three items exceeds $\frac{2}{3}$. If there is no big item, and at most one medium item, then the total size of these three items is at most $\frac{2}{3}$. ∎

In all sections where the case of unit weights is studied, we use the concept *weight* to denote a term which we define and use in our analysis. We also use the function $w(x)$ defined on sizes of items. For a set of items $B$, we let $w(B)$ denote the total weight according to our definition.

The proof has the following structure. After the weight is defined, we will analyze the total weight in bins of $A$, and in bins of an optimal solution. We will define a (modified) weight function on the items, which is based on the original weight function and, on $A$, and on an optimal packing which we consider. We will show that a bin of the optimal packing has weight of at most 1.7, all bins of $A$ (except for $C-2$ bins) have weights of at least 1, and all $3^+$-bins have weights of at least $1+\varepsilon$. This will allow us to derive a contradiction. We use the weighting function defined in Table 2 (for simplicity we define a weight even in the case of a zero-sized item, which cannot exist).

| The interval | Type | $w(x)$ | $\sup_x \frac{w(x)}{x}$ | $\inf_x \frac{w(x)}{x}$ |
|---|---|---|---|---|
| $x \in [0, \frac{1}{12}]$ | small, tiny | $\frac{13}{11}x$ | $\frac{13}{11}$ | $\frac{13}{11}$ |
| $x \in (\frac{1}{12}, \frac{1}{6}]$ | small, semi-small | $\left(\frac{6}{5} - 6\varepsilon\right)x$ | $\frac{6}{5} - 6\varepsilon$ | $\frac{6}{5} - 6\varepsilon$ |
| $x \in (\frac{1}{6}, \frac{1}{4}]$ | medium | $\left(\frac{9}{5} - 12\varepsilon\right)x - \frac{1}{10} + 3\varepsilon$ | $\frac{7}{5}$ | $\frac{6}{5} + 6\varepsilon$ |
| $x \in (\frac{1}{4}, \frac{1}{3}]$ | medium | $\left(\frac{9}{5} - 12\varepsilon\right)x - \frac{1}{10} + 3\varepsilon$ | $\frac{3}{2} - 3\varepsilon$ | $\frac{7}{5}$ |
| $x \in (\frac{1}{3}, \frac{5}{12}]$ | big, semi-big | $\left(\frac{6}{5} - 12\varepsilon\right)x + \frac{1}{10} + 5\varepsilon$ | $\frac{3}{2} + 3\varepsilon$ | 1.44 |
| $x \in (\frac{5}{12}, \frac{1}{2}]$ | big, large | $\frac{13}{11}x + \frac{1}{132} + \frac{1}{10} = \frac{13}{11}x + \frac{71}{660}$ | 1.44 | $\frac{461}{330} \approx 1.39697$ |
| $x \in (\frac{1}{2}, 1]$ | huge | 1 | 2 | 1 |

Table 2: Types of items, weights, and supremum and infimum ratios between weights and sizes of items

**Proposition 4.7** *Table 2 contains upper bounds and lower bounds on $\frac{w(x)}{x}$ for the five first intervals of the definition of $w$. For $\varepsilon \leq \frac{1}{330}$ the slope $\frac{13}{11}$ is no larger than the slope $\frac{6}{5} - 6\varepsilon$.*

We consider some properties of $w$. Note that $w$ is a piecewise linear function. We will bound the total weight of items both using the bins of an optimal solution for $G$ and the bins of $A$.

**Lemma 4.8** *The function $w$ is strictly monotonically increasing for $0 \leq x \leq \frac{1}{2}$.*

**Proof.** Since for $x \in [0, \frac{1}{2}]$ the function $w$ is piecewise linear with non-zero slopes, it is strictly monotonically increasing in the five intervals into which $[0, \frac{1}{2}]$ is partitioned. Table 3 considers the limits of the function at the breakpoints, which are the only points that could be points of discontinuity. It can be seen from the table that the function is indeed strictly monotonically increasing (using $\varepsilon \leq \frac{1}{330}$ for the point $x = \frac{1}{12}$). ∎

| The point | limit from the left | limit from the right |
|---|---|---|
| $x = \frac{1}{12}$ | $\frac{13}{11 \cdot 12} = 0.098485$ | $\frac{1}{10} - \frac{1}{2}\varepsilon$ |
| $x = \frac{1}{6}$ | $\frac{1}{5} - \varepsilon$ | $\frac{1}{5} + \varepsilon$ |
| $x = \frac{1}{3}$ | $\frac{1}{2} - \varepsilon$ | $\frac{1}{2} + \varepsilon$ |
| $x = \frac{5}{12}$ | $\frac{6}{10}$ | $\frac{6}{10}$ |
| $x = \frac{1}{2}$ | $\frac{461}{660} = 0.69848$ | 1 |

Table 3: The limits of $w$ at breakpoints

**Theorem 4.9** *Consider a bin $B^*$. The total weight of items that are packed in $B^*$ is at most $1.7$, and it is smaller for certain types of configurations. Specifically, there are several cases with respect to the huge, big, and medium items packed in the bin, as follows:*

- *If the bin does not contain a huge item then its weight is below $1.7 - 21\varepsilon$. In this case the bin can contain at most six items of size above $\frac{1}{7}$, out of which at most five are medium.*

- *If the bin contains a huge item, a large item, and possibly tiny items as well, then its weight is at most $\frac{1121}{660} \approx 1.6985$.*

- *If the bin contain a huge item, a semi-big item, and possibly small items as well, then its weight is at most $1.7$. In this case the bin contains at most one semi-small item.*

- *If the bin contains a huge item, two medium items, and possibly some small items, then its weight is at most $1.7$. The bin can contain at most one semi-small item. If the bin contains a semi-small item then the total weight is at most $1.7 - 9\varepsilon$. If the medium items satisfy the condition that the size of smaller one of them is at most $1/5$, and the size of the other item is at most $1/4$, then the total weight of the bin is no larger than $1.7 - 8\varepsilon$.*

- *If the bin contains a huge item, a medium item, and possibly some small items, then its weight is at most $1.7 - 2\varepsilon$. The bin can contain at most two semi-small items. If the medium item is of size at most $\frac{1}{5}$, or the bin contains two semi-small items of size above $\frac{1}{7}$, then the total weight is below $1.7 - 21\varepsilon$.*

- *If the bin contains a huge item and no big or medium items (but possibly some small items), then its weight is below $1.7 - 21\varepsilon$.*

**Proof.** We analyze all cases with respect to the contents of $B^*$.

First, consider the case that the bin does not contain huge item. Due to the bounds of Table 2, the total weight is at most $\frac{3}{2} + 3\varepsilon < 1.7 - 21\varepsilon$. The number of items of size above $1/7$ can never exceed 6, and the number of items of size above $1/6$ can never exceed 5.

In the remaining cases the bin contains a huge item. Thus, the bin can contain at most one big item. If the bin contains a large item, the remaining items are tiny. The total weight of items is at most $1 + \frac{13}{11} \cdot \frac{1}{2} + \frac{71}{660} = \frac{1121}{660} \approx 1.698485$.

If the bin contains a semi-big item, the remaining items are small, and there is at most one semi-small item (since the total size of two semi-small items exceeds $1/6$). Let $\frac{1}{3} < s_x \leq \frac{5}{12}$ be the size of the semi-big item. Using $s_x > 1/3$ we get

$$w(B^*) \leq 1 + \left(\frac{6}{5} - 6\varepsilon\right)\left(\frac{1}{2} - s_x\right) + \left(\frac{6}{5} - 12\varepsilon\right)s_x + \frac{1}{10} + 5\varepsilon$$
$$= 1.7 - 6\varepsilon s_x + 2\varepsilon < 1.7 \ .$$

We are left with the case that there are no big items in the bin. There can be at most two medium items. If there are two medium items in the bin, let their sizes be $\frac{1}{3} \geq s_y \geq s_x > \frac{1}{6}$. We have $s_x + s_y < \frac{1}{2}$. The total size of remaining items is less than $\frac{1}{2} - s_x - s_y \leq \frac{1}{6}$ and they are small (but their size can be close to zero) so we have

$$w(B^*) \leq 1 + \left(\frac{6}{5} - 6\varepsilon\right)\left(\frac{1}{2} - s_x - s_y\right) + \left(\frac{9}{5} - 12\varepsilon\right)(s_x + s_y) - \frac{2}{10} + 6\varepsilon$$
$$= \frac{7}{5} + (\frac{3}{5} - 6\varepsilon)(s_x + s_y) + 3\varepsilon < \frac{7}{5} + (\frac{3}{5} - 6\varepsilon)\frac{1}{2} + 3\varepsilon = 1.7 \ .$$

If the bin contains in addition to the huge item and the two medium items also a semi-small item (note that there cannot be more than one such item), then $s_x + s_y < \frac{1}{2} - \frac{1}{12} = \frac{5}{12}$ and

$$w(B^*) \leq \frac{7}{5} + (\frac{3}{5} - 6\varepsilon)(\frac{5}{12}) + 3\varepsilon = \frac{33}{20} + \frac{\varepsilon}{2} \leq 1.7 - 9\varepsilon \ .$$

If $s_x \leq 1/5$ and $s_y \leq 1/4$, then the total weight of the bin is at most

$$w(B^*) \leq \frac{7}{5} + (\frac{3}{5} - 6\varepsilon)(\frac{1}{5} + \frac{1}{4}) + 3\varepsilon = \frac{167}{100} + \frac{3\varepsilon}{10} \leq 1.7 - 8\varepsilon \ .$$

If there is exactly one medium item in the bin, denote its size by $s_y$ where $\frac{1}{6} < s_y \leq \frac{1}{3}$. The remaining items are small and the total weight in this case is bounded as follows.

$$w(B^*) \leq 1 + \left(\frac{6}{5} - 6\varepsilon\right)\left(\frac{1}{2} - s_y\right) + \left(\frac{9}{5} - 12\varepsilon\right)s_y - \frac{1}{10} + 3\varepsilon$$
$$= \frac{3}{2} + (\frac{3}{5} - 6\varepsilon)s_y \leq \frac{3}{2} + (\frac{3}{5} - 6\varepsilon)\frac{1}{3} = 1.7 - 2\varepsilon \ .$$

If this bin contains two semi-small items, both of size above $\frac{1}{7}$, then $s_y < \frac{3}{14}$, and $w(B^*) \leq \frac{3}{2} + (\frac{3}{5} - 6\varepsilon)s_y \leq \frac{3}{2} + (\frac{3}{5} - 6\varepsilon)\frac{3}{14} = \frac{57}{35} - \frac{9}{7}\varepsilon < 1.7 - 21\varepsilon$. If the medium item has size of at most $\frac{1}{5}$, then since $\frac{1}{5} < \frac{3}{14}$ the last calculation is valid as well.

If there are no medium items in the bin, then the bin has one huge item and small items (out of which at most five can be semi-small). The total weight in this case is bounded by $w(B^*) \leq 1 + \left(\frac{6}{5} - 6\varepsilon\right)\frac{1}{2} = 1.6 - 3\varepsilon < 1.7 - 21\varepsilon$. ∎

Next, we consider the total weight of bins of $A$. We start with bins containing three relatively large items.

**Lemma 4.10** *A $3^+$-bin of $A$ which contains items of a total size of at least $\frac{11}{12}$ has a total weight of at least $\frac{13}{12} \geq 1 + 13\varepsilon$. Every regular 11-bin satisfies this property.*

**Proof.** By Table 2, for every item of size $x$ which is not huge $\frac{w(x)}{x} \geq \frac{13}{11}$. The second part follows since by Lemma 4.2, a regular 11-bin always contains a total size of items of at least $\frac{11}{12}$. ∎

**Lemma 4.11** *Let $k_1 \geq 2$ be the minimum integer such that $\rho_{k_1} \leq \frac{1}{12}$, then every $(k_1+1)^+$-bin and every regular $k_1^+$-bin contains a total weight of at least $\frac{13}{12}$. Moreover, with the exception of at most one bin, every $12^+$-bin has a total weight of at least $\frac{13}{12}$.*

**Proof.** We prove the second property. If there are no $12^+$-bins then we are done. Otherwise, let $k_2 \geq 12$ be the minimum integer such that at least one $k_2$-bin exists. We have $\rho_{k_2} \leq \frac{1}{k_2} \leq \frac{1}{12}$, and thus $k_1 \leq k_2$. ∎

In what follows, we consider $k$-bins for $k = 1, 2, \ldots, 10$. For a given value of $k$, we only consider the case $\rho_k > \frac{1}{12}$, that is, the case where there are no tiny items in the bins. (Since if $\rho_k \leq \frac{1}{12}$ then $k_1$ of Lemma 4.11 satisfies $k_1 \leq k$.) Since there is no tiny items in the bin, we have $\frac{w(x)}{x} \geq \frac{6}{5} - 6\varepsilon$.

**Lemma 4.12** *A $3^+$-bin of $A$ which contains three items of size above $\frac{1}{4}$ has a total weight of at least $\frac{21}{20} \geq 1 + 13\varepsilon$. A regular $6^+$-bin of $A$ which contains at least one item of size above $\frac{1}{4}$ has a total weight of at least $\frac{21}{20} \geq 1 + 13\varepsilon$.*

**Proof.** The weight of an item of size $\frac{1}{4}$ is $\frac{7}{20}$. The first claim follows by monotonicity of $w$.

If there exists a regular $k$-bin (for $6 \leq k \leq 10$) which contains at least one item of size above $\frac{1}{4}$ then since the total size of items is at most 1, the size of the smallest item in this bin is at most $\frac{1-\frac{1}{4}}{k-1} = \frac{3}{4}/(k-1) \geq \rho_k$, we have $\rho_k \leq \frac{3}{4(k-1)}$. By Lemma 4.2, the total size of items is at least $1 - \rho_k \geq 1 - \frac{3}{4k-4} = \frac{4k-7}{4k-4} \geq \frac{17}{20}$. Let $s_y > \frac{1}{4}$ be the size of the largest item in the bin. Using Table 2 we have a total weight of at least $(\frac{17}{20} - s_y) \cdot (1.2 - 6\varepsilon) + \frac{461}{330} s_y = \frac{102}{100} - \frac{51}{10}\varepsilon + s_y(\frac{461}{330} - \frac{6}{5} + 6\varepsilon) \geq 1.02 - 5.1\varepsilon + \frac{1}{4}(\frac{65}{330} + 6\varepsilon) = \frac{7057}{6600} - 3.6\varepsilon \geq 1 + 13\varepsilon$. ∎

**Lemma 4.13** *For $6 \leq k \leq 10$, if $\rho_k > \frac{1}{12}$, then every regular $k$-bin contains a total weight of at least $\frac{36}{35} - \frac{36}{7}\varepsilon > 1 + \varepsilon$. If $\rho_5 \leq \frac{1}{7}$, then every regular 5-bin contains a total weight of at least $\frac{36}{35} - \frac{36}{7}\varepsilon \geq 1 + \varepsilon$.*

*If a regular 5-bin contains at least one item of size above $\frac{1}{4}$ then its total weight is at least $\frac{151}{140} - \frac{51}{14}\varepsilon > 1 + 13\varepsilon$.*

**Proof.** Consider a bin with $k$ items where $5 \leq k \leq 10$. By Lemma 4.2, every $6^+$-bin contains a total size of at least $\frac{6}{7}$. Moreover, if $\rho_5 \leq \frac{1}{7}$, then every regular 5-bin contains a total size of at least $\frac{6}{7}$. Using Table 2, in these cases we have a total weight of at least $\frac{6}{7} \cdot (1.2 - 6\varepsilon) = \frac{36}{35} - \frac{36}{7}\varepsilon > 1 + \varepsilon$.

Next, consider a regular 5-bin with an item of size above $\frac{1}{4}$. Let $s_y > \frac{1}{4}$ be the size of the largest item in the bin. Using Table 2 we have a total weight of at least $(\frac{6}{7} - s_y) \cdot (1.2 - 6\varepsilon) + \frac{461}{330} s_y = \frac{36}{35} - \frac{36}{7}\varepsilon + s_y(\frac{461}{330} - \frac{6}{5} + 6\varepsilon) \geq \frac{36}{35} - \frac{36}{7}\varepsilon + \frac{1}{4}(\frac{65}{330} + 6\varepsilon) = \frac{9959}{9240} - \frac{51}{14}\varepsilon > 1 + 13\varepsilon$. ∎

**Corollary 4.14** *Except for at most seven $6^+$-bins, every $6^+$-bin has a total weight of at least $1 + \varepsilon$, and it satisfies the following properties. If the bin contains three items of size above $\frac{1}{4}$, then its total weight is at least $1 + 13\varepsilon$. If the bin contains two items of size above $\frac{1}{4}$, then its total weight is at least $1 + 9\varepsilon$. If the bin contains one item of size above $\frac{1}{4}$, then its total weight is at least $1 + 5\varepsilon$.*

**Lemma 4.15** *Every 1-bin has a total weight of 1. Every regular 2-bin has a total weight of at least 1. If a 2-bin contains two big items, or at least one large item, then its total weight is at least $1 + 2\varepsilon$.*

**Proof.** Since every 1-bin has a huge item, the claim for these bins follows. Since the weight of a big item is at least $\frac{1}{2} + \varepsilon$, a 2-bin with two big items has a weight of at least $1 + 2\varepsilon$.

Next, assume that a 2-bin has an item which is not big, . By Lemma 4.4, given two items packed together in a 2-bin of $A$, the smaller item has size above $\frac{1}{4}$, and thus it is medium, while the larger item is big. Let $s_y \geq s_x$ be the sizes of the two items. We have $\frac{1}{3} < s_y \leq \frac{1}{2}$ and $\frac{1}{4} < s_x \leq \frac{1}{3}$. We have $\rho_2 \leq s_x$ and by Lemma 4.2, $s_y + s_x \geq 1 - \rho_2$ which implies $s_y \geq 1 - 2 \cdot s_x$ and $s_x \geq \frac{1-s_y}{2}$.

If the larger item is large ($s_y > \frac{5}{12}$), then the total weight of the two items is at least $\frac{13}{11} s_y + \frac{71}{660} + (\frac{9}{5} - 12\varepsilon) \cdot s_x - \frac{1}{10} + 3\varepsilon \geq \frac{13}{11} s_y + \frac{71}{660} + (\frac{9}{5} - 12\varepsilon) \cdot \frac{1-s_y}{2} - \frac{1}{10} + 3\varepsilon = \frac{599}{660} - 3\varepsilon + s_y(\frac{186}{660} + 6\varepsilon) \geq \frac{599}{660} - 3\varepsilon + \frac{5}{12}(\frac{186}{660} + 6\varepsilon) = \frac{491}{440} - \frac{\varepsilon}{2} > 1 + 2\varepsilon$ for $\varepsilon \leq \frac{1}{330}$.

If the larger item is semi-big ($\frac{1}{3} < s_y \leq \frac{5}{12}$), then the total weight of the two items is at least $(\frac{6}{5} - 12\varepsilon) s_y + \frac{1}{10} + 5\varepsilon + (\frac{9}{5} - 12\varepsilon) \cdot s_x - \frac{1}{10} + 3\varepsilon \geq (\frac{6}{5} - 12\varepsilon)(1 - 2s_x) + 8\varepsilon + (\frac{9}{5} - 12\varepsilon) \cdot s_x = 1.2 - 4\varepsilon - s_x(\frac{3}{5} - 12\varepsilon) \geq 1.2 - 4\varepsilon - \frac{1}{3}(\frac{3}{5} - 12\varepsilon) = 1$. ∎

**Lemma 4.16** *If $\rho_3 > \frac{1}{12}$, then every regular 3-bin has a total weight of at least $\frac{21}{20} > 1 + 13\varepsilon$.*

**Proof.** If the bin contains at least two big items, then since the weight of any item which is not tiny is at least $\frac{1}{10} - \frac{\varepsilon}{2}$, and the weight of any big item is at least $\frac{1}{2} + \varepsilon$, then the total weight of all three items is at least $\frac{11}{10} + \frac{3}{2}\varepsilon > \frac{21}{20}$.

We are left with the case where there is at most one big item in the bin. If the bin contains a big item then by Lemma 4.5, there is at most one semi-small item, and the third item is medium (or else the bin contains two medium items). Otherwise, by the same lemma, all three items are medium. If the bin contains three medium items, since their total size is at least $\frac{3}{4}$ (by Lemma 4.2), we have a total weight of at least $(\frac{9}{5} - 12\varepsilon)\frac{3}{4} - \frac{3}{10} + 9\varepsilon = \frac{21}{20}$.

For the remaining cases, let $s_y \geq s_x \geq s_z$ be the sizes of the items. We have $s_y > \frac{1}{3}$, and $s_y + s_x + s_z > \max\{1 - \rho_k, \frac{3}{4}\} \geq \max\{1 - s_z, \frac{3}{4}\}$.

If the big item is large, and the two other items are medium, we get a total weight of at least $\frac{13}{11}s_y + \frac{71}{660} + (\frac{9}{5} - 12\varepsilon)(s_x + s_z) - \frac{2}{10} + 6\varepsilon$. Using $s_y > 1 - s_x - 2s_z$ we get at least $\frac{13}{11}(1 - s_x - 2s_z) - \frac{61}{660} + (\frac{9}{5} - 12\varepsilon)(s_x + s_z) + 6\varepsilon = \frac{719}{660} + 6\varepsilon + s_x(\frac{408}{660} - 12\varepsilon) - s_z(\frac{372}{660} + 12\varepsilon)$. Using $s_z \leq s_x$ we get that the weight is at least $\frac{719}{660} + 6\varepsilon + s_x(\frac{36}{660} - 24\varepsilon)$. If $\frac{36}{660} - 24\varepsilon$ is non-negative, using $s_x \geq \frac{1}{6}$, the weight is at least $\frac{719}{660} + 6\varepsilon + \frac{6}{660} - 4\varepsilon = \frac{725}{660} + 2\varepsilon > \frac{21}{20}$. If $\frac{36}{660} - 24\varepsilon$ is negative, using $s_x \leq \frac{1}{3}$, the weight is at least $\frac{719}{660} + 6\varepsilon + \frac{12}{660} - 8\varepsilon = \frac{731}{660} - 2\varepsilon > \frac{21}{20}$ for $\varepsilon \leq \frac{1}{330}$.

If the big item is semi-big, and the two other items are medium, we get a total weight of at least $(\frac{6}{5} - 12\varepsilon)s_y + \frac{1}{10} + 5\varepsilon + (\frac{9}{5} - 12\varepsilon)(s_x + s_z) - \frac{2}{10} + 6\varepsilon = (\frac{6}{5} - 12\varepsilon)(s_x + s_y + s_z) + \frac{3}{5}(s_x + s_z) - \frac{1}{10} + 11\varepsilon \geq (\frac{6}{5} - 12\varepsilon)(1 - s_z) + \frac{3}{5}(2 \cdot s_z) - \frac{1}{10} + 11\varepsilon = \frac{11}{10} - \varepsilon + 12\varepsilon s_z \geq \frac{11}{10} + \varepsilon > \frac{21}{20}$, using $s_z > \frac{1}{6}$.

If the big item is large, and there is a small item, we get a total weight of at least $\frac{13}{11}s_y + \frac{71}{660} + (\frac{9}{5} - 12\varepsilon)s_x - \frac{1}{10} + 3\varepsilon + (\frac{6}{5} - 6\varepsilon)s_z \geq \frac{13}{11}(1 - 2s_z - s_x) + \frac{71}{660} + (\frac{9}{5} - 12\varepsilon)s_x - \frac{1}{10} + 3\varepsilon + (\frac{6}{5} - 6\varepsilon)s_z = \frac{785}{660} + 3\varepsilon + s_z(-\frac{768}{660} - 6\varepsilon) + s_x(\frac{408}{660} - 12\varepsilon) \geq \frac{785}{660} + 3\varepsilon + s_z(-\frac{360}{660} - 18\varepsilon) \geq \frac{785}{660} + 3\varepsilon + \frac{1}{6}(-\frac{360}{660} - 18\varepsilon) \geq \frac{785}{660} + 3\varepsilon - \frac{60}{660} - 3\varepsilon = \frac{725}{660} > \frac{21}{20}$.

If the big item is semi-big, and there is a small item, we get a total weight of at least $(\frac{6}{5} - 12\varepsilon)s_y + \frac{1}{10} + 5\varepsilon + (\frac{9}{5} - 12\varepsilon)s_x - \frac{1}{10} + 3\varepsilon + (\frac{6}{5} - 6\varepsilon)s_z = (\frac{6}{5} - 12\varepsilon)(s_y + s_x + s_z) + \frac{3}{5}s_x + 6\varepsilon s_z + 8\varepsilon \geq (\frac{6}{5} - 12\varepsilon)(1 - s_z) + \frac{3}{5}s_z + 6\varepsilon s_z + 8\varepsilon = (1.2 - 4\varepsilon) + s_z(-0.6 + 18\varepsilon) \geq 1.2 - 4\varepsilon + \frac{1}{6}(18\varepsilon - 0.6) = 1.1 - \varepsilon > \frac{21}{20}$, using $s_z \leq \frac{1}{6}$. ∎

**Lemma 4.17** *If $\rho_4 > \frac{1}{12}$, then every regular 4-bin has a total weight of at least $\min\{\frac{26}{25} + \frac{12}{5}\varepsilon, \frac{21}{20} - \frac{21}{4}\varepsilon\} \geq 1 + 9\varepsilon$.*

**Proof.** We let $s_t \leq s_z \leq s_x \leq s_y$ be the item sizes. We have $s_t + s_z + s_x + s_y \geq \frac{4}{5}$ and $2s_t + s_z + s_x + s_y \geq 1$.

First, we consider the cases where the is no small item in the bin. In this case at most one item is big, since the total size of two big items and two medium items exceeds 1. If all four items are medium, $(\frac{9}{5} - 12\varepsilon)(s_t + s_z + s_x + s_y) - \frac{4}{10} + 12\varepsilon \geq \frac{4}{5}(\frac{9}{5} - 12\varepsilon) - \frac{4}{10} + 12\varepsilon = \frac{26}{25} + 2.4\varepsilon$.

If one item is big, since the weight of a big item is at least $\frac{1}{2} + \varepsilon$ and the weight of a medium item is at least $\frac{1}{5} + \varepsilon$, we get a total weight of at least $1.1 + 4\varepsilon > \frac{26}{25} + \frac{12}{5}\varepsilon$.

If the bin contains at least one small item, then the number of such items is at most three, since the total size of four small items is at most $\frac{2}{3}$. If $s_t \leq \frac{1}{8}$, $s_t + s_z + s_x + s_y \geq \frac{7}{8}$, and using Table 2, the total weight is at least $\frac{7}{8} \cdot (1.2 - 6\varepsilon) = \frac{21}{20} - 5.25\varepsilon$. If the bin contains two big items, then their total weight together with at least one small item is at least $2(\frac{1}{2} + \varepsilon) + \frac{1}{10} - \frac{\varepsilon}{2} = 1.1 + 1.5\varepsilon > \frac{21}{20} - 5.25\varepsilon$.

We are left with the case $\frac{1}{6} \geq s_t > \frac{1}{8}$, and there is at most one big item, that is $s_x \leq \frac{1}{3}$. We have that the weight of the smallest item is at least $\frac{3}{20} - \frac{3}{4}\varepsilon$, and $s_t + s_z + s_x + s_y > \frac{5}{6}$.

Consider the case of one small item, we have three cases. If all additional items are medium, the total weight is at least $(\frac{6}{5} - 6\varepsilon)s_t + (\frac{9}{5} - 12\varepsilon)(s_z + s_x + s_y) - \frac{3}{10} + 9\varepsilon \geq (\frac{6}{5} - 6\varepsilon)s_t + (\frac{9}{5} - 12\varepsilon)(1 - 2s_t) - \frac{3}{10} + 9\varepsilon = (\frac{9}{5} - 12\varepsilon - \frac{3}{10} + 9\varepsilon) + s_t(\frac{6}{5} - 6\varepsilon - 2(\frac{9}{5} - 12\varepsilon)) = (\frac{3}{2} - 3\varepsilon) + s_t(-\frac{12}{5} + 18\varepsilon) \geq (\frac{3}{2} - 3\varepsilon) + \frac{1}{6}(-\frac{12}{5} + 18\varepsilon) = 1.1 > \frac{21}{20} - 5.25\varepsilon$. If two items are medium and the largest one is big, the total weight of these four types of items is at least $\frac{1}{2} + \varepsilon + 2(\frac{1}{5} + \varepsilon) + \frac{3}{20} - \frac{3}{4}\varepsilon = 1.05 + 2.25\varepsilon > \frac{21}{20} - 5.25\varepsilon$.

If there are two small items, we have three cases. If both additional items are medium, the total weight is at least $(\frac{6}{5} - 6\varepsilon)(s_t + s_z) + (\frac{9}{5} - 12\varepsilon)(s_x + s_y) - \frac{2}{10} + 6\varepsilon = (\frac{9}{5} - 12\varepsilon)(s_t + s_z + s_x + s_y) - (\frac{3}{5} - 6\varepsilon)(s_t + s_z) - \frac{2}{10} + 6\varepsilon \geq \frac{5}{6}(\frac{9}{5} - 12\varepsilon) - \frac{2}{6}(\frac{3}{5} - 6\varepsilon) - \frac{2}{10} + 6\varepsilon = 1.1 - 2\varepsilon > \frac{21}{20} - 5.25\varepsilon$. If one item is medium and the other one is semi-big, the total weight is at least $(\frac{6}{5} - 6\varepsilon)(s_t + s_z) + (\frac{9}{5} - 12\varepsilon)s_x - 0.1 + 3\varepsilon + (\frac{6}{5} - 12\varepsilon)s_y + 0.1 + 5\varepsilon = (\frac{6}{5} - 6\varepsilon)(s_t + s_z + s_x + s_y) + (\frac{3}{5} - 6\varepsilon)s_x - 6\varepsilon s_y + 8\varepsilon \geq \frac{5}{6}(\frac{6}{5} - 6\varepsilon) + \frac{1}{6}(\frac{3}{5} - 6\varepsilon) - 6\varepsilon\frac{5}{12} + 8\varepsilon = 1.1 - 0.5\varepsilon > \frac{21}{20} - 5.25\varepsilon$. If one item is medium and the other one is large, the total weight is at least $(\frac{6}{5} - 6\varepsilon)(s_t + s_z) + (\frac{9}{5} - 12\varepsilon)s_x - 0.1 + 3\varepsilon + \frac{13}{11}s_y + \frac{71}{660} = (\frac{6}{5} - 6\varepsilon)(s_t + s_z + s_x + s_y) + (\frac{3}{5} - 6\varepsilon)s_x + \frac{5}{660} + 3\varepsilon - s_y(\frac{6}{5} - 6\varepsilon - \frac{13}{11}) \geq \frac{5}{6}(\frac{6}{5} - 6\varepsilon) + \frac{1}{6}(\frac{3}{5} - 6\varepsilon) + \frac{5}{660} + 3\varepsilon - \frac{1}{2}(\frac{6}{5} - 6\varepsilon - \frac{13}{11}) = \frac{725}{660} > \frac{21}{20} - 5.25\varepsilon$.

Finally, consider the case of three small items. By Lemma 4.6, the fourth item is big. If the largest item is semi-big, the total weight is at least $(\frac{6}{5} - 6\varepsilon)(s_t + s_z + s_x) + (\frac{6}{5} - 12\varepsilon)s_y + 0.1 + 5\varepsilon = (\frac{6}{5} - 6\varepsilon)(s_t + s_z + s_x + s_y) - 6\varepsilon s_y + 0.1 + 5\varepsilon \geq \frac{5}{6}(\frac{6}{5} - 6\varepsilon) - \frac{5}{12} \cdot 6\varepsilon + 0.1 + 5\varepsilon = 1.1 - 2.5\varepsilon > \frac{21}{20} - 5.25\varepsilon$. If the largest item is large, the total weight is at least $(\frac{6}{5} - 6\varepsilon)(s_t + s_z + s_x) + \frac{13}{11}s_y + \frac{71}{660} \geq (\frac{6}{5} - 6\varepsilon)(s_t + s_z + s_x + s_y) + (\frac{13}{11} - 1.2 + 6\varepsilon)s_y + \frac{71}{660} \geq \frac{5}{6}(\frac{6}{5} - 6\varepsilon) + \frac{1}{2}(\frac{13}{11} - 1.2 + 6\varepsilon) + \frac{71}{660} = \frac{725}{660} - 2\varepsilon > \frac{21}{20} - 5.25\varepsilon$, since $\frac{13}{11} - 1.2 + 6\varepsilon \leq 0$ for $\varepsilon \leq \frac{1}{330}$. ∎

The 5-bins need to be considered more carefully. A *basic bin* or a *basic 5-bin* is a regular 5-bin which contains five items of sizes in $(\frac{1}{7}, \frac{1}{5}]$. Other 5-bins are called non-basic.

**Lemma 4.18** *If $\rho_5 > \frac{1}{7}$, then a non-basic regular 5-bin has a total weight of at least $1 + 5\varepsilon$. If a non-basic regular 5-bin has at least two items of size above $\frac{1}{4}$ then it has a total weight of at least $1 + 13\varepsilon$.*

**Proof.** Consider a non-basic 5-bin. All items have sizes above $\frac{1}{7}$. If there are at least two items of size above $\frac{1}{4}$, the total size of the items is at least $\frac{13}{14}$, and their weight is at least $\frac{13}{11} \cdot \frac{13}{14} = \frac{169}{154} > 1 + 13\varepsilon$.

The weight of each item is at least $\frac{1}{7}(\frac{6}{5} - 6\varepsilon) = \frac{6}{35} - \frac{6}{7}\varepsilon$. Let $s_y > \frac{1}{5}$ denote the size of the largest item in the bin. Such an item must exist, since otherwise the bin is basic. If $s_y > \frac{1}{3}$, then the weight of this item is at least $\frac{1}{2} + \varepsilon$ and the total weight of all five items is at least $4(\frac{6}{35} - \frac{6}{7}\varepsilon) + \frac{1}{2} + \varepsilon = \frac{83}{70} - \frac{17}{7}\varepsilon > 1 + 13\varepsilon$.

Consider the case $\frac{1}{5} < s_y \leq \frac{1}{3}$. Let $\ell$ be the total size of the five items. If the bin does not contain a small item, then all items are medium and $\ell \geq \frac{4}{6} + s_y \geq \frac{13}{15}$, and the total weight is at least $(\frac{9}{5} - 12\varepsilon)\ell - \frac{5}{10} + 15\varepsilon \geq (\frac{9}{5} - 12\varepsilon)\frac{13}{15} - \frac{5}{10} + 15\varepsilon = \frac{53}{50} + \frac{23}{5}\varepsilon > 1 + 13\varepsilon$.

If there is at least one small item, let $s_t > \frac{1}{7}$ denote its size. We have $\ell \geq 4s_t + s_y$ and $\ell > 1 - s_t$, and we get $\ell \geq 4s_t + s_y > 4(1 - \ell) + s_y$, or equivalently, $5\ell > 4 + s_y$.

Using Table 2, the total weight is at least

$$\left(\frac{6}{5} - 6\varepsilon\right)(\ell - s_y) + \left(\frac{9}{5} - 12\varepsilon\right)s_y - \frac{1}{10} + 3\varepsilon \geq \left(\frac{6}{5} - 6\varepsilon\right)\ell + \left(\frac{3}{5} - 6\varepsilon\right)s_y - \frac{1}{10} + 3\varepsilon$$
$$\geq \left(\frac{6}{5} - 6\varepsilon\right)\left(\frac{4 + s_y}{5}\right) + \left(\frac{3}{5} - 6\varepsilon\right)s_y - \frac{1}{10} + 3\varepsilon$$
$$= \frac{43}{50} - \frac{9}{5}\varepsilon + s_y(0.84 - 7.2\varepsilon) \geq \frac{43}{50} - \frac{9}{5}\varepsilon + \frac{1}{5}(0.84 - 7.2\varepsilon) = 1.028 - 3.24\varepsilon > 1 + 5\varepsilon.$$

∎

**Corollary 4.19** *Except for at most ten $3^+$-bins, every $3^+$-bin which is not a basic 5-bin has a total weight of at least $1 + \varepsilon$ and the following properties. If the bin contains three items of size above $\frac{1}{4}$, then its total weight is at least $1 + 13\varepsilon$. If the bin contains two items of size above $\frac{1}{4}$, then its total weight is at least $1 + 9\varepsilon$. If the bin contains one item of size above $\frac{1}{4}$, then its total weight is at least $1 + 5\varepsilon$.*

Next, we focus on basic bins. A $(i, 5 - i)$-bin is a basic 5-bin which contains $i$ items of size in $(\frac{1}{7}, \frac{1}{6}]$ and $5 - i$ items of size in $(\frac{1}{6}, \frac{1}{5}]$. By Lemma 4.2, the relevant values of $i$ are $i = 0, 1, 2, 3, 4$.

**Lemma 4.20** *A $(0, 5)$-bin contains items of a total weight of at least $1 + 5\varepsilon$. A $(1, 4)$-bin contains items of a total weight of at least $1 + 3\varepsilon$. A $(2, 3)$-bin contains items of a total weight of at least $1 + \varepsilon$. A $(3, 2)$-bin contains items of a total weight of at least $1 - \varepsilon$. A $(4, 1)$-bin contains items of a total weight of at least $1 - 3\varepsilon$.*

**Proof.** The weight of an item of size above $\frac{1}{6}$ is at least $\frac{1}{5}+\varepsilon$, and thus the first claim follows. Otherwise, consider a basic $(i, 5-i)$ bin for some $1 \leq i \leq 4$, and let $\ell_1$ denote the total size of items of size in $(\frac{1}{7}, \frac{1}{6}]$ and let $\ell_2$ denote the total size of other items. We have $\ell_1 + \ell_2 > \frac{5}{6}$, $\frac{5-i}{6} < \ell_2 \leq \frac{5-i}{5}$, $\frac{i}{7} < \ell_1 \leq \frac{i}{6}$.

We have a total weight of at least $(\frac{6}{5}-6\varepsilon)\ell_1 + (\frac{9}{5}-12\varepsilon)\ell_2 + (5-i)(3\varepsilon - 0.1) = (\frac{6}{5}-6\varepsilon)(\ell_1 + \ell_2) + (\frac{3}{5}-6\varepsilon)\ell_2 + (5-i)(3\varepsilon - 0.1) \geq (\frac{6}{5}-6\varepsilon)\frac{5}{6} + (\frac{3}{5}-6\varepsilon)\frac{5-i}{6} + (5-i)(3\varepsilon - 0.1) \geq (\frac{6}{5}-6\varepsilon)\frac{5}{6} + (5-i)(0.1 - \varepsilon + 3\varepsilon - 0.1) = 1 - 5\varepsilon + 2\varepsilon(5-i) = 1 + 5\varepsilon - 2\varepsilon i$. ∎

Next, we define modified weights as follows. Every item which is packed into a basic 5-bin in $A$, and it is packed in OPT in a bin for which the upper bound on the weight proved in Theorem 4.9 is not 1.7 gets a new weight. If it is medium then its weight is increased by $4\varepsilon$ compared to its original weight, and otherwise is is increased by $\varepsilon$.

We consider each possible bin type of OPT, as they are listed in the statement of Theorem 4.9, and compute the maximum increase of its total weight. A bin can contain at most six items of size in $(\frac{1}{7}, \frac{1}{6}]$, out of which, at most five items have size in $(\frac{1}{6}, \frac{1}{5}]$, thus the weight is increased by at most $21\varepsilon$. Thus for bins whose weight was at most $1 - 21\varepsilon$, the total weight after the increase does not exceed 1.7.

The bins of the second type do not have items of size in $(\frac{1}{7}, \frac{1}{5}]$ so their weights are not modified. The weights of bins of the third type, and bins of the fourth type of weight 1.7 is not modified by definition. Other bins of the fourth type can have the following structures. If there is a semi-small item, then since there is a huge item, then there are at most three items of size in $(\frac{1}{7}, \frac{1}{5}]$, one of which is semi-small, so the weight of the bin increases by at most $9\varepsilon$. If there is no semi-small item, but one medium item has size of at most $\frac{1}{4}$ and the other one of at most $\frac{1}{5}$, the weight was increased by at most $8\varepsilon$. In the fifth type, we only need to consider the case where the medium item has size above $\frac{1}{5}$, and there is at most one semi-small item, then the weight of the bin is increased by $\varepsilon$ and it is now $1.7 - \varepsilon$.

Thus, the only items of size in $(\frac{1}{7}, \frac{1}{5}]$ which are packed into basic 5-bins, whose weight was not increased come from two types of bins in OPT. We call these bin types *bad bins* of OPT. Both these bin types have a huge item, and possibly have tiny items. In addition, the first type of a bad bin has a semi-big item, and the item coming from the basic bin in $A$ has size in $(\frac{1}{7}, \frac{1}{6}]$. The second type of bad bin has a medium item of size in $(\frac{1}{4}, \frac{5}{14}]$, and the item coming from a basic bin in $A$ has a size in $(\frac{1}{6}, \frac{1}{5}]$. This defines a (partial) matching on the items, where two items of sizes in $(\frac{1}{7}, \frac{1}{5}]$, and in $(\frac{1}{4}, \frac{5}{14}]$ respectively, are matched if they are packed together into a bad bin of OPT. In each matched pair, the larger item is called the *larger partner* of the smaller item and the smaller item is called the *smaller partner* of the larger item.

We apply one other modification of weights. Consider 2-bins of $A$ which contain two big items. Recall that in this case the weight of the bin is at least $1 + 2\varepsilon$. For every such big item, if it has a smaller partner, a weight of $\varepsilon$ is transferred from the big item to its smaller partner. Note that the smaller partner must be semi-small. As a result, all regular 2-bins of $A$ now have weights of at least 1.

Recall that the larger partners are of size in $(\frac{1}{4}, \frac{5}{14}]$. Since they are not huge, if such an item is not packed in a 2-bin, then it is packed into a $3^+$-bin in $A$. We reduce the weight of item of size in $(\frac{1}{4}, \frac{5}{14}]$, packed into a $3^+$-bin in $A$ by $4\varepsilon$ if it is medium and by $\varepsilon$ if it is semi-big (since the partners in these cases, if they exist, are medium and semi-small, respectively). Each bin of $A$ can have at most three such items. By Corollary 4.19, a bin of $A$ whose weight is reduced still has a weight of at least $1 + \varepsilon$ after the reduction. For every such item which has a smaller partner, we increase the weight of its smaller partner by the reduced amount ($\varepsilon$ for a smaller partner which is semi-small, and $4\varepsilon$ for a medium smaller partner).

The only items of size in $(\frac{1}{7}, \frac{1}{5}]$ whose weight was not increased are those packed into bad bins of OPT, and their partners are packed into 2-bins in $A$ that have one medium item and one semi-big item. We let these items be denoted by $P_s$ and let the set of their larger partners be denoted by $P_\ell$. If at least one item of $P_\ell$ is medium, then we let $\frac{1}{3} - \alpha \leq \frac{1}{3}$ denote the size of the smallest item in $P_\ell$. Otherwise, all items of $P_\ell$ are semi-big and their partners are semi-small, so there are only semi-small items in basic 5-bins whose weight was not increased.

**Claim 4.21** *If $\alpha$ was defined, then a semi-small item of $P_s$ must be smaller than $\frac{1}{6} - \alpha$ (except for possibly one such item), and a medium item of $P_s$ is smaller than $\frac{1}{6} + \alpha$.*

**Proof.** We have $\rho_2 \leq \frac{1}{3} - \alpha$. The total size of items in a regular 2-bin thus exceeds $\frac{2}{3} + \alpha$. For every regular 2-bin which has one medium item, the other item must have size above $\frac{1}{3} + \alpha$. (The last claim holds for every regular 2-bin with a medium item, unrelated to the existence of partners.) The small partner of the bigger item (if such a partner exists) has size below $\frac{1}{6} - \alpha$ (since they are packed with a huge item in a bin of OPT). There may be one small partner of an item coming from the special bin, which is larger.

Since all items in 2-bins which are medium have sizes of at least $\frac{1}{3} - \alpha$, their small partners have sizes below $\frac{1}{6} + \alpha$. ∎

**Lemma 4.22** *Using the new weights, every basic 5-bin (except for at most one bin) has a weight of at least $1 + \varepsilon$.*

**Proof.** We only need to consider $(3, 2)$ bins and $(4, 1)$ bins. Moreover, since the previous weight of each such bin was at least $1 - 3\varepsilon$, if at least one medium item has an increased weight then we are done, since the weight of the bin is at least $1 - 3\varepsilon + 4\varepsilon = 1 + \varepsilon$. Thus, we only need to consider the case that $\alpha$ was defined (since otherwise all medium items in basic 5-bins have increased weight.

Recall that small items whose weight is unchanged have sizes below $\frac{1}{6} - \alpha$ and medium items whose weight is unchanged have sizes below $\frac{1}{6} + \alpha$ (except for at most one item whose weight is unchanged, called the unique item). Consider a basic 5-bin which contains a medium item (or items) whose weight was not increased, and the unique item is not in this bin. Let $n_c$ denote the number of (semi-small) items whose weight was changed. Their total size is at most $\frac{n_c}{6}$, and since the total size of items in a 5-bin exceeds $\frac{5}{6}$, the total size of the $5 - n_c$ remaining items is at least $\frac{5-n_c}{6}$. For items whose weight was not increased, a medium item has size below $\frac{1}{6} + \alpha$, and a semi-small item has size below $\frac{1}{6} - \alpha$, thus out of the remaining $5 - n_c$ items, the number of medium items must be larger than the number of semi-small items (whose weight was not increased). Therefore, in a $(4, 1)$ bin, all semi-small items must have an increased weight, which gives a total of at least $1 - 3\varepsilon + 4\varepsilon = 1 + \varepsilon$. In a $(3, 2)$ bin, there is at most one semi-small item whose weight was not increased, which gives a total weight of at least $1 - \varepsilon + 2\varepsilon = 1 + \varepsilon$. ∎

**Corollary 4.23** *In the packing $A$, every $3^+$-bin (except for at most eleven bins) has a modified weight of at least $1 + \varepsilon$. Every 1-bin and every 2-bin (except for at most one bin) has a modified weight of at least 1. In $\text{OPT}(G)$, the weight of every bin is at most 1.7.*

We find an upper bound on the number of bins in $A$ whose weight is strictly below $1 + \varepsilon$.

**Lemma 4.24** *The number of regular 1-bins and 2-bins in $A$ is at most $\frac{3}{2}\text{OPT}(G)$.*

**Proof.** The number of items of size above $\frac{1}{4}$ is at least twice the number of 1-bins and 2-bins, since each such bin has two such items (a huge item counts as two such items), while an optimal solution can combine at most three such items in each bin. ∎

We find bounds on the total weight of all items in $G$. In the optimal solution we have $\text{OPT}(G)$ bins, each one of weight of at most 1.7. As for the bins of $A$, all bins have weights of at least 1 (except for at most twelve bins, which we call special). Moreover, the $3^+$-bins of $A$ (except for the special $3^+$-bins) have weights of at least $1+\varepsilon$. Thus, there are at least $cost(A) - 12 - \frac{3}{2}\text{OPT}(G) > (0.2 - \delta)\text{OPT}(G)$ bins of $A$ which have weights of at least $1 + \varepsilon$, since we assumed $cost(A) > (1.7 - \delta)\text{OPT}(G) + 12$. The total weight of the bins of $A$ is at least $cost(A) + \varepsilon \cdot (0.2 - \delta)\text{OPT}(G) - 12$.

Let $W$ be the total weight of items in $G$. We have $W > cost(A) - 12 + \varepsilon((0.2-\delta)\text{OPT}(G)) > (1.7-\delta)\text{OPT}(G) + \varepsilon((0.2-\delta)\text{OPT}(G)) = (1.7 - \delta + \varepsilon(0.2-\delta))\text{OPT}(G)$, and $W \leq 1.7 \cdot \text{OPT}(G)$. Combining the two bounds on $W$ we get $1.7 - \delta + \varepsilon(0.2 - \delta) < 1.7$, or alternatively, $\delta(1+\varepsilon) > 0.2\varepsilon$, contradicting the definition of $\delta$. Letting $\varepsilon = \frac{1}{330}$ and $\delta = \frac{1}{1655}$ we find an upper bound of $1.7 - \delta \approx 1.6993996$ on the PoA.

## 5 The SPoA and SPoS for unit weights

We consider the quality of strong equilibria.

While any SNE is (by definition) a WPO-NE, and every SPO-NE is a WPO-NE, not every SNE is SPO-NE and not every SPO-NE is a SNE. Consider the following simple examples.

Let $J_1$ be the input consisting of four items of the sizes 0.55, 0.56, 0.34 and 0.35. Consider a packing where each one of the two first items is packed in a dedicated bin and the two last items are packed in a third bin. The costs of the items are $1, 1, \frac{1}{2}$ and $\frac{1}{2}$. This packing is not strictly Pareto optimal, since in the packing where the third item is combined with the first one, and the fourth item with the second one, all costs are $\frac{1}{2}$. However, the first packing is a strong equilibrium since the last two items cannot have costs below $\frac{1}{2}$ in any packing, so they have no incentive to join a coalition, while the two other items cannot reduce their costs without the cooperation of the last two items.

Let $J_2$ be the input consisting of six items of the sizes 0.52, 0.54, 0.24, 0.23, 0.22, and 0.21. Consider a packing where the first item, third item, and fourth item are packed in one bin, and the remaining items are packed in another bin. This packing is strictly Pareto optimal, since every item has cost $\frac{1}{3}$, and the only way that each one of the two first items can have such a cost is that it is packed with two of the smaller items. However, the packing is not a SNE, since the four small items can form a coalition and move to a new bin together, reducing their costs to $\frac{1}{4}$. It can be seen that the solution is a solution of optimal social cost, while no such solution is a SNE. In this section we will show that indeed the SPoS is much higher than 1.

### 5.1 The SPoA

One special case of the GSC for unit weights is NFI, since by packing the smallest items one by one, the resulting subset has maximum cardinality. It was proved in [26] that NFI which is a polynomial time algorithm finds a NE. We have shown that NFI is a polynomial time algorithm which finds a SNE. Since the tight asymptotic approximation ratio of NFI is $T_\infty$, this gives a lower bound on the SPoA. (This lower bound is achieved with items of sizes $\frac{1}{t_i-1} + \varepsilon$ for a small $\varepsilon > 0$).

Since the generic algorithm may give outputs that cannot be created by NFI (for example if the input consists of four items of sizes $0.34, 0.35, 0.36, 0.37$, there are three possible outputs of the algorithm, and only one of them can be achieved by NFI), we prove an upper bound of the same value separately, by showing the no execution of the algorithm can output a larger number of bins than that of NFI. Note that some executions may result in a smaller number of bins than that of NFI (which we discuss in more detail in the study of the SPoS). One example is an input with two items of size 0.4 and two items of size 0.6. NFI uses three bins, but an algorithm which chooses two items of different size for the first bin will only use two bins.

**Lemma 5.1** *For an input $I$, let $n_I = \text{NFI}(I)$ be the number of bins created by NFI. Any execution of the algorithm GSC results in at most $n_I$ bins. Thus NFI computes the worst strong equilibrium for every input.*

**Proof.** Assume by contradiction that some execution of GSC created at least $n_I + 1$ bins. Let $B_1, B_2, \ldots, B_{n_I}$ denote the bins of NFI, in the order of creation, and let $\mathcal{B}_1, \mathcal{B}_2, \ldots, \mathcal{B}_{n_I}, \mathcal{B}_{n_I+1}$ denote the first $n_I + 1$ bins of the other output of GSC, in the order of creation. Since $|\mathcal{B}_{n_I+1}| > 0$, $\sum_{j=1}^{n_I} |B_j| > \sum_{j=1}^{n_I} |\mathcal{B}_j|$ so there exists at least one value $i$ such that $1 \leq i \leq n_I$ such that

$\sum_{j=1}^{i} |B_j| > \sum_{j=1}^{i} |\mathcal{B}_j|$. We let $1 \leq i' \leq n_I$ denote a minimal such value. Let $k = \sum_{j=1}^{i'} |B_j|$, $k' = |B_{i'}|$, $\kappa = \sum_{j=1}^{i'} |\mathcal{B}_j|$, $\kappa' = |\mathcal{B}_{i'}|$. Using this notation, $k > \kappa$. Note that $\sum_{j=1}^{i'-1} |B_j| \leq \sum_{j=1}^{i'-1} |\mathcal{B}_j|$ (i.e., $k - k' \leq \kappa - \kappa'$) and thus $k' > \kappa'$. This clearly holds if $i' = 1$, and for the case $i' > 1$ it holds since otherwise $i' - 1$ would satisfy the requirement and $i'$ would not be minimal.

Let $X = \cup_{j=1}^{i'} B_j$. We claim that every $k'$ items of $X$ fit into one bin. This holds since the $k'$ largest items of $X$ are packed into $B_{i'}$. We will show that the set $I \setminus \cup_{j=1}^{i'-1} \mathcal{B}_j$ contains a non-empty subset of $k - \kappa + \kappa'$ items which can fit into one bin. Note that using $k > \kappa$, $k - \kappa + \kappa' > \kappa'$, and using $k - k' \leq \kappa - \kappa'$, $k - \kappa + \kappa' \leq k'$.

Consider the set $Y = X \setminus \cup_{j=1}^{i'-1} \mathcal{B}_j$. We have $Y \subseteq I \setminus \cup_{j=1}^{i'-1} \mathcal{B}_j$, so all items of $Y$ are available for packing after the bins $\mathcal{B}_1, \ldots, \mathcal{B}_{i'-1}$ were created. Recall that any $k'$ items of $Y$ can fit into a bin and therefore any $k - \kappa + \kappa' \leq k'$ items of $Y$ can fit into a bin. In addition, $|Y| \geq |X| - (\kappa - \kappa') = k - \kappa + \kappa'$. Thus, when the items of $\mathcal{B}_{i'}$ are selected by GSC, it is possible to pick a set of $k - \kappa + \kappa' > \kappa'$ items rather than picking $\mathcal{B}_{i'}$ which has only $\kappa'$ items, which is a contradiction. ∎

We have proved the following theorem.

**Theorem 5.2** *The SPoA is equal to $T_\infty \approx 1.69103$.*

## 5.2 The SPoS

We turn to the study of the SPoS and define a different sequence as follows. Let $\tau_1 = 2$, $\tau_2 = 4$, $\tau_3 = 9$, $\tau_4 = 37$, and for $i \geq 4$, $\tau_{i+1} = \tau_i(\tau_i - 1) + 1$. Since $\tau_i \geq 2$ for all $i$, $\tau_{i+1} \geq \tau_i + 1$ and the sequence is monotonically increasing. In fact, we can show that $\tau_{i+1} \geq \tau_i + 2$ and $\tau_{i+1} \geq 2\tau_i$. This holds for $i = 1, 2, 3$, and for $i \geq 4$, $\tau_i \geq 37$, and $\tau_{i+1} = \tau_i(\tau_i - 1) + 1 \geq 36\tau_i + 1 > 2\tau_i + 2$. We next show that $\frac{\tau_{i+1} - 2}{\tau_{i+1}} + \frac{1}{\tau_i} \geq 1$ holds for all $i \geq 1$. This last property is equivalent to $\tau_{i+1} \geq 2\tau_i$. Finally, note that for $i \geq 4$, $\tau_{i+1} - 1$ is divisible by $\tau_i - 1$ and by $\tau_i$. The sequence satisfies $\sum_{i=1}^{\infty} \frac{1}{\tau_i} + \frac{1}{\tau_3} = 1$. Let $\Theta_\infty = \sum_{i=1}^{\infty} \frac{1}{\tau_i - 1} + \frac{1}{\tau_3 - 1} \approx 1.6118624$.

**Theorem 5.3** *The SPoS is at least $\Theta_\infty$.*

**Proof.** We define an input for which the output of NFI is the unique output of GSC. Thus, there is a unique SNE (up to swapping locations of items of equal size).

We define the set of items via an optimal solution. This optimal solution which has $N$ bins, each of which has one item of size $\frac{1}{\tau_i} + \varepsilon$ for $1 \leq i \leq j$ for a fixed value of $j \geq 10$, and an additional item of size $\frac{1}{\tau_3} + \varepsilon$. The value of $\varepsilon$ is chosen so that it satisfies $\varepsilon < \frac{1}{(j+1) \cdot \tau_{j+1}}$. The total size of items in a bin is $1 - \frac{1}{\tau_{j+1}} + (j+1)\varepsilon \leq 1$. The number $N$ is chosen to be divisible by $2(\tau_j - 1)$, and thus it is divisible by any $\tau_i - 1$ for $1 \leq i \leq j$.

Since all items have sizes of at least $\frac{1}{\tau_j} + \varepsilon$, the first bin should contain at most $\tau_j - 1$ items. It is possible to pack $\tau_j - 1$ items of size $\frac{1}{\tau_j} + \varepsilon$, and any other combination is impossible since for $j > 1$ we have $\frac{1}{\tau_{j-1}} + \varepsilon + (\tau_j - 2)(\frac{1}{\tau_j} + \varepsilon) > \frac{1}{\tau_{j-1}} + \frac{\tau_j - 2}{\tau_j} \geq 1$. Thus, the algorithm packs $\frac{N}{\tau_j - 1}$ identical bins. After this packing is completed, the algorithm is faced with the same situation for $j - 1, j - 2, \ldots$. In the case that only one type of items remains (of size $\frac{1}{2} + \varepsilon$), the items must be packed into dedicated bins. We get that the algorithm packs $\sum_{i=1}^{j} \frac{N}{\tau_i - 1} + \frac{1}{\tau_3 - 1}$ bins while an optimal solution has $N$ bins. Letting $j$ tend to infinity we can get arbitrarily close to $\Theta_\infty$. ∎

Next, we prove a tight upper bound. For this we describe a class of specific runs of GSC(which can be applied in polynomial time). For every input $I$, such a run creates a SNE of cost of at most $\Theta_\infty \text{OPT}(I) + 20$. Note that this output is not necessarily the best SNE (while NFI necessarily outputs the worst SNE). It can be seen that the problem of finding the best SNE is strongly NP-hard. Consider the following decision problem: given an instance of bin packing and an integer $m$, does there exist a SNE packing with at most $m$ bins. It is possible to define a reduction from 3-PARTITION to this problem. The input of 3-PARTITION is an integer $B$ and $3m$ items of integers in $(\frac{B}{4}, \frac{B}{2})$, and sum $3B$, and the question is whether they can be partition into subsets of three items so that the sum of each triple is $B$. By scaling the numbers by $B$, we get an instance of bin packing. Note that no four items can be packed into one bin, and therefore if a packing into $m$ bins exists then this packing must be a SNE.

We use the following parameters

- $\delta_1 = \frac{1}{12} \approx 0.83333$,
- $\delta_3 = \frac{3}{868} \approx 0.00345622$,
- $\delta_4 = \frac{1}{40} = 0.025$,
- $\delta_5 = \frac{1}{60} \approx 0.016667$,
- $\delta_6 = \frac{1}{155} \approx 0.006451613$,
- $\delta_7 = \frac{1}{360} \approx 0.0027778$,
- $\delta_8 = \frac{1}{144} \approx 0.009444$,
- $\delta_9 = \frac{11}{1736} \approx 0.0063364$,
- $\delta_{10} = \frac{1}{308} \approx 0.003246753$,
- $\delta_{11} = \frac{1}{630} \approx 0.0015873$,

Note that $10\delta_{11} + \delta_6 = \frac{1}{42}$, $9\delta_{10} + \delta_6 \leq \frac{3}{77}$, $8\delta_9 + \delta_6 = \frac{2}{35}$, and $5\delta_6 + \delta_3 = \frac{1}{28}$. We define the algorithm STEPS, which is a special case of GSC. This algorithm executes NFI most of the time, but in several cases it tries to create bins which are packed in a better way. In such cases the algorithm creates lists out of which the items for the bins which are not packed using NFI are chosen. At the time of creation of the list, the items are not removed from the input, and are still available in steps where the lists are not used (unless the item has been packed). When the algorithm applies NFI, this means that it repeatedly picks a maximum prefix of the (non-decreasingly) sorted list of items, removing it from the input set of items and assigning it to a bin. The item is also removed from all lists which were created by the algorithm.

**Algorithm** STEPS

Let $I$ be an item set.

**Step 1.** Apply NFI. For each new bin, if the last item packed into the bin has size above $\frac{1}{12}$ (so no items of size at most $\frac{1}{12}$ remain), go to step 2.

**Step 2.** Since all remaining items are larger than $\frac{1}{12}$, no bin with more than eleven items can be formed. In this step we form bins with eleven items. Create a list $X_{11}$ of unpacked items of size in $(\frac{1}{12} \approx 0.08333, \frac{1}{12} + \delta_{11} = \frac{107}{1260} \approx 0.084920635]$ and a list $X_6$ of (unpacked) items of size in $(\frac{1}{7} \approx 0.142857, \frac{1}{7} + \delta_6 = \frac{162}{1085} \approx 0.149309]$. Recall that the items of $X_6$ and $X_{11}$ are not removed from $I$ at this time.

As long as $|X_6| > 0$ and $|X_{11}| > 9$, pack 10 items of $X_{11}$ with one item of $X_6$ into a bin, and remove them from these sets and from $I$. The items fit into the bin since their total size is at most $10(\frac{1}{12} + \delta_{11}) + (\frac{1}{7} + \delta_6) = \frac{41}{42} + 10\delta_{11} + \delta_6 \leq 1$. Go to step 3.

**Step 3.** Apply NFI. For each new bin, if the last item packed into the bin has size above $\frac{1}{11}$, go to step 4.

**Step 4.** Since all remaining items are larger than $\frac{1}{11}$, no bin with more than ten items can be formed. In this step we form bins with ten items. Create a list $X_{10}$ of unpacked items of size in $(\frac{1}{11} \approx 0.090909, \frac{1}{11} + \delta_{10} = \frac{319}{3388} \approx 0.09415584]$.

As long as $|X_6| > 0$ and $|X_{10}| > 8$, pack 9 items of $X_{10}$ with one item of $X_6$ into a bin, and remove them from these sets and from $I$. The items fit into the bin since their total size is at most $9(\frac{1}{11} + \delta_{10}) + (\frac{1}{7} + \delta_6) = \frac{74}{77} + 9\delta_{10} + \delta_6 \leq 1$. Go to step 5.

**Step 5.** Apply NFI. For each new bin, if the last item packed into the bin has size above $\frac{1}{10}$, go to step 6.

**Step 6.** Since all remaining items are larger than $\frac{1}{10}$, no bin with more than nine items can be formed. In this step and the next step we form bins with nine items. Create a list $X_9$ of unpacked items of size in $(\frac{1}{10} = 0.1, \frac{1}{10} + \delta_9 = \frac{923}{8680} \approx 0.1063364]$.

As long as $|X_6| > 0$ and $|X_9| > 7$, pack 8 items of $X_9$ with one item of $X_6$ into a bin, and remove them from these sets and from $I$. The items fit into the bin since their total size is at most $8(\frac{1}{10} + \delta_9) + (\frac{1}{7} + \delta_6) = \frac{33}{35} + 8\delta_9 + \delta_6 = 1$. Go to step 7.

**Step 7.** Create a list $X_7$ of unpacked items of size in $(\frac{1}{8} \approx 0.08333, \frac{1}{8} + \delta_7 = \frac{23}{180} \approx 0.127778]$ and a list $X'_9$ of unpacked items of size in $(\frac{1}{10}, \frac{1}{10} + \delta_7 = \frac{37}{360} \approx 0.102778]$. As long as $|X_7| > 2$ and $|X'_9| > 5$, pack three items of $X_7$ with six items of $X'_9$ into a bin, and remove them from these sets and from $I$. The items fit into the bin since their total size is at most $3(\frac{1}{8} + \delta_7) + 6(\frac{1}{10} + \delta_7) = \frac{39}{40} + 9\delta_7 = 1$. Go to step 8.

**Step 8.** Apply NFI. For each new bin, if the last item packed into the bin has size above $\frac{1}{9}$, go to step 9.

**Step 9.** Since all remaining items are larger than $\frac{1}{9}$, no bin with more than eight items can be formed. In this step we form bins with eight items. Create a list $X'_7$ of unpacked items of size in $(\frac{1}{8} \approx 0.08333, \frac{1}{8} + \delta_8 = \frac{19}{144} \approx 0.1319444]$ and a list $X_8$ of unpacked items of size in $(\frac{1}{9}, \frac{1}{9} + \delta_8 = \frac{17}{144} \approx 0.1180555]$.

As long as $|X'_7| > 3$ and $|X_8| > 3$, pack four items of $X'_7$ with four items of $X_8$ into a bin, and remove them from these sets and from $I$. The items fit into the bin since their total size is at most $4(\frac{1}{8} + \delta_8 + \frac{1}{9} + \delta_8) = \frac{17}{18} + 8\delta_8 = 1$. Go to step 10.

**Step 10.** Apply NFI. For each new bin, if the last item packed into the bin has size above $\frac{1}{7}$, go to step 11.

**Step 11.** Since all remaining items are larger than $\frac{1}{7}$, no bin with more than six items can be formed. In this step we form bins with six items. Create a list $X_3$ of unpacked items of size in $(\frac{1}{4} = 0.25, \frac{1}{4} + \delta_3 = \frac{55}{217} \approx 0.2534562]$.

As long as $|X_3| > 0$ and $|X_6| > 4$, pack one item of $X_3$ with five items of $X_6$ into a bin, and remove them from these sets and from $I$. The items fit into the bin since their total size is at most $\frac{1}{4} + \delta_3 + 5(\frac{1}{7} + \delta_6) = 5\delta_6 + \delta_3 + \frac{27}{28} = 1$. Go to step 12.

**Step 12.** Apply NFI. For each new bin, if the last item packed into the bin has size above $\frac{1}{6}$, go to step 13.

**Step 13.** Since all remaining items are larger than $\frac{1}{6}$, no bin with more than five items can be formed. In this step we form bins with five items. Create a list $X_5$ of unpacked items of size in $(\frac{1}{6} \approx 0.16667, \frac{1}{6} + \delta_5 = \frac{11}{60} \approx 0.18333]$, and a list $X'_3$ of unpacked items of size in $(\frac{1}{4}, \frac{1}{4} + \delta_5 = \frac{4}{15} \approx 0.26667]$.

As long as $|X'_3| > 0$ and $|X_5| > 3$, pack one item of $X'_3$ with four items of $X_5$ into a bin, and remove them from these sets and from $I$. The items fit into the bin since their total size is at most $4(\frac{1}{6} + \delta_5) + (\frac{1}{4} + \delta_5) = \frac{11}{12} + 5\delta_5 = 1$. Go to step 14.

**Step 14.** Apply NFI. For each new bin, if the last item packed into the bin has size above $\frac{1}{5}$, go to step 15.

**Step 15.** Since all remaining items are larger than $\frac{1}{5}$, no bin with more than four items can be formed. In this step we form bins with four items. Create a list $X_4$ of unpacked items of size in $(\frac{1}{5} = 0.2, \frac{1}{5} + \delta_4 = \frac{9}{40} = 0.225]$, and a list $X''_3$ of unpacked items of size in $(\frac{1}{4}, \frac{1}{4} + \delta_4 = \frac{11}{40} = 0.275]$.

As long as $|X''_3| > 1$ and $|X_4| > 1$, pack two items of $X''_3$ with two items of $X_4$ into a bin, and remove them from these sets and from $I$. The items fit into the bin since their total size is at most $2(\frac{1}{5} + \frac{1}{4} + 2\delta_4) = \frac{9}{10} + 4\delta_4 = 1$. Go to step 16.

**Step 16.** Apply NFI. For each new bin, if the last item packed into the bin has size above $\frac{1}{3}$, go to step 17.

**Step 17.** Since all remaining items are larger than $\frac{1}{3}$, no bin with more than two items can be formed. In this step we form bins with two items. Create a list $X_2$ of unpacked items of size in ($\frac{1}{3} \approx 0.3334, \frac{1}{3} + \delta_1 = \frac{5}{12} = 0.416667$], and a list $X_1$ of unpacked items of size in ($\frac{1}{2}, \frac{1}{2} + \delta_1 = \frac{7}{12} \approx 0.583334$]. The items of $X_1$ and $X_2$ are not removed from $I$ at this time.

As long as $|X_1| > 0$ and $|X_2| > 0$, pack one item of $X_1$ with one item of $X_2$ into a bin, and remove them from these sets and from $I$. The items fit into the bin since their total size is at most $\frac{1}{2} + \delta_1 + \frac{1}{3} + \delta_1 = \frac{5}{6} + 2\delta_1 = 1$. Go to step 18.

**Step 18.** Apply NFI until no items remain.

The weight function $w : (0,1] \to \mathbb{R}$ is defined as follows. Let $I_k = (\frac{1}{k+1}, \frac{1}{k}]$. For $x \in I_k$, such that $1 \leq k \leq 12$, or $k = t_i - 1$ for some $i$, we let $w(x) = \frac{1}{k}$. Otherwise we let $w(x) = \frac{k+1}{k}x$.

This weight function is similar to the weight function of [5] $\omega : (0,1] \to \mathbb{R}$, defined as follows: For $x \in I_k$, such that $k = t_i - 1$ for some $i$, $\omega(x) = \frac{1}{k}$, and otherwise $\omega(x) = \frac{k+1}{k}x$.

Note that in some steps it is possible to apply maximum matching techniques which allow to construct a potentially better packing. However, the analysis of the current algorithm is easier, and since we already obtain tight bounds of the SPoS, modifying the algorithm would not give a lower asymptotic bound.

We define a new weighting function $w' : I \to \mathbb{R}$. The weight $w'(i)$ of an item $i$ which was packed by NFI (i.e., it was packed in one of the steps 1, 3, 5, 8, 10, 12, 14, 16, 18) is $w(s_i)$.

We let $\gamma_2 = \frac{5}{726}$, $\gamma_4 = \frac{1}{150}$, $\gamma_6^1 = \frac{1}{150}$, $\gamma_6^2 = \frac{11}{1800}$, $\gamma_7 = \frac{2}{189}$, $\gamma_9 = \frac{1}{112}$, $\gamma_{11} = \frac{1}{36}$, $\gamma_{13} = \frac{2}{75}$, $\gamma_{15} = \frac{1}{24}$, and $\gamma_{17} = \frac{1}{4}$.

The weights of the items packed in other steps are as follows. For an item $i$, if $i$ was packed in step $j$ ($j \in \{2, 4, 7, 9, 11, 13, 15, 17\}$), $w'(i) = w(s_i) - \gamma_j$. As for items packed in step 6, for an item of $I_6$ (if $i$ was packed in step 6) $w'(i) = w(s_i) - \gamma_6^1$, and for an item of $I_9$ (packed in step 6), $w'(i) = w(s_i) - \gamma_6^2$. Note that for any $i$, $w'(i) > 0$, since if $w'(i) < w(s_i)$ then $s_i > \frac{1}{12}$ and $w(s_i) \geq \frac{1}{12}$, and if $w'(i) = w(s_i) - \gamma_{17}$ then $s_i > \frac{1}{3}$, so $w'(i) > \frac{1}{4}$.

**Lemma 5.4** *The weight according to $w'$ of each bin created in steps 2, 4, 6, 7, 9, 11, 13, 15, 17 is 1.*

**Proof.** The weight according to $w$ of any item of size in $(\frac{1}{k+1}, \frac{1}{k}]$ (for $k \leq 12$) is $\frac{1}{k}$.

According to $w$, the weight of the 11 items packed into a bin of step 2 is $\frac{10}{11} + \frac{1}{6}$. The weight according to $w'$ is $\frac{10}{11} + \frac{1}{6} - 11\gamma_2 = 1$.

According to $w$, the weight of the 10 items packed into a bin of step 4 is $\frac{9}{10} + \frac{1}{6}$. The weight according to $w'$ is $\frac{9}{10} + \frac{1}{6} - 10\gamma_4 = 1$.

According to $w$, the weight of the 9 items packed into a bin of step 6 is $\frac{8}{9} + \frac{1}{6}$. The weight according to $w'$ is $\frac{8}{9} + \frac{1}{6} - \gamma_6^1 - 8\gamma_6^2 = 1$.

According to $w$, the weight of the 9 items packed into a bin of step 7 is $\frac{6}{9} + \frac{3}{7}$. The weight according to $w'$ is $\frac{6}{9} + \frac{3}{7} - 9\gamma_7 = 1$.

According to $w$, the weight of the 8 items packed into a bin of step 9 is $\frac{4}{8} + \frac{4}{7}$. The weight according to $w'$ is $\frac{4}{8} + \frac{4}{7} - 8\gamma_9 = 1$.

According to $w$, the weight of the 6 items packed into a bin of step 11 is $\frac{5}{6} + \frac{1}{3}$. The weight according to $w'$ is $\frac{7}{6} - 6\gamma_{11} = 1$.

According to $w$, the weight of the 5 items packed into a bin of step 13 is $\frac{4}{5} + \frac{1}{3}$. The weight according to $w'$ is $\frac{17}{15} - 5\gamma_{13} = 1$.

According to $w$, the weight of the 4 items packed into a bin of step 15 is $\frac{2}{4} + \frac{2}{3}$. The weight according to $w'$ is $\frac{7}{6} - 4\gamma_{15} = 1$.

According to $w$, the weight of the two items packed into a bin of step 17 is $1 + \frac{1}{2}$. The weight according to $w'$ is $\frac{3}{2} - 2\gamma_{17} = 1$. ∎

Next, we define a constant number of special bins, and set the weight (according to $w'$) of all items packed into these bins to be zero. We say that a bin is a transition bin of NFI if its items do not all belong to one interval $I_k$. In steps 3,5,8,12,14, the only transition bin can be the last bin of each such step. In step 10, there can also be a transition bin with items of $I_7 \cup I_8$, and in step 16 there can also be a transition bin with items of $I_4 \cup I_3$.

The special bins consist of the last bins created in each one of the steps 1, 3, 5, 8, 10, 12, 14, 16, (if any bins are created in those steps), and transition bins which are not the last bins from steps 10 and 16 (if such bins exist). If in step 18 there may exist a bin which contains a single item of $I_2$, which we also define to be special (if it exists). We set the weights of all items packed in special bins to zero, and therefore their weights according to $w'$ are no larger than other items of the same classes (whose weight is positive).

**Lemma 5.5** *Let $D$ be the number of items created in steps 1, 3, 5, 8, 10, 12, 14, 16, 18. The total weight of items in these bins according to $w'$ is at least $D - 14$.*

**Proof.** In each one of these steps, the algorithm applies NFI on a subset of items. Let $D_i$ be the number of bins for step $i$, excluding the last bin. Using [23], the number of bins in the output of NFD applied to the contents of the bins of step 1 is $D_1$ as well. Using [5], the total weight of these items according to $\omega$ is at least $D_1 - 3$. The items in the bins packed in step 1 (except for the last bin) have sizes of at most $\frac{1}{12}$, for every such size $x$, $w(x) = \omega(x)$, and this is also the weight of the item according to $w'$.

In the other steps, every bin which is not a special bin contains items of one interval $I_k$. It must contain exactly $k$ such items, so the total weight is at least 1. In the last step, except for the special bin, every bin either contains an item of $I_1$ (possibly in addition to an item of $I_2$) or two items of $I_2$, so its total weight is at least 1.

Thus we have weight of at least $D_i$ in step $i$, for $i = 3, 5, 8, 12, 14, 18$, at least $D_i - 1$ for $i = 10, 16$, and at least $D_1 - 3$ in step 1. There are at most 11 special bins, thus the total weight is at least $D - 14$. ∎

Thus, the total weight according to $w'$ of all input items is at least the number of bins in the output minus 14.

**Theorem 5.6** *For any bin of an optimal packing OPT, the total weight of items packed in it is at most $\Theta_\infty$. This is true for all bins except for at most 72 bins, each of which has a weight of at most $T_\infty$. The total weight of the items is at most $\Theta_\infty \text{OPT} + 6$, and the SPoS is at most $\Theta_\infty$.*

**Proof.** Since $w(s_i) \geq w'(i)$ for every item $i$, we analyze the weight with respect to $w$ in some of the cases. We analyze possible bins based on their contents. For an item $i$ where $w'(i) < w(s_i)$ we say that $i$ has reduced weight. This reduced weight can be a result of being packed in a special bin of steps 1,3,6,8,10,12,14,16, and 18, or of being packed in one of the other steps.

We use the following property.

**Claim 5.7** *The total weight according to $w$ of a set of items packed into a bin is at most 1.691.*

**Proof.** The weight function $\omega$ used in [5] is no smaller than $w$ for every item, and the property was proved for this function. Note that many related (but not identical) weight functions have been defined [31, 24, 19], some of which (but not all of which) can be used in the proof instead of the one of [5]. ∎

Consider a bin $B$ packed by OPT. Since for $x \leq \frac{1}{2}$, $\frac{w(x)}{x} \leq \frac{3}{2}$, the following claim holds.

**Claim 5.8** *If $B$ contains no item of $I_1$ then $w(B) \leq \frac{3}{2}$.*

In what follows we use the property: if $x \leq \frac{1}{k}$ then $\frac{w(x)}{x} \leq \frac{k+1}{k}$.

**Claim 5.9** *If $B$ contains an item of $I_1$, and all additional items have size of at most $\frac{1}{5}$, then $w(B) \leq \frac{8}{5} = 1.6$.*

**Proof.** The weight of the large item is 1, and the weight of the remaining items is at most $\frac{6}{5} \cdot \frac{1}{2}$. ∎

**Claim 5.10** *If $B$ contains an item of $I_1$, no items of $I_2$, and an item of $I_3$, and all additional items have size of at most $\frac{1}{9}$, then $w(B) < \Theta_\infty$.*

**Proof.** The weight of the large items is $\frac{4}{3}$, and the weight of the remaining items is at most $\frac{10}{9} \cdot \frac{1}{4}$. The total is at most $\frac{29}{18} \approx 1.61111 < \Theta_\infty$. ∎

**Claim 5.11** *If $B$ contains an item of $I_1$, an item of $I_3$, an item of $I_7$, and all additional items have size of at most $\frac{1}{10}$, then $w(B) < \Theta_\infty$.*

**Proof.** The weight of the large items is $\frac{31}{21}$. If there is an item of $I_{10}$, the total size of remaining items is below $\frac{3}{88} < \frac{1}{29}$, and their total weight is at most $\frac{30}{29} \cdot \frac{3}{88}$. This gives a total weight of at most $\frac{215903}{133980} \approx 1.611457$.

Otherwise, if there is an item of $I_{11}$, the total size of remaining items is below $\frac{1}{24}$, and their total weight is at most $\frac{25}{24} \cdot \frac{1}{24}$. This gives a total of at most $\frac{71429}{44352} \approx 1.6105$.

If all additional items have size of at most $\frac{1}{12}$, the weight of the remaining items is at most $\frac{13}{12} \cdot \frac{1}{8}$. The total is $\frac{361}{224} \approx 1.611607 < \Theta_\infty$. ∎

**Claim 5.12** *If $B$ contains an item of $I_1$, no items of $I_2 \cup I_3$, and at item of $I_4$, then $w(B) \leq 1.61$.*

**Proof.** If all additional items have size of at most $\frac{1}{5}$, then the weight of the large items is $\frac{5}{4}$, and the weight of the remaining items is at most $\frac{6}{5} \cdot \frac{3}{10}$, which results in a total weight of at most $1.61$. Otherwise, there can be at most one additional item of $I_4$, in which case the weight of the large items is $\frac{3}{2}$, the remaining items have total size below $\frac{1}{10}$ so the total weight is at most $\frac{3}{2} + \frac{11}{10} \cdot \frac{1}{10} = 1.61$. ∎

We partition the remaining types of bins into *good* and *bad* bins.

A bin which contains an item of $I_1$ and an item of $I_2$ is called good if one of these two items is sufficiently large, that is, either the larger item has size above $\frac{1}{2} + \delta_1$ or if the other item has size above $\frac{1}{3} + \delta_1$. Otherwise this bin is called bad.

**Lemma 5.13** *If $B$ contains an item of $I_1$ and an item of $I_2$ then $w'(B) \leq \frac{8}{5}$, no matter whether it is good or bad.*

**Proof.** If $B$ is good, then the total size of the two large items is at least $\frac{11}{12}$. The remaining items have weight of at most $\frac{1}{12} \cdot \frac{13}{12}$. The total weight is at most $1 + \frac{1}{2} + \frac{13}{144} \approx 1.5902778$.

If $B$ is bad, then the item of $I_1$ was on the list $X_1$ of step 17, and the item of $I_2$ was on the list $X_2$. The step is completed only once one of the two list is empty, thus the weight of at least one of these two items according to $w'$ is smaller by $\gamma_{17}$ than its weight according to $w$. Thus $w'(B) \leq T_\infty - \gamma_{17} < \frac{3}{2}$. ∎

We are left with the cases where the bin contains an item of $I_1$, an item of $I_3$, and an item of size above $\frac{1}{9}$. Since the remaining size in the bin is less than $\frac{1}{4}$, this last item can be of one of the intervals $I_4, I_5, I_6, I_7, I_8$. This item is called the *third item*.

In the case that the size of third item is in $I_7$, the remaining items have total size of less than $\frac{1}{8}$. The case where the largest item in the remaining set of items (the *fourth item*) has size of at most $\frac{1}{10}$ is already covered, so there are two cases to cover, where the fourth item is of $I_8$ and the case where it is of $I_9$.

**Case 1.** The third item is of $I_4$. We define a good bin as a bin where the third item has size above $\frac{1}{5} + \delta_4$ or the item of $I_3$ has size above $\frac{1}{4} + \delta_4$. Other such bins are called bad.

Note that every bad bin of this case has an item of size at in $(\frac{1}{4}, \frac{1}{4} + \delta_4]$ and an item of size at in $(\frac{1}{5}, \frac{1}{5} + \delta_4]$. There is at most one bad bin where none of these two items has a reduced weight according to $w'$, which holds since step 15 terminates when $X_4$ has at most one item or $X_3''$ has at most one item (if there were at least two such bins then $|X_4| \geq 2$ and $|X_3''| \geq 2$), and every such item which was removed earlier has a reduced weight as well.

**Lemma 5.14 (C.1)** *Every bin $B$ which satisfies the condition of this case (except for at most one bin) has $w'(B) \leq \Theta_\infty$, no matter whether it is good or bad.*

**Proof.** If $B$ is good, then the total size of the three large items is at least $\frac{1}{2} + \frac{1}{4} + \frac{1}{5} + \delta_4 = \frac{39}{40}$. The remaining items have weight of at most $\frac{1}{40} \cdot \frac{41}{40}$. The total weight is at most $\frac{7723}{4800} \approx 1.6089583$.

If $B$ is bad, then $B$ has an item of reduced weight (which is true for all such bins except for at most one bin). The item with reduced weight is packed in a special bin or in one of the steps 11,13, or 15. The total size of the three large items is at least $\frac{19}{20}$, so $w(B) \leq 1 + \frac{1}{3} + \frac{1}{4} + \frac{1}{20} \cdot \frac{21}{20} = \frac{1963}{1200}$. The difference with the weight according to $w'(B)$ is at least $\gamma_{13} = \frac{2}{75}$. Thus we have $w'(B) \leq \frac{1963}{1200} - \frac{2}{75} = \frac{1931}{1200} \approx 1.6091667$. ∎

**Case 2.** The third item is of $I_5$. We define a good bin as a bin where the third item has size above $\frac{1}{6} + \delta_5$ or the item of $I_3$ has size above $\frac{1}{4} + \delta_5$. Other such bins are called bad.

Note that every bad bin has an item of size at in $(\frac{1}{6}, \frac{1}{6} + \delta_5]$ and an item of size at in $(\frac{1}{4}, \frac{1}{4} + \delta_5]$. There are at most three bins where none of these two items has a reduced weight according to $w'$ at the termination of step 13 (though if an item of $I_3$ remains then its weight can be still reduced later), since step 13 terminates when $X_5$ has at most three items or $X_3'$ is empty.

**Lemma 5.15 (C.2)** *Every bin $B$ which satisfies the condition of this case (except for at most three bins) has $w'(B) \leq \Theta_\infty$, no matter whether it is good or bad.*

**Proof.** If $B$ is good, then the total size of the three large items is at least $\frac{1}{2} + \frac{1}{4} + \frac{1}{6} + \delta_5 = \frac{14}{15}$. The remaining items have weight of at most $\frac{1}{15} \cdot \frac{16}{15}$. The total weight is at most $\frac{361}{225} \approx 1.60444$.

If $B$ is bad, then it has an item of reduced weight (which is true for all bins except for at most three). The item with reduced weight is packed in a special bin or in one of the steps 11,13, or 15. The total size of the three large items is $\frac{11}{12}$, so $w(B) \leq 1 + \frac{1}{3} + \frac{1}{5} + \frac{1}{12} \cdot \frac{13}{12} = \frac{1169}{720}$. Thus $w'(B) \leq w(B) - \gamma_{13}$ and we have $w'(B) \leq \frac{1169}{720} - \frac{2}{75} = \frac{5749}{3600} \approx 1.596944$. ∎

**Case 3.1** The third item is of $I_7$ and the fourth item is of $I_8$. We define a good bin as a bin where the third item has size above $\frac{1}{8} + \delta_8$ or the fourth item has size above $\frac{1}{9} + \delta_8$. Other such bins are called bad.

Note that every bad bin has an item of size at in $(\frac{1}{8}, \frac{1}{8} + \delta_8]$ and an item of size at in $(\frac{1}{9}, \frac{1}{9} + \delta_8]$. We note that there are at most three bins where none of these two items has a reduced weight according to $w'$ at the termination of step 9.

**Lemma 5.16 (C.3.1)** *Every bin $B$ which satisfies the condition of this case (except for at most three bins) has $w'(B) \leq \Theta_\infty$, no matter whether it is good or bad.*

**Proof.** If $B$ is good, then the total size of the four large items is at least $\frac{1}{2} + \frac{1}{4} + \frac{1}{8} + \frac{1}{9} + \delta_8 = \frac{143}{144}$. The remaining items have weight of at most $\frac{1}{144} \cdot \frac{145}{144}$. The total weight is at most $\frac{233431}{145152} \approx 1.6081831$.

If $B$ is bad, with an item of reduced weight (which is the case for all such bins except for at most three bins), the item with reduced weight is packed in a special bin or in one of the steps 7 or 9. The total size of the four large items is at least $\frac{71}{72}$, so $w(B) \leq 1 + \frac{1}{3} + \frac{1}{7} + \frac{1}{8} + \frac{1}{72} \cdot \frac{73}{72} = \frac{58615}{36288}$. Since $\gamma_7 > \gamma_9 = \frac{1}{112}$, $w'(B) \leq \frac{58615}{36288} - \frac{1}{112} = \frac{58291}{36288} \approx 1.6063437$. ∎

**Case 3.2** The third item is of $I_7$ and the fourth item is of $I_9$. We define a good bin as a bin where the third item has size above $\frac{1}{8} + \delta_7$ or the fourth item has size of above $\frac{1}{10} + \delta_7$. Other such bins are called bad.

Note that every bad bin has an item of size at in $(\frac{1}{8}, \frac{1}{8} + \delta_7]$ and an item of size at in $(\frac{1}{10}, \frac{1}{10} + \delta_7]$. We note that there are at most five bins where none of these two items has a reduced weight according to $w'$ at the termination of step 7.

**Lemma 5.17 (C.3.2)** *Every bin $B$ which satisfies the condition of this case (except for at most five bins) has $w'(B) \leq \Theta_\infty$, no matter whether it is good or bad.*

**Proof.** If $B$ is good, then the total size of the four large items is at least $\frac{1}{2} + \frac{1}{4} + \frac{1}{8} + \frac{1}{10} + \delta_7 = \frac{44}{45}$. The remaining items have weight of at most $\frac{1}{45} \cdot \frac{46}{45}$. The total weight is at most $\frac{22822}{14175} \approx 1.6100176$.

Consider the case that $B$ is bad and has an item of reduced weight which was packed in a special bin or in one of the steps 6 or 7 (which holds for all such bins except for at most five). The total size of the four large items is at least $\frac{39}{40}$, so $w(B) \leq 1 + \frac{1}{3} + \frac{1}{7} + \frac{1}{9} + \frac{1}{40} \cdot \frac{41}{40} = \frac{162583}{100800}$. Using $\gamma_7 > \gamma_6^2 = \frac{11}{1800}$, $w'(B) \leq \frac{58615}{36288} - \frac{11}{1800} = \frac{161967}{100800} \approx 1.606815$. ∎

**Case 4.1** The third item is of $I_8$, and so is the fourth item.

**Lemma 5.18 (C.4.1)** *Every bin $B$ which satisfies the condition of this case has $w'(B) \leq \Theta_\infty$.*

**Proof.** In this case the remaining space is $\frac{1}{36}$. The results of [5] for the case that all items are of size at most $\frac{1}{35}$ (where the weight function is equal to our function for such items) imply that the supremum weight is achieved by the total weight of greedy sequence, which is exactly the sequence $\tau_4, \tau_5, \ldots$. This last weight is at most $\Theta_\infty$ minus $1 + \frac{1}{3} + 2 \cdot \frac{1}{8}$, which is exactly the total weight of the four large items. ∎

**Case 4.2** The third item is of $I_8$, and the remaining items have sizes of at most $\frac{1}{9}$.

**Lemma 5.19 (C.4.2)** *Every bin $B$ which satisfies the condition of this case has $w'(B) \leq \Theta_\infty$.*

**Proof.** We consider several cases. If there is an item of $I_9$, the total size of remaining items is at most $1 - \frac{1}{2} - \frac{1}{4} - \frac{1}{9} - \frac{1}{10} = \frac{7}{180}$. Since $\frac{7}{180} \in I_{25}$, the total weight of the items is at most $1 + \frac{1}{3} + \frac{1}{8} + \frac{1}{9} + \frac{7}{180} \cdot \frac{26}{25} = \frac{14489}{9000} \approx 1.6098889$. Otherwise, the total size of remaining items is at most $1 - \frac{1}{2} - \frac{1}{4} - \frac{1}{9} = \frac{5}{36}$. Since all remaining items have size of at most $\frac{1}{10}$, the total weight of the items is at most $1 + \frac{1}{3} + \frac{1}{8} + \frac{5}{36} \cdot \frac{11}{10} = \frac{29}{18} \approx 1.61111$. ∎

**Case 5.1** The third item is of $I_6$, and the remaining items have size of at most $\frac{1}{12}$.

We define a good bin as a bin where the second item has size above $\frac{1}{4} + \delta_3$ or the third item has size above $\frac{1}{7} + \delta_6$. Other such bins are called bad.

Note that every bad bin has an item of size at in $(\frac{1}{4}, \frac{1}{4} + \delta_3]$ and an item of size at in $(\frac{1}{7}, \frac{1}{7} + \delta_6]$. There are at most four bins where none of these two items has a reduced weight according to $w'$ at the termination of step 11.

**Lemma 5.20 (C.5.1)** *Every bin $B$ which satisfies the condition of this case (except for at most four bins) has $w'(B) \leq \Theta_\infty$, no matter whether it is good or bad.*

**Proof.** If $B$ is good, then the total size of the three large items is at least $\frac{1}{2} + \frac{1}{4} + \frac{1}{7} + \delta_3 = \frac{389}{434}$ (since $\delta_6 > \delta_3$). If all additional items have size of at most $\frac{1}{13}$, then the items have weight of at most $1 + \frac{1}{3} + \frac{1}{6} + \frac{45}{434} \cdot \frac{14}{13} = \frac{9093}{5642} \approx 1.61166$. Otherwise, if there is an item of $I_{12}$, the total size of remaining items is at most $\frac{151}{5642} < \frac{1}{37}$. The total weight in this case is at most $\frac{3}{2} + \frac{1}{12} + \frac{38}{37} \cdot \frac{151}{5642} \approx 1.61082$.

Consider the case that $B$ is bad and has an item of reduced weight which was packed in a special bin or in one of the steps 2, 4, 6, or 11 (which holds for all such bins except for at most four). The total size of the three large items is at least $\frac{25}{28}$ and their weight is $\frac{3}{2}$, so $w(B) \leq \frac{3}{2} + \frac{3}{28} \cdot \frac{13}{12} = \frac{181}{112}$. Since $\min\{\gamma_2, \gamma_4, \gamma_6^1, \gamma_{11}\} = \frac{1}{150}$, we have $w'(B) \leq \frac{181}{112} - \frac{1}{150} = \frac{13519}{8400} \approx 1.6094$. ∎

**Case 5.2** The third item is of $I_6$, and the fourth item is of one of the intervals $I_9, I_{10}, I_{11}$.

We say that the bin is good if the third item has size above $\frac{1}{7} + \delta_6$ or both the second item and the fourth item are relatively large, that is, the second item has size above $\frac{1}{4} + \delta_3$, and the fourth item of the interval $I_j$ ($j \in \{9, 10, 11\}$) has size above $\frac{1}{j+1} + \delta_j$.

**Lemma 5.21 (C.5.2.1)** *Every bin good $B$ which satisfies the condition of this case has $w(B) \leq \Theta_\infty$.*

**Proof.** Let $j$ be the index such that the fourth item is of $I_j$. Note that $\delta_j + \delta_3 > \delta_6$ for $j = 9, 10$ and $\delta_{11} + \delta_3 < \delta_6$.

For $j = 9, 10$, the total size of the four large items is at least $\frac{1}{2} + \frac{1}{4} + \frac{1}{7} + \frac{1}{j+1} + \delta_6$, and their total weight is $\frac{3}{2} + \frac{1}{j}$.

If $j = 9$, the total size of the remaining items is at most $\frac{3}{4340}$, so all these items have size below $\frac{1}{1446}$, and the total weight is at most $\frac{3}{4340} \cdot \frac{1447}{1446} + \frac{3}{2} + \frac{1}{9} \leq \frac{30345283}{18826920} \approx 1.6118028$.

If $j = 10$, the total size of the remaining items is at most $\frac{467}{47740}$, so all these items have size below $\frac{1}{102}$, and the total weight is at most $\frac{467}{47740} \cdot \frac{103}{102} + \frac{3}{2} + \frac{1}{10} \leq \frac{7839269}{4869480} \approx 1.609878$.

For $j = 11$, the total size of the four large items is at least $\frac{1}{2} + \frac{1}{4} + \frac{1}{7} + \frac{1}{12} + \delta_3 + \delta_{11}$, and their total weight is $\frac{3}{2} + \frac{1}{11}$. The total size of the remaining items is at most $\frac{5131}{273420}$, so all these items have size below $\frac{1}{53}$, and the total weight is at most $\frac{5131}{273420} \cdot \frac{54}{53} + \frac{3}{2} + \frac{1}{11} \approx 1.61003$. ∎

Next, we analyze bad bins. In this case there are several types of bad bins and we analyze each one of them separately. In all bad bins, the third item has size below $\frac{1}{7} + \delta_6$. The types of bad bins are as follows. The first type is a bin where the second item has size of at most $\frac{1}{4} + \delta_3$, and the fourth item has size of at most $\frac{1}{j+1} + \delta_j$. The second type is a bin where the second item has size of at most $\frac{1}{4} + \delta_3$, but the fourth item has size above $\frac{1}{j+1} + \delta_j$. The third type is a bin where the second item has size above $\frac{1}{4} + \delta_3$, and the fourth item has size of at most $\frac{1}{j+1} + \delta_j$.

**Lemma 5.22 (C.5.2.2)** *For $j = 9$, every bad bin $B$ of the first type which satisfies the condition of this case (except for at most seven bins) has $w'(B) \leq \Theta_\infty$.*

**Proof.** We show that except for a constant number of bins, it is either the case that the third item has reduced weight, or both the second and fourth items have reduced weight. Specifically, there are either at most four bins where the third item does not have reduced weight or at most seven bins where one of the two other items does not have reduced weight.

If there are at most four bins where the third item does not have reduced weight, then the items in the other bins were packed in in a special bin or in one of the steps 2, 4, 6, or 11, and the weight was reduced by at least $\min\{\gamma_2, \gamma_4, \gamma_6^1, \gamma_{11}\}$. Otherwise, there are at least five bins where the third item does not have reduced weight, then it is always satisfied that $|X_6| \geq 5$ after step 6, so this step terminated with $|X_9| \leq 7$ and step 11 terminated with $|X_3| = 0$. This means that there are no bins where the second item does not have reduced weight, and there are at most seven bins where the fourth item does not have reduced weight. In the other bins, for items which are not packed in special bins, the weight of the second item was reduced in step 11, and the weight of the fourth item is step 6.

The total size of other items is at most $1 - \frac{1}{2} - \frac{1}{4} - \frac{1}{7} - \frac{1}{10} = \frac{1}{140}$, and we get $w(B) \leq 1 + \frac{1}{3} + \frac{1}{6} + \frac{1}{9} + \frac{1}{140} \cdot \frac{141}{140} = \frac{285469}{176400}$, and since $\min\{\gamma_2, \gamma_4, \gamma_6^1, \gamma_{11}, \gamma_6^2 + \gamma_{11}\} = \frac{1}{150}$, we get $w'(B) \leq \frac{285469}{176400} - \frac{1}{150} = \frac{284293}{176400} \approx 1.61163822$. ∎

**Lemma 5.23 (C.5.2.3)** *For $j = 10, 11$, every bad bin $B$ of the first type which satisfies the condition of this case (except for at most nine bins for each value of $j$) has $w'(B) \leq \Theta_\infty$.*

**Proof.** The proof that there are either at most four bins where the third item does not have reduced weight or at most nine bins (eight bins if $j = 10$) where one of the two other items does not have reduced weight is similar to the previous case.

For a bin $B$, if the third item has a reduced weight, then it was packed in a special bin or in one of the steps 2, 4, 6, or 11. If the second item has reduced weight after step 11, then it was packed in a special bin or in this step. If the fourth item has reduced weight after step 4, then it was packed in a special bin, in step 2, or in step 4.

The total size of items except for the three large items is at most $1 - \frac{1}{2} - \frac{1}{4} - \frac{1}{7} = \frac{3}{28}$. All items have size of at most $\frac{1}{10}$, so $w(B) \leq \frac{3}{2} + \frac{11}{10} \cdot \frac{3}{28} = \frac{453}{280}$, and since $\min\{\gamma_2, \gamma_4, \gamma_6^1, \gamma_{11}, \gamma_2 + \gamma_{11}, \gamma_4 + \gamma_{11}\} = \frac{1}{150}$, we get $w'(B) \leq \frac{453}{280} - \frac{1}{150} = \frac{6767}{4200} \approx 1.61119$. ∎

**Lemma 5.24 (C.5.2.4)** *Every bad bin $B$ of the second type which satisfies the condition of this case (except for at most four bins) has $w'(B) \leq \Theta_\infty$.*

**Proof.** Except for at most four bins, after step 11 it is either the case that the second item has reduced weight or the third item does. If $B$ is bad and has an item of reduced weight which was packed in a special bin or in one of the steps 2, 4, 6, or 11.

The total size of remaining items is at most $1 - \frac{1}{2} - \frac{1}{4} - \frac{1}{7} - \frac{1}{j+1} - \delta_j$. For $j = 9$, $j = 10$, $j = 11$, this is at most $\frac{1}{1240}$, at most $\frac{1}{77}$, and at most $\frac{1}{45}$, respectively.

In the three cases, the upper bounds on $w(B)$ is at most $\frac{3}{2} + \frac{1}{j}$ plus the weight of the smaller items, which gives $\frac{3}{2} + \frac{1}{9} + \frac{1}{1240} \cdot \frac{1241}{1240} = \frac{22306369}{13838400} \approx 1.611918$ for $j = 9$, $\frac{3}{2} + \frac{1}{10} + \frac{1}{77} \cdot \frac{78}{77} = \frac{95644}{59290} \approx 1.61315568$ for $j = 10$, and $\frac{3}{2} + \frac{1}{11} + \frac{1}{45} \cdot \frac{46}{45} = \frac{71887}{44550} \approx 1.61362514$ for $j = 11$.

Using $\min\{\gamma_2, \gamma_4, \gamma_6^1, \gamma_{11}\} = \frac{1}{150}$ we have $w'(B) \leq \frac{71887}{44550} - \frac{1}{150} = \frac{7159}{4455} \approx 1.606958$. ∎

**Lemma 5.25 (C.5.2.5)** *Every bad bin $B$ of the third type which satisfies the condition of this case (except for at most nine bins for each value of $j$) has $w'(B) \leq \Theta_\infty$.*

**Proof.** Except for at most nine bins, after step 6 it is either the case that the third item has reduced weight or the fourth item does. If $B$ is bad and has an item of reduced weight which was packed in a special bin or in one of the steps 2, 4, or 6. In this case the weight of the third item or the weight of the fourth item was reduced.

The total size of the items except for the three largest items is at most $1 - \frac{1}{2} - \frac{1}{4} - \delta_3 - \frac{1}{7} = \frac{45}{434}$. These items have sizes of at most $\frac{1}{9}$, so $w(B) \leq \frac{3}{2} + \frac{10}{9} \cdot \frac{45}{434} = \frac{701}{434} \approx 1.6152$, and since $\min\{\gamma_2, \gamma_4, \gamma_6^1, \gamma_6^2, \gamma_{11}\} = \frac{11}{1800}$, $w'(B) \leq \frac{701}{434} - \frac{11}{1800} = \frac{628513}{390600} \approx 1.609096$. ∎

We found that except for at most 72 bins, the total weight of items in a bin is at most $\Theta_\infty$. The total weight of items is therefore at most $\Theta_\infty(\text{OPT} - 72) + 72 T_\infty$. Since $T_\infty - \Theta_\infty \leq \frac{2}{25}$, we get a total weight below $\Theta_\infty \cdot \text{OPT} + 6$.

Let $ALG$ denote the number of bins created by the algorithm. We have $w(I) \leq \Theta_\infty \text{OPT} + 6$ and $w(I) \geq ALG - 14$, thus $ALG \leq \Theta_\infty \text{OPT} + 20$, which implies the upper bound on the SPoS. ∎

## 6 The SPO-PoA for unit weights

In this section we analyze the SPO-PoA. We start with an upper bound.

**Theorem 6.1** *The SPO-PoA is at most 1.628113.*

**Proof.** Consider a NE packing $A$ for an input $I$ which is strictly Pareto optimal. We show that the number of bins in $A$ is at most $1.6281130267 \cdot \text{OPT}(I) + \mathcal{C}$, where $\mathcal{C}$ is a constant which does not exceed 1000.

We use the next weighting function (recall that since we are dealing with unit weight, we can use the term weight for a function which we define) on the items packed into $2^+$-bins. Let $\varepsilon \geq \zeta \geq \gamma \geq \beta \geq \alpha \geq \mu > 0$ be constants, some of which are determined later. These parameters are also called *bonuses*. We require $\frac{1}{180} \leq \mu \leq \frac{1}{40}$, $\frac{1}{63} \leq \alpha \leq \frac{1}{30}$, $\beta \leq \frac{1}{20}$, $\gamma \leq \frac{1}{15}$. We let $\xi = \frac{5}{18}$ and $\zeta = \frac{1}{5}$.

$$\omega(x) = \frac{13}{12}x + \begin{cases} 0, & \text{if } 0 < x \leq \frac{1}{12} \\ \mu, & \text{if } \frac{1}{12} < x \leq \frac{1}{8} \\ \alpha, & \text{if } \frac{1}{8} < x \leq \frac{1}{6} \\ \beta, & \text{if } \frac{1}{6} < x \leq \frac{1}{4} \\ \gamma, & \text{if } \frac{1}{4} < x \leq \frac{1}{3} \\ \zeta, & \text{if } \frac{1}{3} < x \leq \frac{1}{2} \\ \xi, & \text{if } \frac{1}{2} < x \leq 1 \end{cases}$$

Note that for $0 < x \leq \frac{1}{3}$, $\omega(x) \leq 1.4 \cdot x$.

Let $\tau_{12}$ be the number of the $12^+$-bins. We show that the total weight of items in those bins is at least $\tau_{12} - 1$, no matter what the values of parameters are.

**Lemma 6.2** *The total weight of $12^+$-bins is at least $\tau_{12} - 1$.*

**Proof.** Let $\iota \geq 12$ be the smallest integer such that a $j$-bin with a total size of items below $\frac{12}{13}$ exists in $A$, and let $\mathcal{B}^\iota$ be such a bin of minimum load. If $j$ does not exist then the claim holds trivially, since the weight of every item is at least $\frac{13}{12}$ times its size. Otherwise, $\mathcal{B}^\iota$ contains at least one item $x$ of size below $\frac{1}{13}$. There can be no additional $\iota^+$-bin with a total size below $\frac{12}{13}$, since $x$ can benefit from moving into such a bin. Thus, each $12^+$-bins (with the exception of at most one bin) has a total weight of 1. ∎

Let $\tau_1$ be the number of 1-bins. In every NE packing all 1-bins except for at most one bin contain an item of size above $\frac{1}{2}$. The weight of items in 1-bins was not defined yet and we define the weight of every item in a 1-bin which has size above $\frac{1}{2}$ to be 1, and otherwise its weight is zero. Thus the total weight of items in 1-bins is at least $\tau_1 - 1$.

For every $2 \leq j \leq 11$ we define a special bin and regular bins as in Section 4. By Lemma 4.2, the total size of items in each regular $k$-bin is above $\max\{1 - \rho_k, \frac{k}{k+1}\}$, and by definition, all these items have size of at least $\rho_k$. A regular $j$-bin is called *light* if it does not contain any items of size above $\frac{1}{j}$. If a light $j$-bin contains at least one item of size at most $\frac{1}{j+1}$, then it is called *very light*, and otherwise (if it is light but not very light) it is called *standard*. Standard $j$-bins contain $j$ items of size in $(\frac{1}{j+1}, \frac{1}{j}]$. Other regular bins are called *heavy*. Note that all 1-bins are light, and there is at most one very light 1-bin, thus all 1-bins except for at most one such bin are standard.

**Lemma 6.3** *For any $2 \leq j \leq 11$, the number of very light $j$-bins is at most $j^2 + j - 1$.*

**Proof.** Assume by contradiction that $A$ contains at least $j^2 + j$ very light bins, and consider $j^2 + j$ such bins. Those bins contain at least $j^2 + j$ items of size at most $\frac{1}{j+1}$ (at least one per bin), and the size of each item of the remaining $(j-1)(j^2+j)$ items is at most $\frac{1}{j}$. We create an alternative packing in which only the items of these bins are packed differently as follows. We create $j$ bins of $j+1$ items of size at most $\frac{1}{j+1}$ and $j^2 - 1$ bins, each with $j$ of the remaining items. All items coming from $j$ bins are now packed into $j^+$-bins, and there are $j$ $(j+1)$-bins. No item has a larger cost and there are $j(j+1)$ items with a smaller cost, which contradicts the Pareto optimality of $A$. ∎

For all values of $j$ such that $1 \leq j \leq 11$, the number of very light $j$-bins and special bins together is at most 561. In what follows we neglect such bins in our calculations, and assume that only standard and heavy bins exist.

**Lemma 6.4** *For any $2 \leq j \leq 10$, if $\rho_j \leq \frac{1}{12}$, then every regular $j$-bin has total weight of at least 1.*

**Proof.** By Lemma 4.2, every regular $j$-bin has a total size at least $1 - \rho_j \geq \frac{11}{12}$. If the bin contains at least two items of size above $\frac{1}{12}$ then the total weight is at least $\frac{13}{12} \cdot \frac{11}{12} + 2\mu = \frac{143}{144} + \frac{1}{90} > 1$. If the bin contains at least one item of size above $\frac{1}{8}$ then the total weight is at least $\frac{13}{12} \cdot \frac{11}{12} + \alpha = \frac{143}{144} + \frac{1}{63} > 1$.

Otherwise, the bin must contains at most one item of size in $(\frac{1}{12}, \frac{1}{8}]$, and the other items (at most nine items) have size of at most $\frac{1}{12}$. This is impossible since the total size does not exceed $9 \cdot \frac{1}{12} + \frac{1}{8} < \frac{11}{12}$. ∎

In what follows, we consider only the case $\rho_j > \frac{1}{12}$ for $2 \leq j \leq 10$, since otherwise all $j$-bins have sufficient weight.

**Lemma 6.5** *Every regular 11-bin except for at most 23 bins has total weight of at least 1.*

**Proof.** If the total size of items in a bin is at least $\frac{12}{13}$, or the total size is at least $\frac{11}{12}$ and it has an item of bonus $\alpha$ or two items of bonus $\mu$ then we are done similarly to the previous lemma. Thus, in order to have an 11-bin of weight below 1 we must have $\rho_{11} \in (\frac{1}{13}, \frac{1}{12}]$, since in the case $\rho_{11} < \frac{1}{13}$ the total size of items in every 11-bin is at least $\frac{12}{13}$, and otherwise if $\rho_{11} > \frac{1}{12}$ then every bin has 11 items with bonus at least $\mu$.

By Lemma 4.2, every 11-bin must contain an item of size above $\frac{1}{12}$. We only need to consider bins where there is exactly one such item, and its size does not exceed $\frac{1}{8}$. Such a bin must be heavy since standard bins have 11 items with bonus $\mu$. If the total size of items in a bin is at least 0.92 then the total weight is at least $\frac{13}{12} \cdot 0.92 + \mu > 1$. Thus if there exists a regular (heavy) bin of weight below 1 then $\rho_{11} \geq 0.08$.

Assume by contradiction that there are at least 24 heavy bins of weight below 1. Every such bin contains ten items of size in $[0.08, \frac{1}{12}]$ and one item of size in $(\frac{1}{12}, 0.12]$. We create an alternative solution where only the items of these 24 bins are packed differently. There are 12 bins which contain two larger items and nine smaller items (the total size of these items is at most 0.99) and 11 bins containing 12 smaller items each. In this transformation, all items previously packed into 11-bins are packed into (possibly different) 11-bins, while some items of 11-bins are now packed into 12-bins, which contradicts the Pareto optimality of $A$. ∎

Consider the sets $F_1$ and $F_2$ of items packed into 1-bins (which are all standard), and *standard* 2-bins in $A$, respectively.

**Lemma 6.6** *There is at most one item of $F_2$ that is packed into the same bin as an item of $F_1$ in* OPT.

**Proof.** Assume by contradiction that there are at least two such items $x_1$ and $x_2$. Let $y_i$ be the member of $F_1$ packed with $x_i$ in OPT. We modify the packing $A$ so that only the packing of the (three or four) bins of $x_1, x_2, y_1, y_2$ is changed. The item $x_i$ is packed with $y_i$. If $x_1$ and $x_2$ are packed into different standard 2-bins of $A$, the items which were packed with them are combined into a bin (each one of those items has size at most $\frac{1}{2}$, since these bins are standard). As a result, the items which were packed into 2-bins are still packed into (possibly different) 2-bins, but the change is beneficial for $y_1$ and $y_2$. Thus $A$ is not Pareto optimal, a contradiction. ∎

If there exists a pair of items, one of $F_1$ and one of $F_2$ packed together in OPT, we neglect this bin in our calculations. We set the bonus of any item of size at most $\frac{1}{2}$ packed into a heavy 2-bin into zero.

**Lemma 6.7** *If there exist at least 12 standard 3-bins with a total size of items below $\frac{12}{13}$, then there exist at most 11 standard 4-bins with a total size of items below $\frac{12}{13}$.*

**Proof.** Assume by contradiction that there are 12 standard 3-bins and 12 standard 4-bins, each one with a total size below $\frac{12}{13}$. We modify the packing of these 24 bins as follows. For each 4-bin, remove the largest item in it and pack every such four items in a bin (which results in three new bins). This is possible since each such item has size of at most $\frac{1}{4}$. For a 4-bin which had total size $z \leq \frac{12}{13}$ of items and the largest item was removed, the current total size is at most $\frac{3}{4}z \leq \frac{9}{13}$. Each such gap is filled with an item coming from a standard 3-bin. Specifically, remove the smallest item from each 3-bin and insert it into one of the former 4-bins (which temporarily have only three items). The size of each such item is at most $\frac{1}{3} \cdot \frac{12}{13} = \frac{4}{13}$, so as a result, the total size of items in each such bin is at most 1, and it contains four items. The remaining 24 items are packed into eight bins, which is possible since items of standard 3-bins have size of at most $\frac{1}{3}$. The 24 bins were thus replaced with 23 bins. ∎

**Lemma 6.8** *Every heavy or standard 2-bin has a weight of at least 1.*

**Proof.** The total size of items in a standard 2-bin is at least $\frac{2}{3}$. If the bin is standard, the total weight is at least $\frac{2}{3} \cdot \frac{13}{12} + 2\zeta \geq 1$. Otherwise, the total weight is at least $\frac{2}{3} \cdot \frac{13}{12} + \xi = 1$. ∎

**Lemma 6.9** *Every heavy 3-bin has a weight of at least 1, and every standard 3-bin has weight of at least 1, if $\gamma \geq \frac{1}{16}$.*

**Proof.** The weight of a standard 3-bin is at least $\frac{3}{4} \cdot \frac{13}{12} + 3\gamma \geq 1$. A heavy bin has an item of bonus at least $\zeta$, so the total weight is at least $\frac{3}{4} \cdot \frac{13}{12} + \zeta \geq 1$. ∎

**Lemma 6.10** *If the conditions* $\min\{2\gamma + 2\mu, \gamma + 2\beta + \mu\} \geq \frac{5}{96}$, $\gamma + \beta + 2\alpha \geq \frac{7}{72}$, *and* $\gamma + 3\beta \geq \frac{2}{15}$ *hold, then every heavy 4-bin has a weight of at least 1, and every standard 4-bin has weight of at least 1, if* $\beta \geq \frac{1}{30}$.

**Proof.** The weight of a standard bin is at least $\frac{4}{5} \cdot \frac{13}{12} + 4\beta \geq 1$. If a heavy bin has an item of bonus at least $\zeta$, then we are done. Otherwise, the largest item has a bonus of $\gamma$ (and size of at most $\frac{1}{3}$).

If $\frac{1}{12} < \rho_4 \leq \frac{1}{8}$, then all regular bins have total size above $1 - \rho_4 \geq \frac{7}{8}$. The bin must contain another item of bonus at least $\beta$, since otherwise the total size of items is at most $3 \cdot \frac{1}{6} + \frac{1}{3} < \frac{7}{8}$. If there are two items of bonus $\gamma$, then the other two have bonuses of at least $\mu$. Otherwise, either both remaining items have bonuses of $\alpha$, or if there is an item of bonus $\mu$, the size of the remaining item is above $\frac{7}{8} - \frac{1}{3} - \frac{1}{4} - \frac{1}{8} = \frac{1}{6}$, and has a bonus of at least $\beta$. The weight before the bonuses is at least $\frac{13}{12} \cdot \frac{7}{8} = \frac{91}{96}$, thus we require $2\gamma + 2\mu \geq \frac{5}{96}$, $\gamma + \beta + 2\alpha \geq \frac{5}{96}$, $\gamma + 2\beta + \mu \geq \frac{5}{96}$.

If $\frac{1}{8} < \rho_4 \leq \frac{1}{6}$, then all regular bins have total size above $1 - \rho_4 \geq \frac{5}{6}$. Once again, the bin must contain another item of bonus at least $\beta$ (since otherwise the total size of items is at most $\frac{5}{6}$). The weight before the bonuses is at least $\frac{13}{12} \cdot \frac{5}{6} = \frac{65}{72}$, thus we require $\gamma + \beta + 2\alpha \geq \frac{7}{72}$.

Otherwise $\rho_4 > \frac{1}{6}$ and all regular bins have total size above $\frac{4}{5}$. The weight before the bonuses is at least $\frac{13}{12} \cdot \frac{4}{5} = \frac{13}{15}$, thus we require $\gamma + 3\beta \geq \frac{2}{15}$. ∎

**Lemma 6.11** *If the conditions* $\min\{2\beta + \alpha + 2\mu, \gamma + \beta + 3\mu, \gamma + 2\alpha + 2\mu\} \geq \frac{5}{96}$ *and* $\beta + 4\alpha \geq \frac{7}{72}$ *hold, then every heavy 5-bin has a weight of at least 1, and every standard 5-bin has weight of at least 1, if* $\beta \geq \frac{7}{360}$.

**Proof.** The weight of a standard bin is at least $\frac{5}{6} \cdot \frac{13}{12} + 5\beta \geq 1$.

If a heavy bin has an item of bonus at least $\zeta$, then we are done. Otherwise, the largest item has a bonus of $\beta$ or $\gamma$.

If the total size of items is in $(\frac{7}{8}, \frac{12}{13}]$, consider first the case that the largest item has a bonus of $\beta$. If there is another item of bonus $\beta$, then the third item must have a bonus of at least $\alpha$, while the remaining two items have bonuses of at least $\mu$. Otherwise, all remaining items have bonuses of $\alpha$. Thus we have $2\beta + \alpha + 2\mu \geq \frac{5}{96}$ and $\beta + 4\alpha \geq \frac{5}{96}$.

If the largest item has bonus $\gamma$, if there is another item of bonus at least $\beta$, then the remaining three items have bonuses of at least $\mu$, and $\gamma + \beta + 3\mu \geq \frac{5}{96}$. Otherwise, there are at least two items of bonus $\alpha$, so $\gamma + 2\alpha + 2\mu \geq \frac{5}{96}$.

If the total size of items is in $(\frac{5}{6}, \frac{7}{8}]$, then $\rho_5 > \frac{1}{8}$ and every item has a bonus of at least $\alpha$. We get $4\alpha + \beta \geq \frac{7}{72}$. ∎

**Lemma 6.12** *If the conditions* $\min\{\alpha + \beta + 4\mu, \gamma + 5\mu\} \geq \frac{5}{96}$ *and* $\beta + 5\alpha \geq \frac{1}{14}$ *hold, then every heavy 6-bin has a weight of at least 1, and every standard 6-bin has weight of at least 1.*

**Proof.** The weight of a standard bin is at least $\frac{6}{7} \cdot \frac{13}{12} + 6\alpha \geq 1$, since $\frac{5}{96} \leq \beta + 5\alpha \leq 6\alpha$.

If a heavy bin has an item of bonus at least $\zeta$, then we are done. Otherwise, the largest item has a bonus of $\beta$ or $\gamma$. If the total size of items is in $(\frac{7}{8}, \frac{12}{13}]$, consider first the case that the largest item has a bonus of $\beta$. At least one additional item has a bonus of at least $\alpha$, so $\alpha + \beta + 4\mu \geq \frac{5}{96}$. Otherwise, we get $\gamma + 5\mu \geq \frac{5}{96}$.

If the total size of items is in $(\frac{6}{7}, \frac{7}{8}]$, then $\rho_7 > \frac{1}{7}$ and every item has a bonus of at least $\alpha$. We get $\frac{6}{7} \cdot \frac{13}{12} + 5\alpha + \beta \geq 1$, which gives $5\alpha + \beta \geq \frac{1}{14}$. ∎

**Lemma 6.13** *If* $\alpha + 6\mu \geq \frac{5}{96}$, *then every heavy 7-bin has a weight of at least 1, and every standard 7-bin has weight of at least 1.*

**Proof.** The weight of a standard bin is at least $\frac{7}{8} \cdot \frac{13}{12} + 7\alpha \geq \frac{91}{96} + \alpha + 6\mu \geq 1$.

A heavy bin has at least one item of bonus at least $\alpha$, and six items of bonus at least $\mu$. Thus $\alpha + 6\mu \geq \frac{5}{96}$. ∎

**Lemma 6.14** *If the condition $\mu \geq \frac{1}{216}$ holds, then every heavy $8^+$-bin has a weight of at least 1, and every standard $8^+$-bin has weight of at least 1.*

**Proof.** The weight of any such bin is at least $\frac{8}{9} \cdot \frac{13}{12} + 8\mu \geq 1$. ∎

**Lemma 6.15** *Consider a bin of OPT which does not contain an item of $F_1$. The total weight of items in this bin is at most $1.614$.*

**Proof.** Every item of size $x \leq \frac{1}{3}$ has $\omega(x) \leq 1.4 \cdot x$. Thus, if the bin contains no larger items then we are done.

Otherwise, the bin can contain one item of bonus $\xi$ or one item of bonus $\zeta$ or two items of bonus $\zeta$ or two items of bonuses $\zeta$ and $\xi$. The remaining space for other items in the four cases is at most $\frac{1}{2}$, $\frac{2}{3}$, $\frac{1}{3}$, and $\frac{1}{6}$, respectively. The supremum total weights in the four cases are therefore $\frac{1}{2} \cdot \frac{13}{12} + \xi + 1.4 \cdot \frac{1}{2} < 1.6$, $\frac{1}{3} \cdot \frac{13}{12} + \zeta + 1.4 \cdot \frac{2}{3} < 1.6$, $\frac{2}{3} \cdot \frac{13}{12} + 2\zeta + 1.4 \cdot \frac{1}{3} < 1.6$, and $\frac{5}{6} \cdot \frac{13}{12} + \xi + \zeta + 1.4 \cdot \frac{1}{6} < 1.614$. ∎

**Lemma 6.16** *Consider a bin of OPT which contains a regular item of $F_1$. The total weight of items in this bin is at most $1 + \frac{13}{24} + \Psi$, where $\Psi = \max\{\gamma + \beta, \gamma + \alpha + \mu, 2\beta + \alpha, \beta + \alpha + 2\mu, 3\alpha + \mu, \alpha + 4\mu\}$.*

**Proof.** Since the bin contains an item of $F_1$, its weight is 1 and the remaining space for other items is below $\frac{1}{2}$. The bin does not contain an item of $F_2$ with bonus $\zeta$. The remaining items of size in $(\frac{1}{3}, \frac{1}{2}]$ have bonuses of zero. Thus, we consider combinations of items of size in $(\frac{1}{12}, \frac{1}{3}]$. The total weight of items, neglecting bonuses, is at most $1 + \frac{1}{2} \cdot \frac{13}{12}$.

The possible worst case combinations of bonuses are as follows: $(\gamma, \beta)$, $(\gamma, \alpha, \mu)$, $(\beta, \beta, \alpha)$, $(\beta, \alpha, \mu, \mu)$, $(\alpha, \alpha, \alpha, \mu)$, $(\alpha, \mu, \mu, \mu, \mu)$. ∎

We use two sets of values for the parameters, one for the case that there are at least 12 standard 3-bins, and one for the case that there are at least 12 standard 4-bins. The values, found using Matlab, are as follows.

In the first case $\mu \approx 0.00562739467$, $\alpha \approx 0.0183189656$, $\beta \approx 0.02394636$, $\gamma = \frac{1}{16}$, and in the second case, $\mu \approx 0.0135621337$, $\alpha \approx 0.01597222$, $\beta = \frac{1}{30}$, $\gamma \approx 0.047534878$.

Using the last Lemma, we find an upper bound of $1.62811302699218$ on the SPO-PoA (in the second case the bound is lower, below $1.625$). ∎

For the lower bound we define an additional sequence as follows. Let $\nu_1 = 9$ and for $i \geq 2$, $\nu_{i+1} = \nu_i(\nu_i - 1) + 1$. We have that $\nu_{i+1} - 1$ is divisible by $\nu_i$. Since $\nu_i > 2$ for all $i$, $\nu_{i+1} > \nu_i + 1$ and the sequence is monotonically increasing. In fact $\nu_{i+1} > 2\nu_i$ for all $i$. The sequence satisfies $\sum_{i=1}^{\infty} \frac{1}{\nu_i} = \frac{1}{8}$, $\sum_{i=1}^{r} \frac{1}{\nu_i} = \frac{1}{8} - \frac{1}{\nu_{r+1}-1}$, and $\sum_{i=1}^{\infty} \frac{1}{\nu_i - 1} \approx 0.13908$. We let $\Gamma_\infty = \sum_{i=1}^{\infty} \frac{1}{\nu_i - 1}$.

**Theorem 6.17** *The SPO-PoA is at least $1.6167808$.*

**Proof.** Consider the following input. Let $M \geq 3$ be an integer, let $N$ be a large integer divisible by $\nu_M - 1$ (and thus it is divisible by $\nu_j - 1$ for $1 \leq j \leq M$), and let $\varepsilon > 0$ be a small constant such that $\varepsilon < \frac{1}{(M+2)\nu_M^2}$. Note that $\nu_2 - 1$ is divisible by 9 and by 8, and that for any $i \geq 1$, $\nu_{i+1} - 1$ is divisible by $\nu_i - 1$ and by $\nu_i$.

Let $\chi_1 = 1/2 + \varepsilon$, $\chi_2 = 1/4 + \varepsilon$, $\chi_3 = 1/4 - 2\varepsilon$, $\chi_4 = 1/8 + 4\varepsilon$, for any $5 \leq j \leq M$, $\chi_j = \frac{1}{\nu_{j-4}} + \varepsilon$ (that is, $\chi_5 = 1/9 + \varepsilon$, $\chi_6 = 1/73 + \varepsilon$ etc.). The numbers of items are as follows: $5N$ items of size $\chi_1$, $5N$ items of size $\chi_2$, $2N$ items of size $\chi_3$, and $3N$ items of size $\chi_j$, for $4 \leq j \leq M$. Since $\varepsilon < \frac{1}{1000}$, all items sizes are in $(0, 1)$.

An optimal solution has $5N$ bins in total. There are $2N$ bins, each with three items of size $\chi_1$, $\chi_2$, $\chi_3$ (whose total size is 1), and $3N$ bins with $M - 1$ items each, one of each size except for $\chi_3$. The total size of items in each one of the last $3N$ bins is $\frac{7}{8} + 6\varepsilon + \sum_{j=1}^{M-4} \frac{1}{\nu_j} + \varepsilon = \frac{7}{8} + (M+2)\varepsilon + \frac{1}{8} - \frac{1}{\nu_{M-3}-1} < 1 + \frac{1}{\nu_M} - \frac{1}{\nu_{M-3}} < 1$.

We consider the following packing with $M$ types of bins. For $1 \leq j \leq M-4$, there are $\frac{3N}{\nu_j-1}$ bins with $\nu_j - 1$ items of size $\chi_{j+4}$. In addition, there are $5N$ bins, each with one item of size $\chi_1$, $\frac{5N}{3}$ bins, each with three items of size $\chi_2$, and $N$ bins, each with two items of size $\chi_3$ and three items of size $\chi_4$. The total size of items in the first type of bins is $\frac{\nu_j-1}{\nu_j} + (\nu_j-1)\varepsilon \leq 1 - \frac{1}{\nu_j} + \frac{1}{\nu_M} < 1$. The total size of items in the last type of bins is $\frac{7}{8} + 8\varepsilon < 1$. This packing consists of $\frac{23N}{3} + 3N \sum_{j=1}^{M-4} \frac{1}{\nu_j-1}$ bins. As $M$ grows, the number of bins approaches $\frac{23N}{3} + 3N \cdot \Gamma_\infty$. The ratio between the costs approaches $1.6167808$.

It is left to show that this packing is a NE and strictly Pareto optimal. We prove by induction that for $j = M-4, M-3, \ldots, 1$, the items of the $(\nu_j - 1)$-bins cannot benefit from moving to another bin, and if there is an alternative packing where no item increases its cost, these items must be packed similarly in that packing (that is, $\nu_j - 1$ items of size $\chi_{j+4} = \frac{1}{\nu_j} + \varepsilon$ per bin, without any additional items).

Consider a specific value of $j \geq 1$. All bins with at least $\nu_j$ items have load above $\frac{\nu_j-1}{\nu_j}$, and therefore no item of size $\chi_{j+4}$ can move to such a bin. For the case $j > 1$, by the induction hypothesis, no smaller items can be combined in an alternative packing with the items of size $\chi_{j-4}$. If $j > 1$, The minimum sized items out of the items of sizes $\{\chi_1, \ldots, \chi_{j-1}\}$ have size $\frac{1}{\nu_{j-1}} + \varepsilon$. However, even replacing one item of size $\chi_j$ with an item of size $\chi_{j-1}$ results in load $(\nu_j - 2) \cdot (\frac{1}{\nu_j} + \varepsilon) + \frac{1}{\nu_{j-1}} + \varepsilon > \frac{(\nu_j-2)+2}{\nu_j} \geq 1$, where the strict inequality holds using $\nu_j \geq 2\nu_{j-1}$.

Consider $j = 1$. The items of size $\frac{1}{9} + \varepsilon$ cannot be combined in a bin with any smaller item. We show that they cannot be combined with an item of size $\chi_3$ or a larger item. Even if we replace one item of size $\chi_5$ with an item of size $\chi_3$, we get a load of $7(\frac{1}{9} + \varepsilon) + \frac{1}{4} - 2\varepsilon > 1$. An item of size $\chi_3$ must be packed with at least four items. If three items of size at least $\chi_3$ are packed in a bin, then it is not possible to add two items of size $\chi_4$ (since $3(\frac{1}{4} - 2\varepsilon) + 2(\frac{1}{8} + 4\varepsilon) > 1$). Recall that items of size $\chi_5$ or smaller items cannot be combined with items of size $\chi_3$, thus, every bin which has an item of size $\chi_3$ must contain at least three items of size $\chi_4$. Thus, the only way to comply with these requirements is to pack the items of size $\chi_3$ in pairs, each pair with three items of size $\chi_4$. This shows that in an alternative packing the items of sizes $\chi_3$, $\chi_4$, and $\chi_5$ must be packed in the same types of bins as in the packing above. Moreover, all bins except for 1-bins and 3-bins have load above $\frac{7}{8}$, so items of sizes $\chi_3$ and $\chi_4$ cannot deviate.

It is not difficult to see that items in 3-bins and 1-bins cannot deviate, and the items of size $\chi_2$ must be packed in triples. We find that any alternative packing has the same structure, and thus no item decreases its cost, that is, no alternative packing can exist. ∎

## 7  The exact convergence time for unit weights

In [26] it was shown that the number of steps for convergence is $O(n^2)$. In this section we find the exact worst case number of steps, which turns out to be $\Theta(n^{3/2})$. Note that [34] showed using methods from [20] that for the case of general weights (in fact, for proportional weights) the number of steps can be exponential.

**Theorem 7.1** *Let $n$ be an integer, consider the integers $i, j$ such that $0 \leq j \leq i-1$ and $n = i(i+1)/2 - j$. Clearly $i = \Theta(\sqrt{n})$. The maximum number of steps until convergence is at most $\frac{i(i+1)(i-1)}{3} + j - ij = \Theta(n^{\frac{3}{2}})$, and there exists an input of $n$ items where this bound can be achieved.*

**Proof.** We start with a lower bound. Consider a set of $n$ items, each of size $\frac{1}{n}$, and an initial packing where each one of the items is packed in its own bin. Let a *staircase packing* be a packing where for every $1 \leq \eta \leq i$, $\eta \neq j$, there is exactly one bin with $\eta$ items.

We show using induction on $i$ that there exists a sequence of exactly $i(i+1)(i-1)/6 - j(j-1)/2$ steps which results in a staircase packing. We first show it for the case $j = 0$ by induction on $i$. For $i = 1$, in every packing there is exactly one bin with one item, and this packing is a staircase packing. For a given value of $i$, $n = i(i+1)/2$. We consider a subset of $n' = n - i = i(i-1)/2$ items. By the induction hypothesis it is possible to obtain a packing such that for any $1 \leq \eta \leq i-1$ there is a bin with $\eta$ items. Considering the set of $n$ items, we get that for any $2 \leq \eta \leq i-1$ there is a bin with $\eta$ items, and there are $i+1$ bins with a single item. By the hypothesis, this packing is obtained in $i(i-1)(i-2)/6$ steps. Let $B_\eta$ denote a specific bin with $\eta$ items for $1 \leq \eta \leq i-1$, where the bin $B_1$ is chosen arbitrarily. The $i$ other items packed in dedicated bins are called *free items*. For the $k$-th free item ($1 \leq k \leq i-1$), it is moved from its bin, to the bins $B_1$, $B_2$, ..., $B_{i-k}$, in this order. $B_{i-k}$ will now contain $i-k+1$ items and will not be used again in this process. After all these steps $B_\eta$ (for $1 \leq \eta \leq i-1$) will contain $\eta + 1$ items. The $i$-th free item remains packed in its own bin, so as a result, for any $1 \leq \eta \leq i$ there is a bin with $\eta$ items. The number of additional steps for the free items is $\sum_{k=1}^{i-1}(i-k) = i(i-1)/2$. The total number of steps is $i(i-1)(i-2)/6 + i(i-1)/2 = i(i+1)(i-1)/6$.

To show the claim for the case for $j \neq 0$, we use the previous claim. Assume $n = i(i+1)/2 - j$ where $0 < j < i$. In this case create a staircase packing of a subset of $n' = i(i-1)/2$ items, leaving $i - j$ free items. The $k$-th free item ($1 \leq k \leq i - j$), is moved from its bin, to the bins $B_1$, $B_2$, ..., $B_{i-k}$, in this order. The bin $B_{i-k}$ will contain $i - k + 1$ items as a result and will not be used for later steps. After this is done for $i - j$ items, $B_\eta$ will contain $\eta + 1$ items for $j \leq \eta \leq i - 1$, and for $1 \leq \eta \leq j - 1$, $B_\eta$ still contains $\eta$ items. Thus, for every $1 \leq \eta \leq i$, $\eta \neq j$, there is exactly one bin with $\eta$ items and this is exactly a staircase packing as required. The number of additional steps is $\sum_{k=1}^{i-j}(i-k) = i(i-1)/2 - j(j-1)/2$. The total number of steps is $i(i+1)(i-1)/6 - j(j-1)/2$.

Once a staircase packing is achieved, we show that it is possible to reach a packing where all items are packed in one bin together using exactly $i(i+1)(i-1)/6 - ij + j(j+1)/2$ steps. We define a phase as follows. In the beginning of a phase there are bins with different numbers of items. Let $J = \{j_1 < j_2 < \cdots < j_{|J|}\}$ be the set of numbers of items before some phase, and let the bin $B_\eta$ for $\eta \in J$ be the bin with $\eta$ items. If $|J| > 1$, we repeatedly take an item from $B_{j_1}$, and move it to $B_{j_2}$ then to $B_{j_3}$ and so forth until it reaches $B_{j_{|J|}}$. A phase ends when all items of $B_{j_1}$ were moved. If $j = 0$ then initially $J = \{1, \ldots, i\}$, there are $i - 1$ phases, and the number of steps is $\sum_{k=1}^{i-1} k(i-k) = i^2(i-1)/2 - (i-1)i(2i-1)/6 = i(i-1)(i+1)/6$. Otherwise, initially $J = \{1, \ldots, i\} - \{j\}$, there are $i - 2$ phases, and the number of steps is $\sum_{k=1}^{j-1} k(i-1-k) + \sum_{k=j+1}^{i-1} k(i-k) = \sum_{k=1}^{i-1} k(i-k) - \sum_{k=1}^{j-1} k - j(i-j) = i^2(i-1)/2 - i(i-1)(2i-1)/6 - j(j-1)/2 - j(i-j) = i(i+1)(i-1)/6 - ij + j/2 + j^2/2$. The total number of steps is therefore

$$\frac{i(i+1)(i-1)}{6} - \frac{j(j-1)}{2} + \frac{i(i-1)(i+1)}{6} + \frac{j}{2} - ij + \frac{j^2}{2} = \frac{i(i+1)(i-1)}{3} + j - ij \ .$$

For the upper bound, consider an input $I$ of $n = i(i+1)/2 - j$ items for $0 \leq j \leq i-1$, an initial configuration and a sequence of moves. Let $p_{\min}$ denote the smallest item size in $I$. Let $\varepsilon = \min\{p_{\min}, 1/n\}$, and let $I'$ be the input where $s_i = \varepsilon$ for $1 \leq i \leq n$. For the input $I'$ there cannot be invalid moves, since all items can be packed into one bin.

**Claim 7.2** *The initial configuration and the sequence of moves of $I$ are valid for $I'$ as well.*

**Proof.** Since no item was increased, all configurations of $I$ are valid for $I'$. Since the cost of an item in a packing depends only on numbers of items in its bin and not on their sizes, modifying the sizes may only increase sets of beneficial deviations, that is, every move which was beneficial and possible for $I$ remains such for $I'$ and the sequence of moves is still valid. ∎

In what follows, we will consider only sequences of moves for $I'$. In particular, we consider only sequences with a maximum number of moves. Such a sequence must exist since from the results of [26] every sequence of moves has finite length.

**Claim 7.3** *Every sequence with a maximum number of moves starts with the configuration where every item is packed in a separate bin, and ends with the configuration that all items are packed in one bin.*

**Proof.** Consider a sequence of $\ell$ moves. Assume that there is a bin $B$ with $k \geq 2$ items in the initial configuration, and let $\phi \in B$. Modify the configuration such that instead of $B$ the starting configuration has the two bins $B \setminus \{\phi\}$ and $\{\phi\}$ (other bins remain unchanged). Next, add a step in the beginning of the sequence of moves where $\phi$ moves to join the items of $B \setminus \{\phi\}$. This is an improving step since $\phi$ reduces its cost from 1 to $\frac{1}{k}$. This results in a sequence of $\ell + 1$ steps, which contradicts maximality.

Next, assume that after the sequence of moves there are at least two non-empty bins, containing $k_1$ and $k_2$ items respectively, where $k_1 \leq k_2$. Let $\psi$ be an item packed in the first bin. Add a move of $\psi$ to the second bin in the end of the sequence. This is an improving step since $\psi$ reduces its cost from $\frac{1}{k_1}$ to $\frac{1}{k_2+1} \leq \frac{1}{k_1+1} < \frac{1}{k_1}$. This results in a sequence of $\ell + 1$ steps, which contradicts maximality. ∎

Let $k > 0$ be an integer. We define a *level $k$ small step* to be a move where an item moves from a $k$-bin to another $k$-bin. A step is called a *small step* if there is an integer $k$ such that the step is a level $k$ small step. Given the set of sequences of steps of maximum length we focus on sequences where the prefix of small steps has maximum length.

**Claim 7.4** *Assume that after a prefix of the sequence of steps is applied there are at least two $k$-bins. Then there is at least one level $k$ small step in the remainder of the sequence.*

**Proof.** Assume by contradiction that there is no level $k$ small step in the remaining part of the sequence. Since the sequence of steps terminates only when all items are packed in one bin, there is at least one item in the union of the $k$-bins which will perform a move (in fact, all the items of all the $k$-bins except for one such bin will do that). Consider the first step after the current configuration was reached which involves one a $k$-bin (either an item moving to the bin or moving out of it).

There are two possible moves. If an item $\psi$ moves from a $k$-bin into a bin with $k' > k$ items, we modify the sequence as follows. First $\psi$ moves to the another $k$-bin, and then it moves to the bin with $k'$ items. The second step is still beneficial for $\psi$ since in the second step it moves from a $(k+1)$-bin to a bin with $k' \geq k+1$ items. This modification augments the length of the sequence by 1, which contradicts maximality.

If an item $\phi$ moves from a bin with $\tilde{k} < k$ items to one of the $k$-bins, we modify the sequence as follows. First choose an arbitrary item from one of $k$-the bins and move it to another $k$-bin. Then, move $\phi$ to the bin out of which an item was just moved (which currently has $k-1$ items). This last move is beneficial since $\tilde{k} \leq k - 1$. This modification augments the length of the sequence by 1, which contradicts maximality. ∎

**Claim 7.5** *Consider the prefix of small steps. After this prefix is performed, every bin has a different number of items.*

**Proof.** Assume by contradiction that at this time there are two $k$-bins. Using Claim 7.4, there will be a level $k$ small step later in the sequence, which will be the first move which involves $k$-bins. Since all items are identical, it is possible to perform such a step immediately instead of at a later time. This does not change the number of steps in the sequence, and it increases the length of the prefix of small steps, which contradicts maximality of the prefix. ∎

**Claim 7.6** *Consider the prefix of small steps. After this prefix is performed, there is one bin of each number of items in $\{1, 2, \ldots, i\} \setminus \{j\}$, that is, a staircase packing is created.*

**Proof.** We prove an invariant which is kept as long as only small steps are done. Let $b_k$ be the number of bins with $k$ items, and recall that initially $b_1 = n$ and $b_\ell = 0$ for $0 < \ell \leq n$. Assume that at a given time, $k_m$ is the maximum integer such that $b_{k_m} > 0$. We say that a number $1 \leq k \leq k_m - 1$ is bad if $b_k = 0$, and otherwise it is good. That is, a number $k$ is bad if there are no $k$-bins, but there exists at least one $(k+1)^+$-bin. If $b_k \geq 2$ then we say that $k$ is *very good*. Two bad numbers are called *consecutive* bad numbers if all numbers between them are good, that is, if $k_1$ and $k_2$ such that $k_1 < k_2 < k_m$ are both bad ($b_{k_1} = 0$ and $b_{k_2} = 0$), and for all $k'$ such that $k_1 < k' < k_2$, $b_{k'} > 0$.

The invariant is as follows. For every pair of bad consecutive numbers $1 \leq k_1 < k_2 < k_m$, there exists a number $k_1 < \tilde{k} < k_2$ such $\tilde{k}$ is very good. Recall that we only consider small steps and consider the change resulting from a single level $k$ small step. Every level $k$ small step implies that before this step there are at least two $k$-bins and so $k$ is very good. Note that if $k = k_m$ then the value $k_m$ increases by 1, and the only number which can become bad as a result of the step is $k$.

Assume first that $k$ remains very good. No bad numbers are created, and since no bin stops beings very good then the invariant holds (even if some bin stops being bad). If $k$ remains good, but not very good, then still no new bad numbers are created and we only need to consider the case that $k$ was the only very good number between two consecutive bad numbers. Let these two numbers be $k_1 < k < k_2$. If $k_2 > k+1$ and $k_1 < k-1$, then the numbers of $k_1$-bins and $k_2$-bins are unchanged and the numbers $k_1, k_2$ remain consecutive bad numbers between which we need to show that a very good number exists after the step. Since $k+1$ was good, as a result of the move $b_{k+1} \geq 2$, and since $k_1 < k+1 < k_2$, there is a very good number between $k_1$ and $k_2$, as required. If $k_1 = k-1$ but $k_2 > k+1$ then $k_1$ becomes good. If $k_1$ was the minimum bad number then we are done. Otherwise, let $k_3 < k_1$ be a bad number such that $k_3$ and $k_1$ were consecutive bad numbers. We now have that $k_3$ and $k_2$ are consecutive bad numbers and $b_{k+1} \geq 2$ so $k+1$ is a very good number between them. If $k_1 < k-1$ but $k_2 = k+1$ then $k_2$ becomes good. If $k_2$ was the maximum bad number then we are done. Otherwise, let $k_4 > k_2$ be a bad number such that $k_2$ and $k_4$ were consecutive bad numbers. We now have that $k_3$ and $k_4$ are consecutive bad numbers and $b_{k-1} \geq 2$ so $k-1$ is a very good number between them. Finally, if both $k_1 = k-1$ and $k_2 = k+1$ hold, then the only case of interest is when $k_1$ was not the minimum bad number and $k_2$ was not the maximum bad number. We let $k_3 < k_1$ be a bad number such that $k_3$ and $k_1$ were consecutive bad numbers, and let $k_4 > k_2$ be a bad number such that $k_4$ and $k_2$ were consecutive bad numbers. Now $k_3$ and $k_4$ are consecutive bad numbers. There is a very good number in $(k_3, k_1)$ which is now a very good number between $k_3$ and $k_4$.

Finally, we consider the case where $k$ becomes bad. If there previously was a bad number $k_2$ such that $k_2 > k$, we distinguish two cases. If $k_2 > k+1$, then $k$ and $k_2$ becomes a consecutive bad pair of numbers, and $k+1$ becomes a very good number between them. Otherwise, $k_2 = k+1$ becomes good. If $k_2$ was the maximum bad number then we are done, and otherwise, let $k_4 > k_2$ be such that $k_2$ and $k_4$ were consecutive bad numbers. Instead, $k$ and $k_4$ are now consecutive bad numbers, and the very good number between them is the same one which was very good between $k_2$ and $k_4$. The proof is symmetric for the case that previously was a bad number $k_1$ such that $k_1 < k$.

To complete the proof, consider the configuration after the prefix of small steps. Since every bin has a different number of items, there are no very good numbers, there is at most one bad number. If there exists a bin with at least $i+1$ items, and there is just one bad number, then there are at least $(i+1)(i+2)/2 - i = i(i+1)/2 + 1 > n$ items. If there is no bin with $i$ items, then there are at most $i(i-1)/2 < n$ items. Thus, there is a bin with $i$ items, and since there is at most one bad number, the bad number must be $j$ if $j \neq 0$, and otherwise there is no bad number. Therefore, the packing at this time is a staircase packing. ∎

**Claim 7.7** *The number of steps in the prefix of small steps is at most $i(i+1)(i-1)/6 - j(j-1)/2$.*

**Proof.** We use the potential function as in [26] which is the sum of squares of number of items in the bins. In the beginning every item is in a dedicated bin, so the potential is equal to $n = i(i+1)/2 - j$. Consider a level $k$ small step. The potential function increases by exactly 2 in this step, since the only change is that instead of two $k$-bins, there is a $(k-1)$-bin a $(k+1)$-bin, and the increase in the potential is exactly $(k+1)^2 + (k-1)^2 - 2k^2 = 2$.

Since a staircase packing is achieved in the prefix of small steps, the value of the potential after this prefix is $\sum_{k=1}^{i} k^2 - j^2 = i(i+1)(2i+1)/6 - j^2$. Thus, the number of steps cannot exceed half the difference between the final potential and the initial potential, which is $(i(i+1)(2i+1)/6 - j^2 - (i(i+1)/2 - j))/2 = i(i+1)(i-1)/3 - j(j-1))/2 = i(i+1)(i-1)/6 - j(j-1)/2$. ∎

**Claim 7.8** *The number of steps in the remainder of the sequence after the prefix of small steps is at most $i(i+1)(i-1)/6 - ij + j(j+1)/2$.*

**Proof.** In this case we define a different potential function. Sort the bins in non-increasing order according to numbers of items. Let the index of an item be the index of the bin into which it is packed. The potential of a packing is sum of indices of items.

The potential is clearly positive at all times. The final potential is $n$, since all items are packed in one bin. Consider a step in which an item moves from a $k_1$-bin $B_v$ to a $k_2$-bin $B_u$ (where $k_2 \geq k_1$). Since all items are identical, we assume that this is the $k_1$-bin of maximum index, and the $k_2$ bin of minimum index. This holds even if $k_1 = k_2$, since in this case there are at least two bins with this number of items. Since the bins are sorted by non-increasing order according to numbers of items we have $v > u$. As a result of the move, $B_u$ now has $k_1 - 1$ items, and $B_v$ now has $k_2 + 1$ items. By definition, if $v > 1$ then $B_{v-1}$ has at least $k_2 + 1$ items. Similarly, if $B_{u+1}$ exists then it has at most $k_1 - 1$ items, so the sorted order is still valid. The change in the potential in this step is the change in the index of the bin of the moving item, which is $v - u \geq 1$.

If $j = 0$, then the potential before the remainder of the sequence of moves is performed is $\sum_{k=1}^{i} k(i-k+1) = (i+1)i(i+1)/2 - i(i+1)(2i+1)/6 = i(i+1)(i+2)/6$ while $n = i(i+1)/2$, so the number of steps is at most $i(i+1)(i+2)/6 - i(i+1)/2 = i(i+1)(i-1)/6$.

If $j > 0$, then the potential before the remainder of the sequence of moves is performed is $\sum_{k=1}^{i-j} k(i-k+1) + \sum_{k=i-j+1}^{i-1} k(i-k) = i \sum_{k=1}^{i-1} k + \sum_{k=1}^{i-j} k - \sum_{k=1}^{i-1} k^2 = i^2(i-1)/2 + (i-j)(i-j+1)/2 - i(i-1)(2i-1)/6 = i(i-1)(i+1)/6 + i^2/2 + j^2/2 - ij + i/2 - j/2$.

In each step the function decreases by at least 1, so the number of steps is at most $i(i-1)(i+1)/6 + i^2/2 + j^2/2 - ij + i/2 - j/2 - i(i+1)/2 + j = i(i+1)(i-1)/6 - ij + j/2 + j^2/2$. ∎

Summing up the maximum number of steps in the prefix and in the remainder we get $i(i+1)(i-1)/6 - j(j-1)/2 + i(i+1)(i-1)/6 - ij + j(j+1)/2 = i(i+1)(i-1)/3 + j - ij$. ∎

## 8  Concluding remarks

In this paper we have studied the (asymptotic) PoA and PoS for several types of equilibria, and have provided a fairly complete analysis of these measures and their interrelations. In addition to the game-theoretical applications, another motivation for the study of equilibria is the role of equilibria as fixed points in local search based algorithms.

Following the traditional analysis of bin packing with respect to asymptotic approximation we used these measures as well. This type of study allows us to find the relation between outputs of algorithms and equilibria. Alternatively, it is possible to consider the absolute PoA and PoS. All the lower bounds presented in this work are valid for this case, and since there exists an optimal solution which is NE, WPO-NE, and SPO-NE, the absolute PoS, WPO-PoS, and SPO-PoS, are all equal to 1. However, the situation regarding the absolute SPoS is different, as it is possible to show a lower bound of 1.6119 (by modifying the lower bound shown above for small inputs), implying that the absolute SPoS is strictly higher than the SPoS.

For the case of general weights, given the relation between FF and the PoA, if it turns out that 1.7 is the absolute approximation ratio of FF, then this will be a tight bound for the absolute PoA, since a lower bound of 1.7 on the absolute PoA was given in [26]. That example is in fact a SNE (though not the best one for the given instance) and a WPO-NE, but it is not a SPO-NE.

We would like to note that SPO-NE and WPO-NE were not studied in the past for the case of proportional weights. Due to Proposition 2.4, the WPO-PoA is equal to the PoA, and since the WPO-PoS and SPO-PoS are equal to 1 for any set of weights, this holds for proportional weights as well. Additionally, it is possible to show that the SPO-PoA is no smaller than the SPoA; in [17] it is shown that every input can be modified so that it has a unique SNE. Once the SNE is unique, GSC has a unique output and in particular, the most loaded bin must exist in any strictly Pareto optimal packing. Using induction, the unique SNE is the unique SPO-NE as well. Thus, the situation for proportional weights is different from the case of unit weights where the SPO-PoA is much lower than the SPoA.